\documentclass[aps,epsf,preprint]{revtex4}
\usepackage{latexsym}
\usepackage{amsfonts}
\usepackage{graphicx} 
\usepackage{epsfig}
\usepackage{caption}
\usepackage{color}
\usepackage{color}

\begin{document}

\title{Physics of the Jagla Model as the Liquid-Liquid Coexistence Line
  Approaches Horizontal}

\author{Jiayuan Luo,$^1$ Limei Xu,$^2$ C. Austen Angell,$^3$
  H. Eugene Stanley,$^1$ and Sergey V. Buldyrev$^{1,4}$}

\bigskip
\bigskip

\affiliation{$^1$Center for Polymer Studies and Department of Physics, Boston
  University, Boston, MA 02215 USA
  \\$^2$International Center for Quantum Materials, Peking University,
  Beijing 100871, China 
  \\$^3$Department of Chemistry and Biochemistry, Arizona State
  University, Tempe, AZ 85287  
  \\$^4$Department of Physics, Yeshiva University, 500 West 185th Street,
  New York, NY 10033 USA}
  
\date{26 April 2012 --- lxasb26apr.tex} 


\begin{abstract}

  The slope of the coexistence line of the liquid-liquid phase
  transition (LLPT) can be positive, negative, or zero. All three
  possibilities have been found in Monte-Carlo simulations of a
  modified spherically symmetric two-scale Jagla model. Since the
  liquid-liquid critical point (LLCP) frequently lies in a region of
  the phase diagram that is difficult to access experimentally, it is
  of great interest to study critical phenomena in the supercritical
  region. We therefore study the properties of the Widom line, which
  is defined in the one-phase region above the critical point as the
  locus of maximum correlation length as function of the ordering
  field at constant thermal field.  Asymptotically close to the
  critical point, the Widom line coincides with the loci of the
  response function extrema, because all response functions can be
  asymptotically expressed as functions of the diverging correlation
  length. We find that the method of identifying the Widom line as the
  loci of heat capacity maxima becomes unfruitful when the slope of
  the coexistence line approaches zero in the $T$-$P$ plane. In this
  case the specific heat displays no maximum in the one-phase region
  because for a horizontal phase coexistence line, according to the
  Clapeyron equation, the enthalpy difference between the coexisting
  phases is zero, and thus there can be no contribution to enthalpy
  fluctuations from the critical fluctuations. The extension of the
  coexistence line beyond the critical point into the one-phase region
  must in this case be performed using density fluctuations; the line
  of compressibility maxima remains well defined, though it bifurcates
  into a symmetrical pair of lines.  These findings agree well with
  the linear scaling theory of the LLCP by Anisimov and collaborators.

\end{abstract}

\maketitle 

\section{introduction}

The liquid-liquid phase transition (LLPT), defined as a transition
between two liquid states of different densities, called low density
liquid (LDL) and high density liquid (HDL), has received considerable
attention not only due to its rarity in nature, but also due to its
importance in our fundamental understanding of the liquid state of
matter
\cite{poole1,poole2,XuPNAS2005,Mishima98xx,FranzeseNature2001,Brazhkin_book_2002,Paschek05,YamadaXX,XuJPCB2010}. The
LLPT was observed in many systems, such as elemental
\cite{KatayamaNature2000, BhatNature2007}, ionic
\cite{ShengNatureMaterials2007}, molecular \cite{KuritaScience2004}, and
covalent \cite{SenPRL2006} liquids.  In some cases, the LLPT can
terminate at a liquid-liquid critical point (LLCP).  Systems such as
liquid water, silicon, silica, and germanium, possess analogous
thermodynamic and dynamic anomalies
\cite{AngellJPC73,SpeedyJCP76,LiuPRL2005,KatayamaNature2000,zhenyuPablo1,XuPNAS2005,XuNP2009,OguniJCP1983}.
The critical phenomena near the LLCP are of crucial importance for the
understanding of the anomalous properties in these systems
\cite{Mishima98xx,XuPNAS2005,PooleJPCM2005,LiuPRL2005}. However the
detection of the LLCP or the LLPT can be difficult due to the fact that
in many cases, the LLCP is deeply  buried in the supercooled region,
where crystallization may occur before we reach the
LLCP~\cite{Mishima98xx}. For example, in the case of water, it has been
hypothesized that the putative LLCP is
 the cause of water anomalies
\cite{sergeyReview,BuldyrevPhysicaA2003,XuPNAS2005}, but the existence
of a LLCP for bulk water in the deeply supercooled region has not been
directly verified by experiments due to crystallization, even though
indications of the existence of the LLCP have been found both in
pressure induced melting experiments \cite{Mishima98x} and in
nanoconfined water \cite{LiuPRL2005}.

According to scaling theory, asymptotically near the critical point
all response functions can be expressed in terms of the correlation
length \cite{StanleyOxford71}. Different response functions diverge at
the critical point, and display maxima in the one-phase region along
constant pressure $P$ paths or constant temperature $T$
paths \cite{Sengers1985,XuPNAS2005,PooleJPCM2005}. The loci of
different response function maxima in the $T$-$P$ plane are different, but
they converge in the vicinity of the critical point to a single line,
called the Widom line \cite{PooleJPCM2005,XuPNAS2005, XuPRE2006}. 
Theoretically, Widom line is defined as the locus of 
maximum correlation length $\xi$ as function of ordering field at 
constant thermal field in the one-phase region.
Approaching the critical point, the
magnitude of the response functions increases, and becomes infinite at
the critical point. This fact provide a new way of locating the
critical point: instead of locating the critical point through the
coexistence line below the critical temperature $T_c$, we may locate
the critical point through the Widom line in the one-phase region from
the higher temperature side by studying the behavior of the loci of 
response function maxima~\cite{XuPNAS2005}. Thus it is important 
to find a general model
system with an accessible LLCP which would permit detailed examination
of the response functions in the vicinity of the LLCP.

The Jagla model of liquids is a simplified model consisting of
particles interacting via a spherically symmetric two-scale potential
with both repulsive and attractive ramps
\cite{Jagla99,XuPNAS2005,zhenyuPablo1,Kumar05}. With
a special choice of parameters, the Jagla model has an accessible LLCP
{\it above\/} the melting line \cite{XuPNAS2005}, allowing us to
explore the behavior near the LLCP in equilibrium liquid states. In
this case, the coexistence line between LDL and HDL is positively
sloped, which means that when cooled down along the same isobar, the
system changes from LDL to HDL. This behavior is opposite to that of
water, where experiments \cite{LiuPRL2005} and simulations
\cite{PooleJPCM2005} show that the coexistence line might be
negatively sloped, and an isobaric cooling path transforms the system
from HDL to LDL \cite{XuPNAS2005}.

Gibson and Wilding found that by changing the parameters of the Jagla
potential it is possible to reduce the slope of the coexistence line to
zero \cite{gibsonWilding}. In this paper we use modified Jagla models to
investigate the behavior of the Widom line as the slope of the
coexistence line changes from positive to horizontal. In Sec.~II we
introduce the modified Jagla model and the simulation method. In
Sec.~III we present our simulation results.  In Sec.~IV we compare our
simulation results to the linear scaling theory of the critical
point. In Sec.~V we further investigate the relationship between the
LLCP, the Widom line, and the glass transition for systems with
different coexistence line slopes. In Sec.~IV we summarize our study.

\section{model}

\begin{figure}[htbp]
\begin{minipage}[t]{0.6\linewidth}
\includegraphics[width=0.9\linewidth]{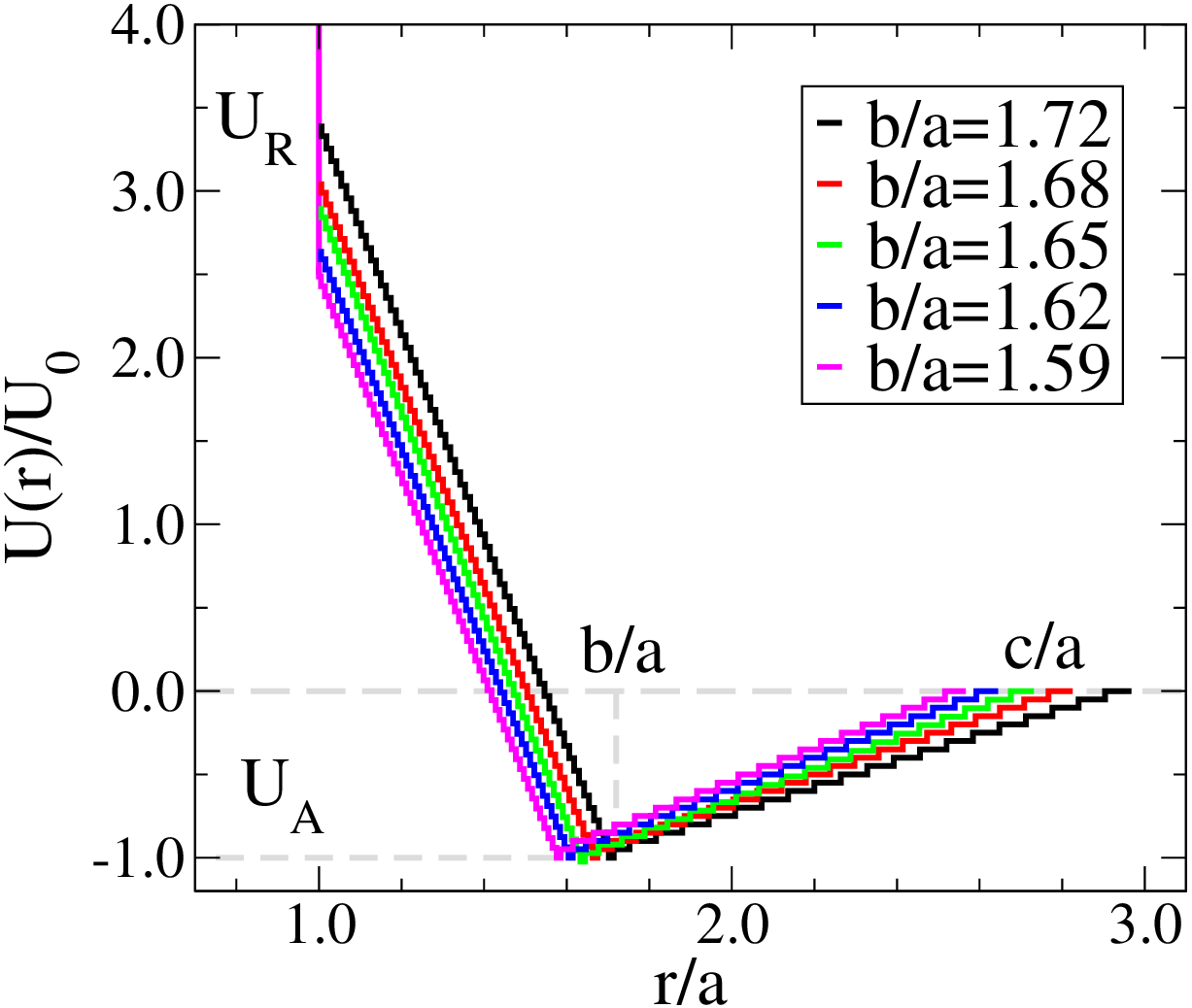}
\end{minipage}
\captionsetup{format=plain,justification=RaggedRight}
\caption{Family of modified spherically symmetric two-scale Jagla ramp potentials [Tab.~\ref{TableModel}].
 The two length scales of Jagla potential are the hard-core distance $r=a$ and soft-core distance
  $r=b$. The long range cutoff is $r=c$. 
  We keep the potential minimum $U_A$ constant, and its depth serves as the unit of energy $U_0=|U_A|$,
  while the hard-core potential $U_R$ varies.
  This convention is different from Ref. \cite{gibsonWilding}, in which $U_R$ is kept constant $U_R=0.69U_0$. 
  The discretized versions of the modified Jagla potential from $b/a=1.72$ to $1.59$ are shown. We use
  discretization steps, $n_1=60$ for $a \leq r<b$ and $n_2=20$ for $b \leq r<c$. 
%  Our choice of $b/a=1.72$ is identical to the discretized Jagla model studied by Xu et al. \cite{XuPNAS2005}, except the discretization steps are changed.
 }
\label{Model}
\end{figure}

Here we study the two length-scale Jagla model with both repulsive and
attractive ramps \cite{Jagla99}. In this model,
particles interact with a spherically symmetric pair potential
\begin{equation}
  U(r) = \left\{
\begin{array}{cc}
\infty &\quad \textrm{$r < a$}  \\
U_A+(U_R-U_A)(b-r)/(b-a) &\quad \textrm{$a \leq r < b$}\\
U_A(c-r)/(c-b) &\quad \textrm{$b \leq r < c$}\\
0 &\quad \textrm{$r \geq c$}
\end{array}
\right.
\label{UJagla}
\end{equation}
where $a$ is the hardcore distance, $b$ is the soft-core distance, and
$c$ is the long-distance cutoff [Fig.~\ref{Model}]; $U_A=-U_0$ is the
minimal potential energy reached at soft-core distance $r=b$, and $U_R$
is the potential energy at the top of the repulsive ramp at hardcore
distance $r=a$.

We implement a family of Jagla potentials with different parameters,
simultaneously decreasing $b$ and $c$---essentially following the
Gibson-Wilding procedure \cite{gibsonWilding}, the only difference being that
we keep $U_A$ constant. The parameters of different models are presented
in Table~\ref{TableModel}.

We perform discrete molecular dynamics (DMD) simulations by discretizing
the ramp into a series of step functions. The discrete Jagla potentials
are
\begin{equation}
  U_k(r) = \left\{
\begin{array}{cc}
\infty &\quad \textrm{$r < a$}  \\
U_R &\quad \textrm{$a \leq r < a+ \frac{1}{2}\Delta r_1$} \\
U_R-k\Delta U_1 &\quad 
     \textrm{$a+(k-\frac{1}{2})\Delta r_1 \leq r <
       a+(k+\frac{1}{2})\Delta r_1, \ \ \  
             1 \le k \le n_1-1$}\\
U_A    &\quad \textrm{$b-\frac{1}{2} \Delta r_1 \leq r < b+\frac{1}{2}
  \Delta r_2$}\\ 
U_A+k\Delta U_2 &\quad \textrm{$b+(k-\frac{1}{2})\Delta r_2 \leq
  r<b+(k+\frac{1}{2})\Delta r_2, \ \ \ 1 \le k \le n_2-1 $}\\ 
0 &\quad \textrm{$r \geq c-\frac{1}{2}\Delta r_2$}
\end{array}
\right.
\label{UDMD}
\end{equation}
where $n_1=60$ and $n_2=20$, $\Delta r_1=(b-a)/n_1$, $\Delta
U_1=(U_R-U_A)/n_1$, and $\Delta r_2=(c-b)/n_2$, $\Delta U_2=U_0/n_2$.

We use $a$ as the unit of length, particle mass $m$ as the unit of mass,
and $U_0$ as the unit of energy.  The simulation time $t$ is measured in
units of $a\sqrt{m/U_0}$, temperature $T$ in units of $U_0/k_B$,
pressure $P$ in units of $U_0/a^{3}$, density $\rho\equiv N/L^{3}$ in
units of $a^{-3}$, isothermal compressibility $K_T$ in units of
$a^3/U_0$, and isobaric specific heat $C_P$ in units of $k_B$.

Our results are based on simulations of a liquid system of $N=1728$
molecules with periodic boundary conditions. Constant volume-temperature
(NVT) and constant pressure-temperature (NPT) simulations are
implemented in this study.

The temperature of the system is controlled by rescaling the velocities
of all particles in the NVT simulations so that the average kinetic
energy per particle approaches the desired value $3K_BT_0/2$, where $T_0$
is the temperature of the thermostat,
\begin{equation}
T'= \bar{T}(1 - \kappa_T \tau_t) + T_0 \kappa_T \tau_t,
\end{equation}
where $\kappa_T=0.2 \ [\sqrt{m/U_0}/a]$ is a heat exchange coefficient,
$\tau _t$ is the time interval between two successive rescaling, $T'$ is
the new temperature, and $\bar{T}$ is the average temperature during the
time interval $\tau_t$. We select $\tau_t$ as the time during which
$N$ collisions between particles occur.

For the NPT simulations, the Berendsen barostat algorithm rescales the
coordinates $\vec{r_j}$ and box dimensions $\vec{L}$ after each $\tau_p$
unit of time,
\begin{equation}
r_j' = r_j + r_j \kappa_p (\bar{P} - P_0)
\end{equation}
\begin{equation}
\vec{L} = \vec{L} + \vec{L} \kappa_p (\bar{P} - P_0),
\end{equation}
where $P_0$ is the desired pressure, $\bar{P}$ is the average pressure
during time interval $\tau_p=1000\tau_t$, and $\kappa_p=0.02
[a^{3}/U_0]$ is the rescaling coefficient.

We perform cooling or heating simulations at a constant cooling/heating
rate, $q \equiv \Delta T/ \Delta t$, where the $T$ decreases/increases
by $\Delta T$ over time $\Delta t$. We measure $q$ in units of
$q_0=\sqrt{U_0^3/(ma^2k_B^2)}$.

\section{Simulation Results}

\subsection{Coexistence line}

\begin{figure}[htbp]
\begin{minipage}[t]{0.6\linewidth}
\includegraphics[width=0.9\linewidth]{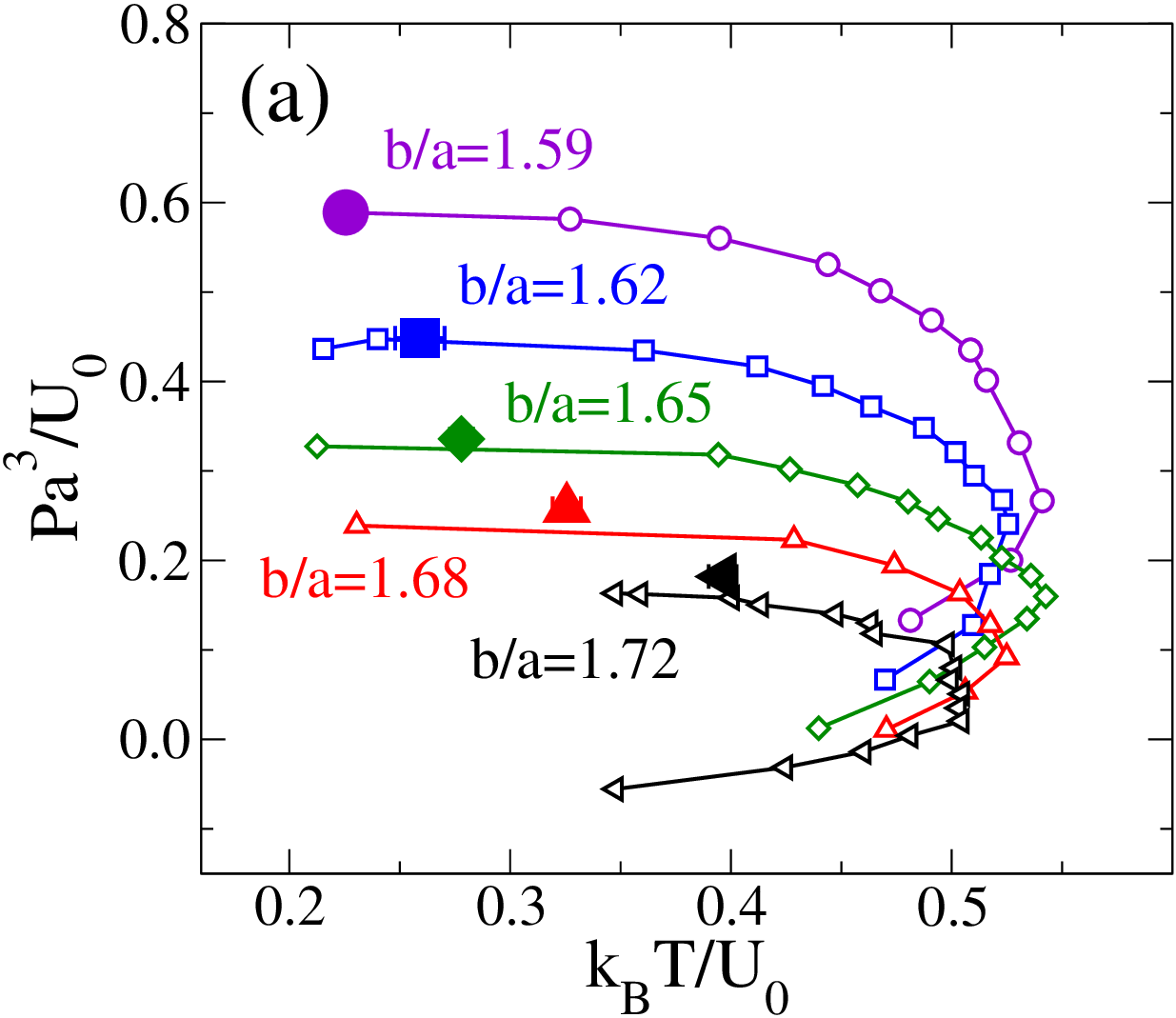}
\end{minipage}
\begin{minipage}[t]{0.6\linewidth}
\includegraphics[width=0.9\linewidth]{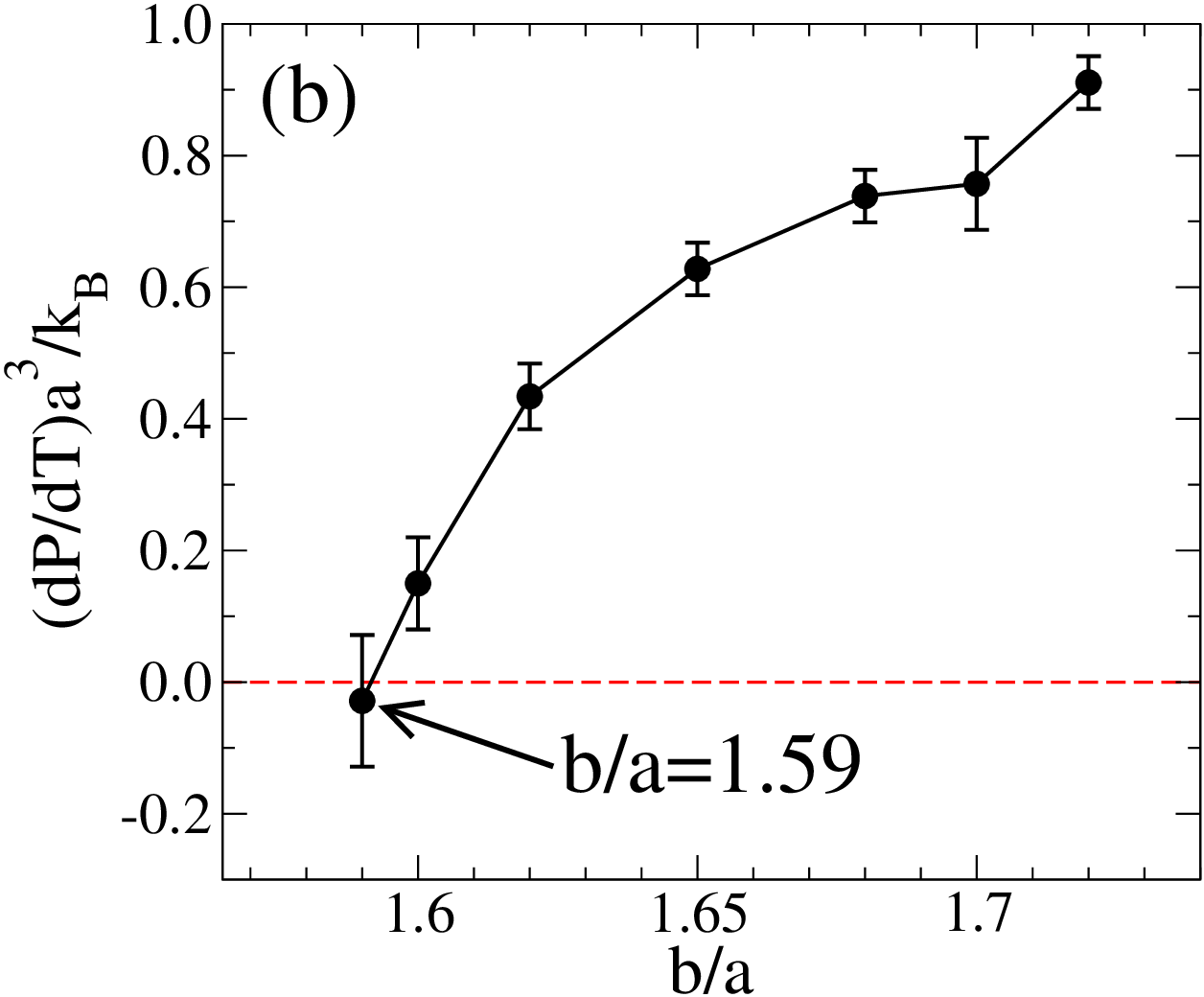}
\end{minipage}
\captionsetup{format=plain,justification=RaggedRight}
\caption{The LLCP, temperature of maximum density (TMD) line and the slope of the LDL-HDL coexistence line for a selection of modified Jagla potentials.  
  (a) The critical point (solid symbols with error bar approximately the symbol size) and the TMD line (open symbols) for systems with $b/a=1.72, 1.68, 1.65, 1.62, 1.59$. 
  One can see that the LLCP monotonically shifts to higher pressure
  and lower temperature as $b/a$
  decreases. The density anomaly region bounded by the TMD line, 
  expands in the $T$-$P$ diagram with decreasing
  $b/a$, and the LLCP moves from above the TMD line towards the density anomaly region and 
  for $b/a=1.59$, the LLCP locates right on top of the TMD line.  
  (b) The slope of the LDL-HDL coexistence lines decreases as the $b/a$ value
  decreases. When $b/a$ approaches $1.59$, the slope decreases to
  zero.}
\label{Co_slope}
\end{figure}

We first explore the phase diagram of each model with different $b/a$
via slow cooling using constant volume simulations. The LLCP corresponds
to the highest temperature crossing of isochores in the $T$-$P$ phase
diagram. The temperature of maximum density (TMD) line is the locus of
state points at which pressure reaches a minimum along each isochore as
a function of $T$ \cite{XuPRE2006}. Figure~\ref{Co_slope}(a) shows that
the LLCP monotonically shifts to lower temperature and higher pressure
as $b/a$ decreases. The region of the density anomaly (the region
bounded by the TMD line) expands and shifts together with the LLCP to
higher pressures as $b/a$ decreases. This behavior coincides with the
behavior reported in Ref.~\cite{gibsonWilding}. (The numerical differences in
$P$ and $T$ arise from the fact that we define the unit of energy as
$U_0\equiv -U_A$, while Ref.~\cite{gibsonWilding} uses $U_0\equiv U_R/0.69$.)
When $b/a<1.59$ the system crystallizes spontaneously near the LLCP
within a short simulation time, and we are not able to obtain the LLCP
and coexistence line.

We calculate the slope of the LLPT coexistence line $dP/dT$ for
systems with different $b/a$ [Fig.~\ref{Co_slope}(b)] by the Maxwell
construction. One can see that the slope decreases from positive to
approximately zero as $b/a$ decreases to 1.59, in agreement with
Ref.~\cite{gibsonWilding}. For large $b/a$, the LLCP lies clearly
above the density anomaly region bounded by the TMD line,
corresponding to the case of $dP/dT>0$, while for $b/a=1.59$, the LLCP
lies on the TMD line corresponding to the case of $dP/dT=0$
[Fig.~\ref{Co_slope}(a)].  Theoretically, if $dP/dT<0$, the LLCP
should be inside the density anomaly region~\cite{SastryPRE2006}, as
is confirmed by the linear scaling theory, which we will address in
detail in Sec.~VI.

\subsection{Widom line}

For the first order phase transition, the order parameter, entropy, or
density is discontinuous on crossing the coexistence line.  At the
critical point, where the coexistence line terminates, the critical
fluctuations of $C_P$ and $K_T$ diverge, and show maxima in the one
phase region close to the critical point. In this section, we study the
behavior of $C_P$ maxima and $K_T$ maxima lines near the LLCP for models
with different coexistence line slopes.

\subsubsection{Isobaric specific heat $C_P$}

\begin{figure}[htbp]
\begin{minipage}[t]{0.35\linewidth}
\includegraphics[width=0.9\linewidth]{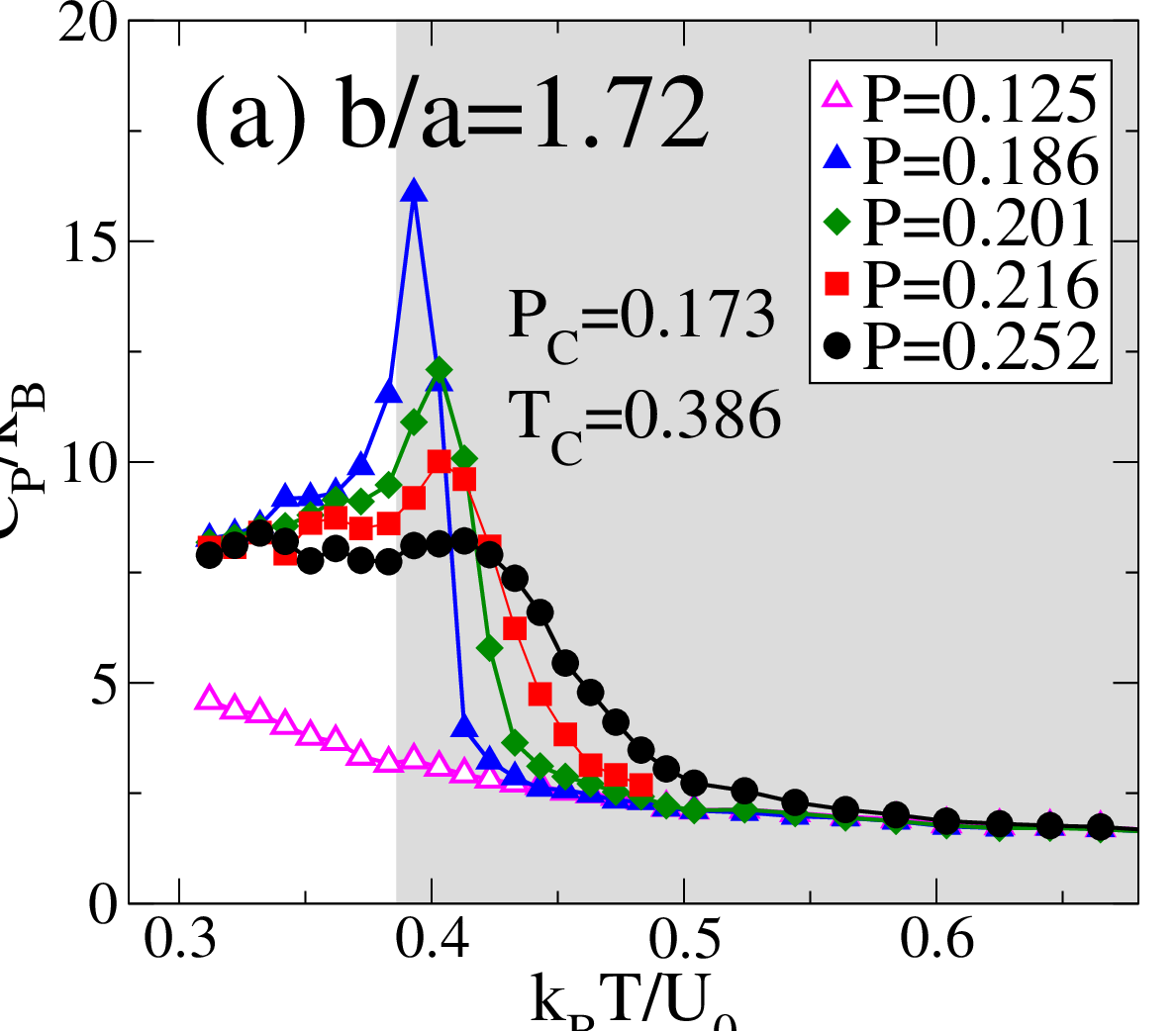}
\end{minipage}\hspace{0.1cm}
\begin{minipage}[t]{0.35\linewidth}
\includegraphics[width=0.9\linewidth]{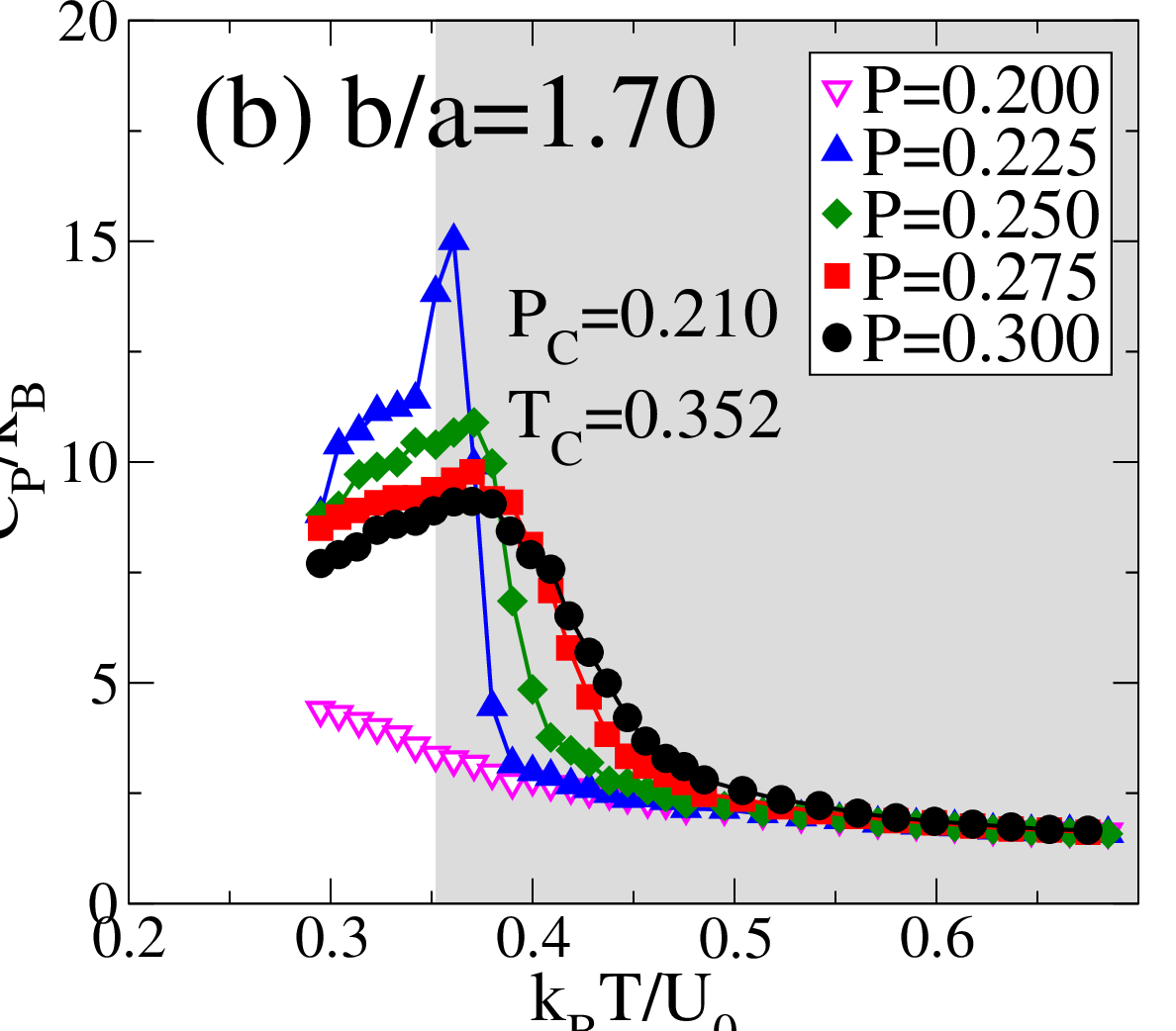}
\end{minipage}\\
\begin{minipage}[t]{0.35\linewidth}
\includegraphics[width=0.9\linewidth]{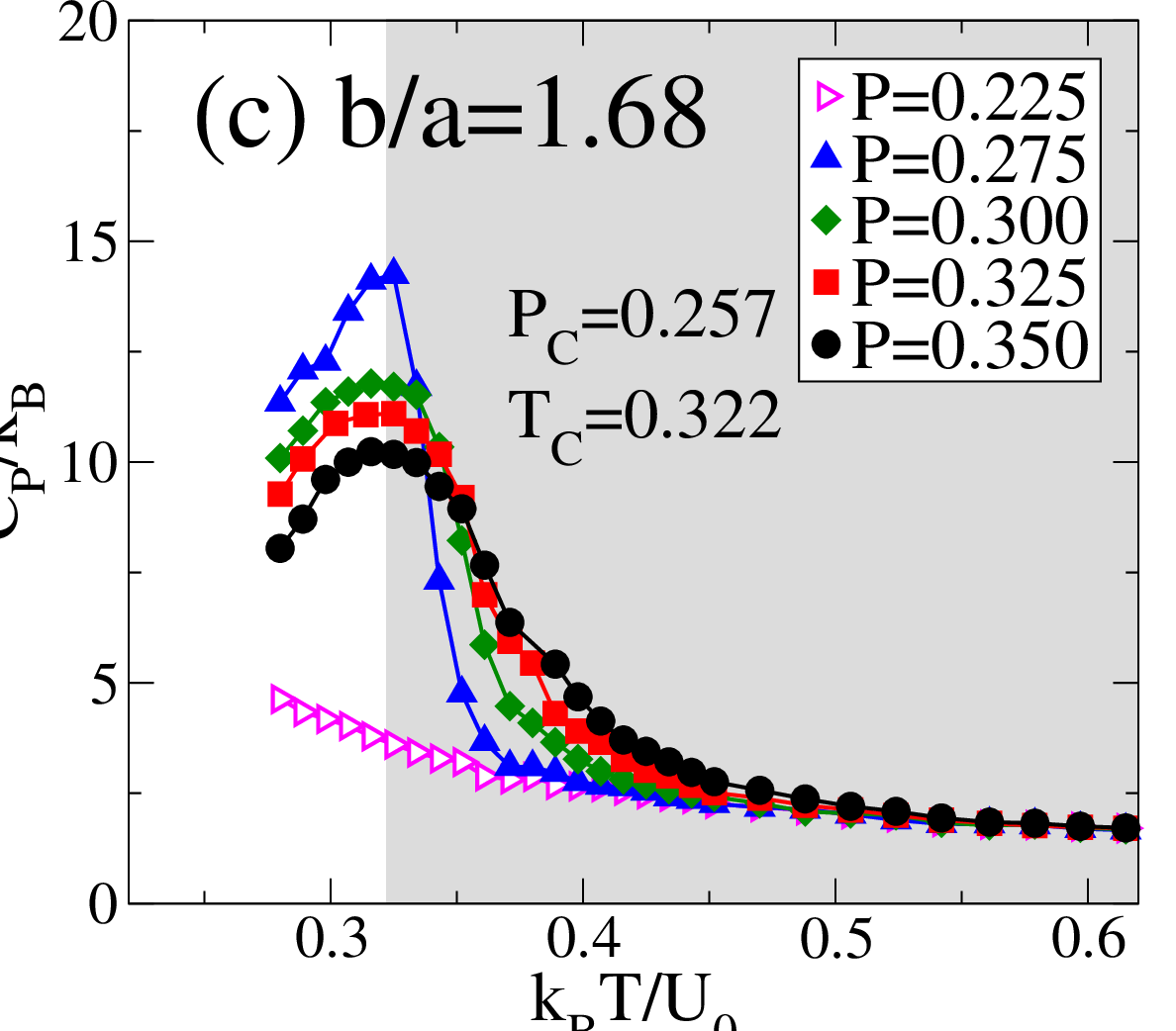}
\end{minipage}\hspace{0.1cm}
\begin{minipage}[t]{0.35\linewidth}
\includegraphics[width=0.9\linewidth]{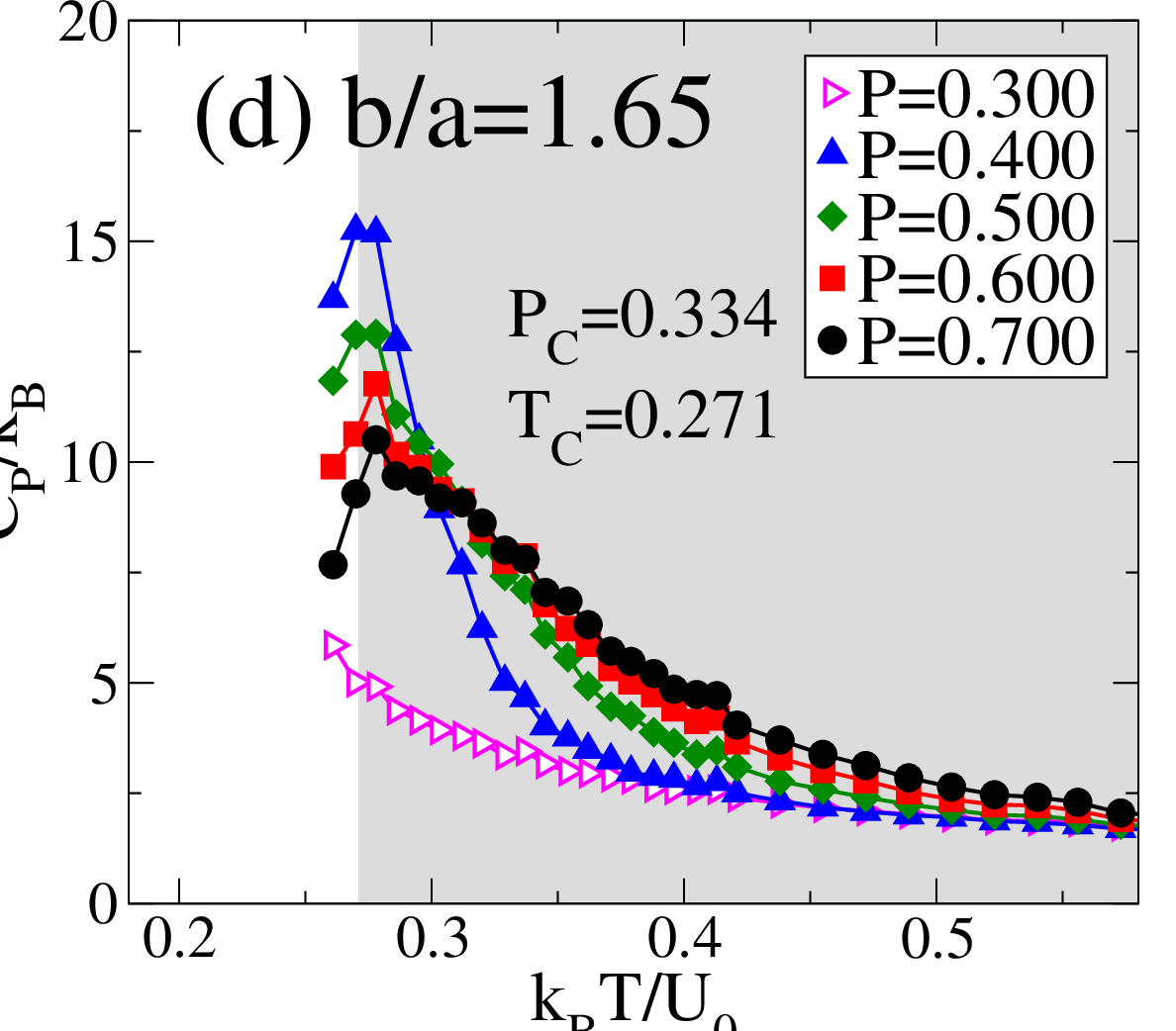}
\end{minipage}\\
\begin{minipage}[t]{0.35\linewidth}
\includegraphics[width=0.9\linewidth]{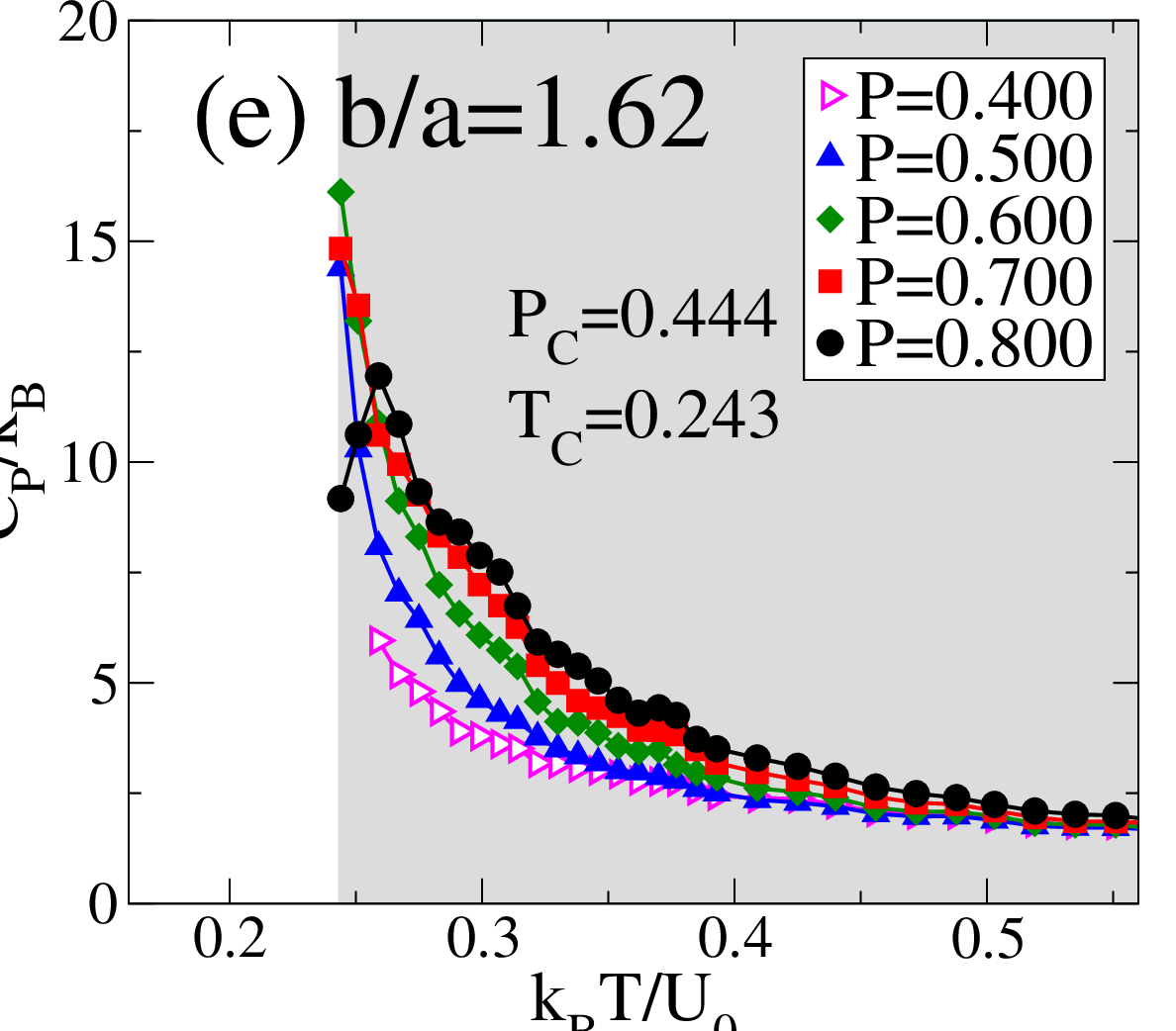}
\end{minipage}\hspace{0.1cm}
\begin{minipage}[t]{0.35\linewidth}
\includegraphics[width=0.9\linewidth]{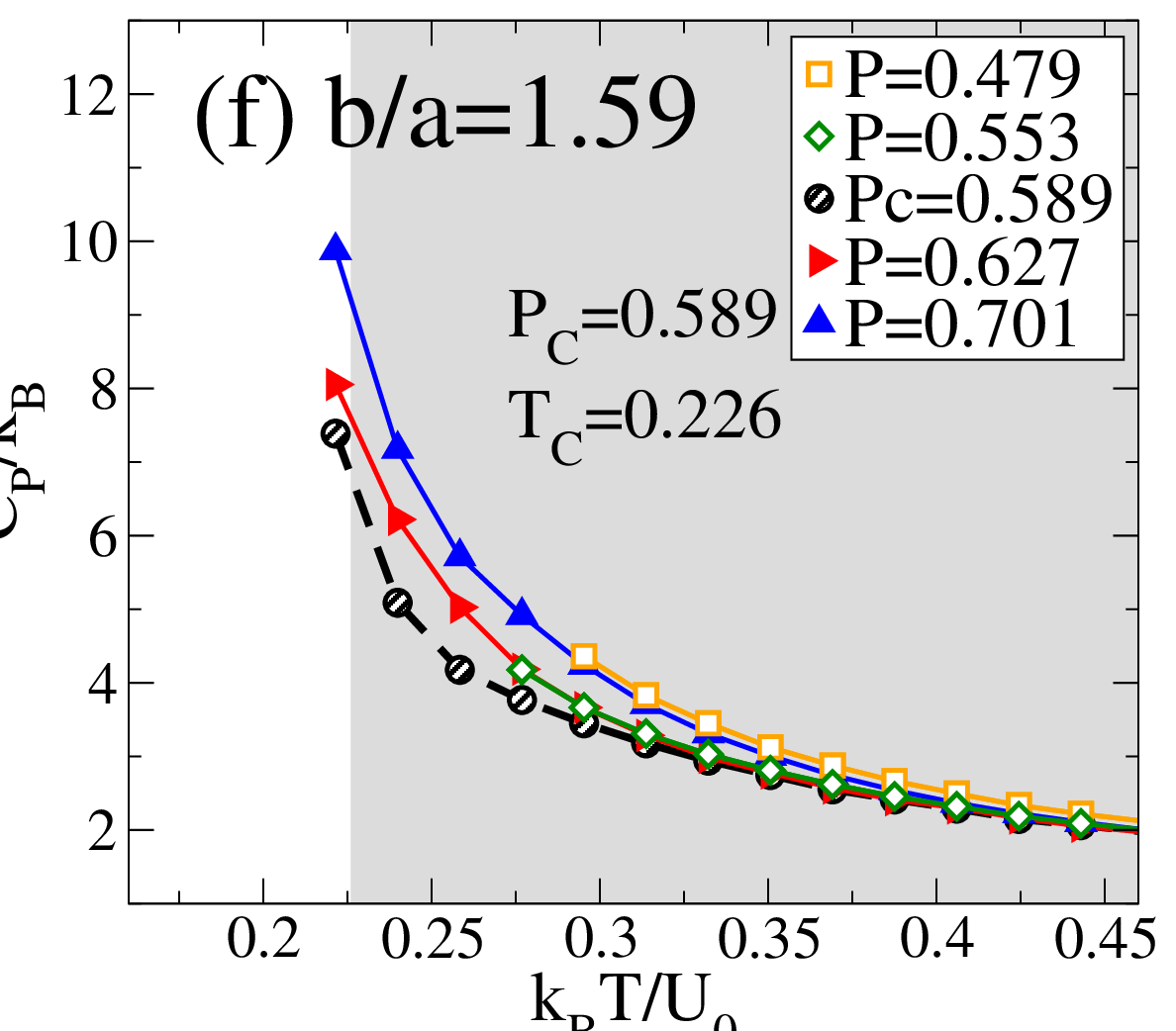}
\end{minipage}\\
\captionsetup{format=plain,justification=RaggedRight}
\caption{Specific heat $C_P$ for systems with different $b/a$.
  Only equilibrated results are shown. The gray area indicates the $T>T_c$ region. 
  (a-d) For $b/a \geq 1.65$, one can see that $C_P$ shows maxima at pressures $P>P_c$, 
  and as $P_c$ is approached, the increase in $C_P$ starts at lower $T$ but becomes sharper. 
  As the pressure increases, the $C_P$ peak moves to higher $T$, 
  indicating a positive slope of the $C_P$ maxima locus, which follows the coexistence line for these models. 
  For pressure $P<P_c$, $C_P$ monotonically increases without any maximum. 
  (e) For $b/a=1.62$, $C_P$ monotonically increases 
  without showing any peak also for $P>P_c$, except at the highest pressure studied $P=0.800$.
  The system enters a glassy state at lower $T$ where no equilibrium data can be obtained.
  (f) For $b/a=1.59$ with horizontal coexistence line, 
  no $C_P$ maxima can be found for the equilibrium states with $T \geq T_c$ 
  for both $P>P_c$ and $P<P_c$.}
\label{CP_all}
\end{figure}

We first explore the behavior of the $C_P$ maxima line for modified
Jagla models with different $b/a$.  For $b/a \geq 1.62$, the coexistence
line has a positive slope [Fig.~\ref{Co_slope}(b)]. Upon cooling along
constant pressure above the critical pressure $P_c$ in the one-phase
region, we observe peaks in $C_P$ for the cases $b/a=1.72$, 1.70, 1.68,
and 1.65 [Fig.~\ref{CP_all}(a-d)]. As the LLCP is approached the
magnitude of the $C_P$ peaks increases, and at the LLCP they diverge in
an infinite system. When $P<P_c$, we observe a continuous increase in
$C_P$ without any peak when the coexistence line is crossed. Thus we can
locate the LLCP by locating the terminal point of the $C_P$ maxima line.

We note that the $C_P$ peaks move toward higher $T$ at higher $P$,
indicating a positively sloped line [Fig.~\ref{CP_all}(a,b)]. This is
consistent with the fact that the Widom line is the extension of the
coexistence line into the one-phase region, and for these values of
$b/a$ the coexistence line is positively sloped. However the slopes of
the $C_P$ maxima lines increase as the slopes of the coexistence lines
decrease and eventually, at $b/a=1.65$, the $C_P$ maxima line becomes
nearly vertical, clearly showing that the $C_P$ maximum is no longer
serving its original purpose, as will be explained in Sec.~IV below.
For $b/a=1.62$, $C_P$ monotonically increases without showing any peak
for pressures $P>P_c$, except at the highest pressure studied $P=0.800$
[Fig.~\ref{CP_all}(e)].

When $b/a=1.59$ with a horizontal coexistence line, $C_P$ monotonically
increases with decreasing $T$ along a constant pressure path both below
and above $P_c$ [Fig.~\ref{CP_all}(f)]. When $b/a=1.62$, there are no
$C_P$ maxima in the equilibrium region with $T \geq T_c$, but $C_P$
behaves symmetrically either below or above $P_c$.

\begin{figure}[htbp]
\begin{minipage}[t]{0.52\linewidth}
\includegraphics[width=0.9\linewidth]{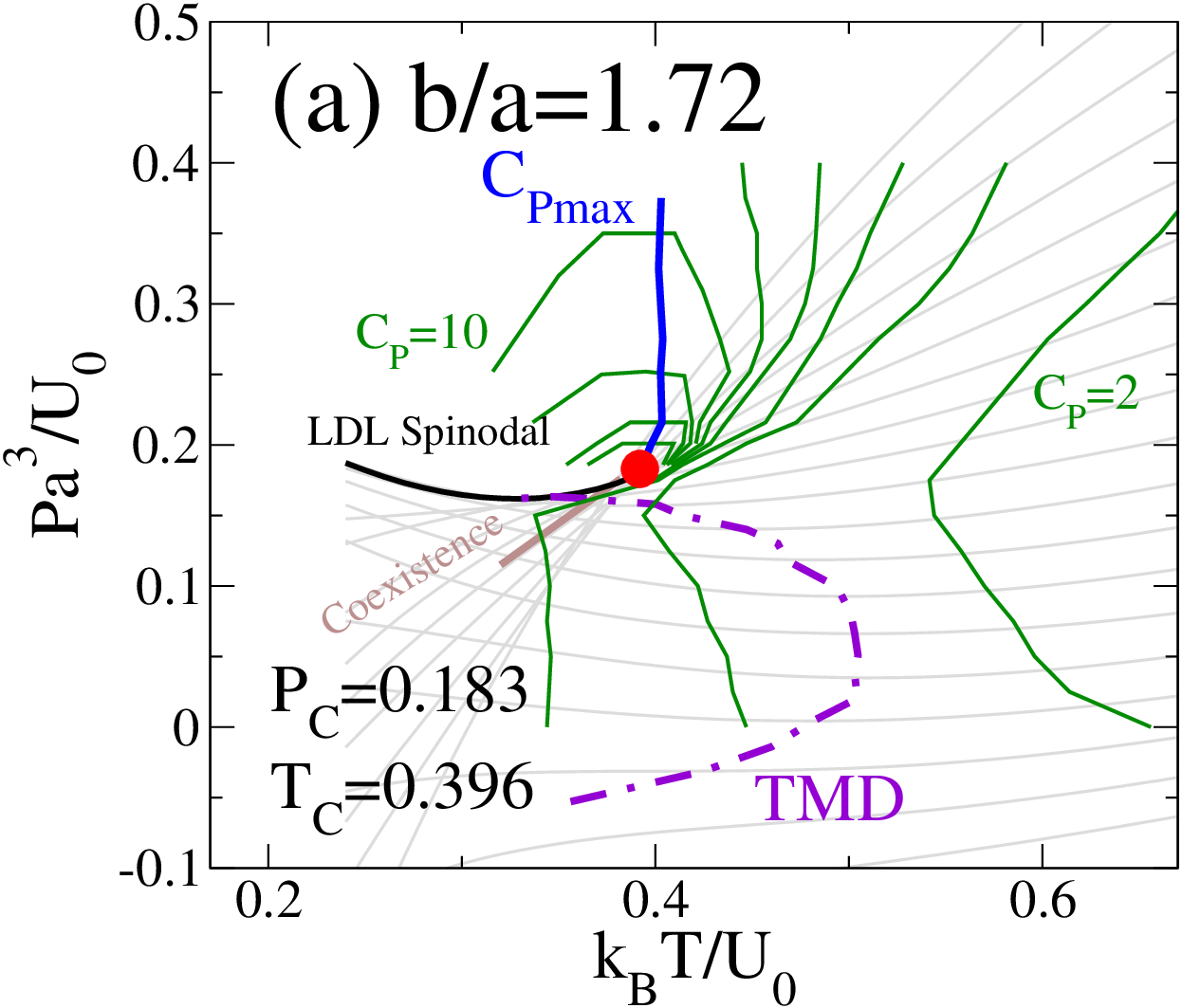}
\end{minipage}
\begin{minipage}[t]{0.52\linewidth}
\includegraphics[width=0.9\linewidth]{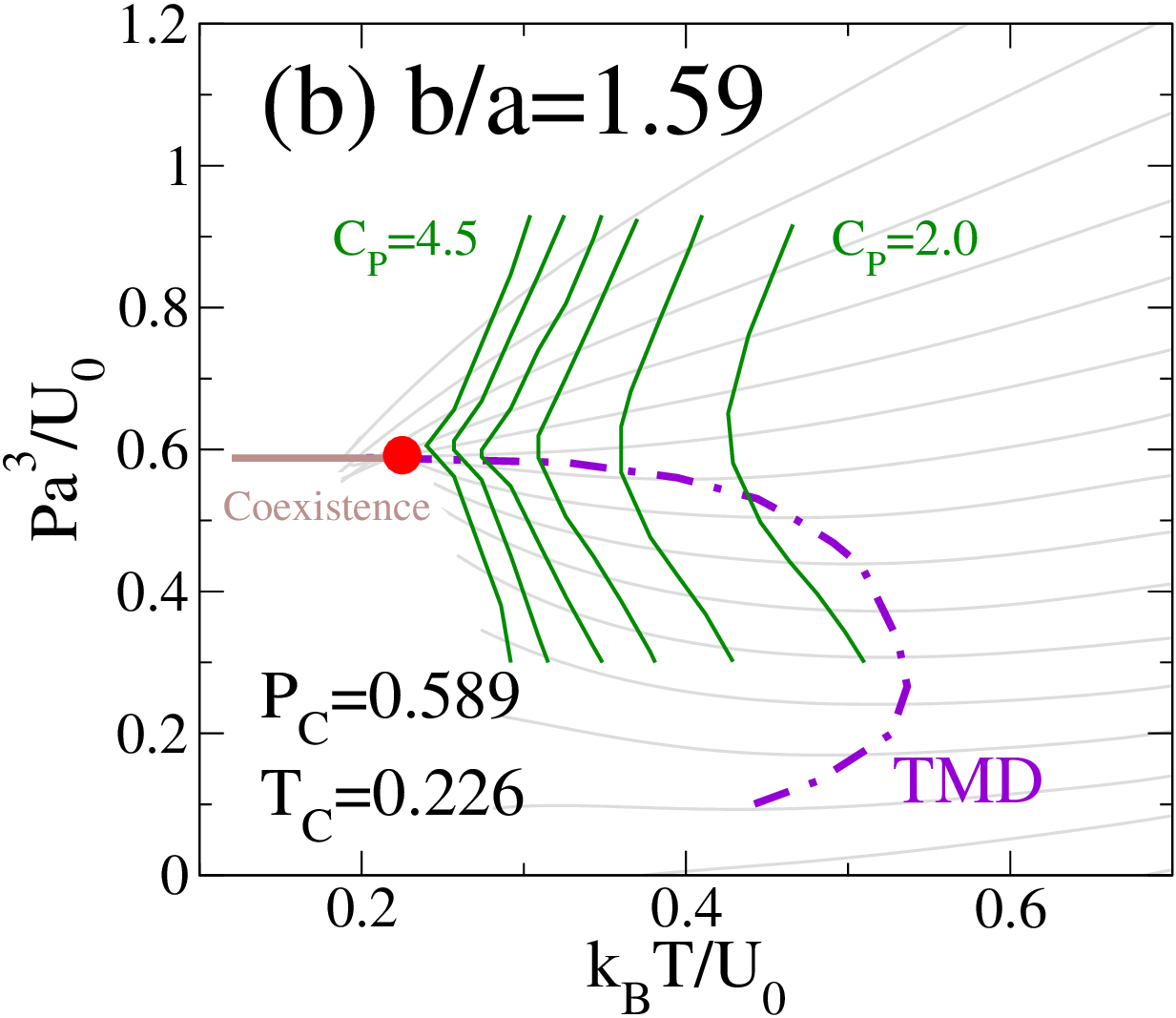}
\end{minipage}\\
\captionsetup{format=plain,justification=RaggedRight}
\caption{Phase diagram with specific heat $C_P$ for positively sloped
  coexistence line case ($b/a=1.72$) and horizontal coexistence line
  case ($b/a=1.59$).  Isochores (solid gray), TMD line (dash purple),
  and LLCP (red solid circle) are shown.  (a) $b/a=1.72$, the lines of
  equal $C_P$ (solid green), change from $C_P=2$ far away from the LLCP
  to $C_P=10$ close to the LLCP with interval $\Delta C_P=1$.  $C_P$
  maxima locus crosses the lines of equal $C_P$ at the points of their
  maximal pressure, and follows the coexistence line into the one-phase
  region, then sharply turns upwards to higher pressures and becomes
  almost vertical.  (b) For $b/a=1.59$, the lines of equal $C_P$ (solid
  green), change from $C_P=2.0$ far away from the LLCP to $C_P=4.5$
  close to the LLCP with interval $\Delta C_P=0.5$. No $C_P$ maxima can
  be observed before the system either goes into glassy states or
  crystallizes. However, one notes that the $C_P$ is symmetric with
  respect to the critical pressure.}
\label{CP_equi}
\end{figure}

We plot the lines of equal $C_P$ for two extreme cases, $b/a=1.72$ with
a positively sloped coexistence line [Fig.~\ref{CP_equi}(a)] and
$b/a=1.59$ with a horizontal coexistence line
[Fig.~\ref{CP_equi}(b)]. When $b/a=1.72$, the lines of equal $C_P$ form
loops in the $T$-$P$ plane and cross the $C_P$ maxima line at their maxima
points. The locus of $C_P$ maxima extends the coexistence line into the
one-phase region in the vicinity of the critical point. Then it sharply
turns upwards to higher pressures and becomes approximately
vertical. For $b/a=1.59$, there are no $C_P$ maxima. The equal $C_P$
lines extend away from the critical point symmetrically without any
loops. At low $T$, we reach the simulation limit due to either
crystallization for $P<P_c$ or due to entering a glassy state for
$P>P_c$, where no equilibrium results can be obtained for the analysis.

We note that the magnitude of $C_P$ drops significantly when the
coexistence line is horizontal with $b/a=1.59$, compared to when
$b/a=1.72$. This is because, when the coexistence line is horizontal,
the difference in enthalpy $H$ between LDL and HDL is zero according to
the Clapeyron equation of thermodynamics,
\begin{equation}
\frac{dP}{dT}=\frac{\Delta H}{T \Delta V}
\end{equation}
In this case, the enthalpy fluctuations that determine the magnitude of
the specific heat gain no strength from the critical fluctuations,
except from the weak $PV$ term.

\subsubsection{Isothermal compressibility $K_T$} 

\begin{figure}[htbp]
\begin{minipage}[t]{0.35\linewidth}
\includegraphics[width=0.9\linewidth]{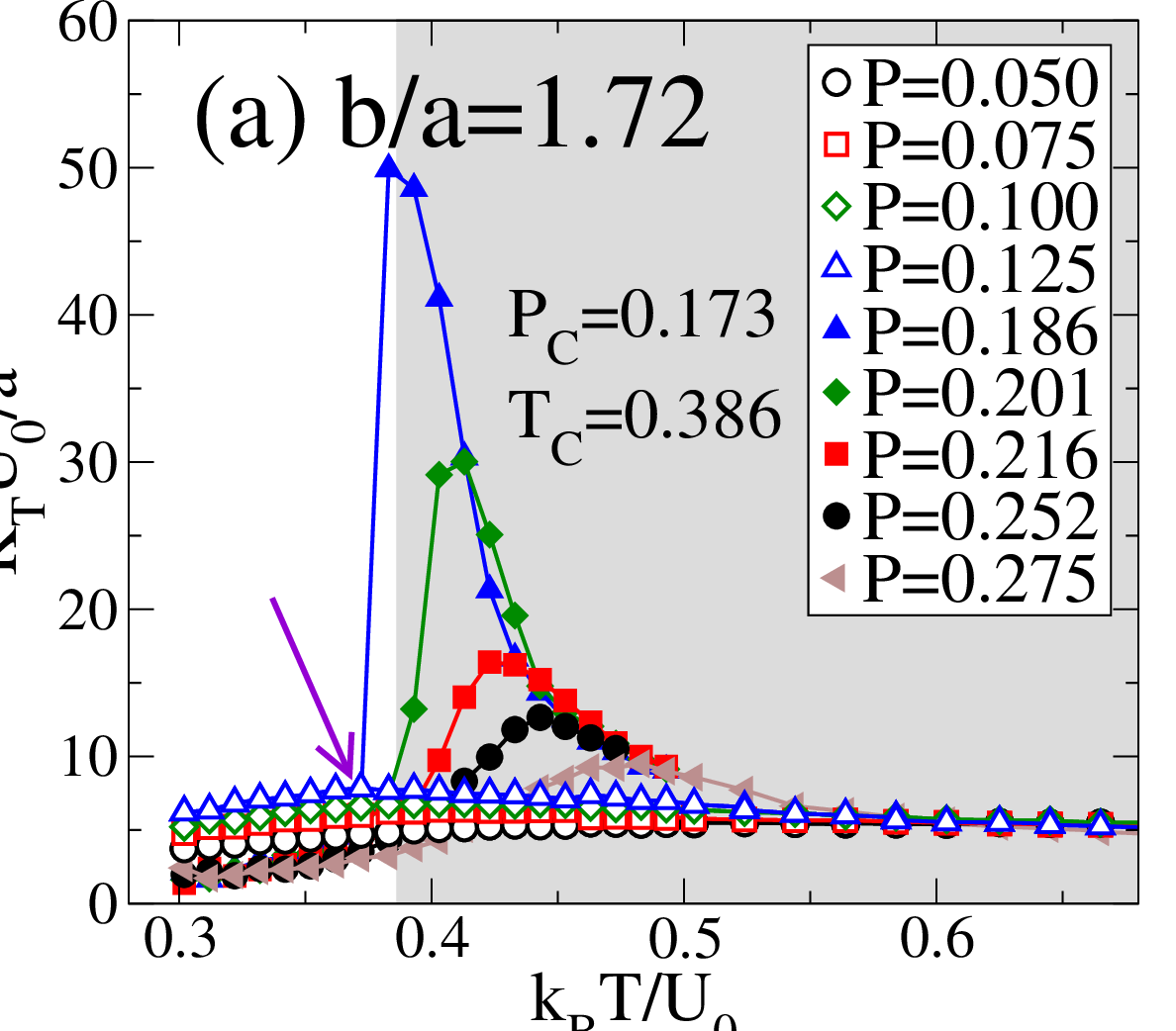}
\end{minipage}\hspace{0.1cm}
\begin{minipage}[t]{0.35\linewidth}
\includegraphics[width=0.9\linewidth]{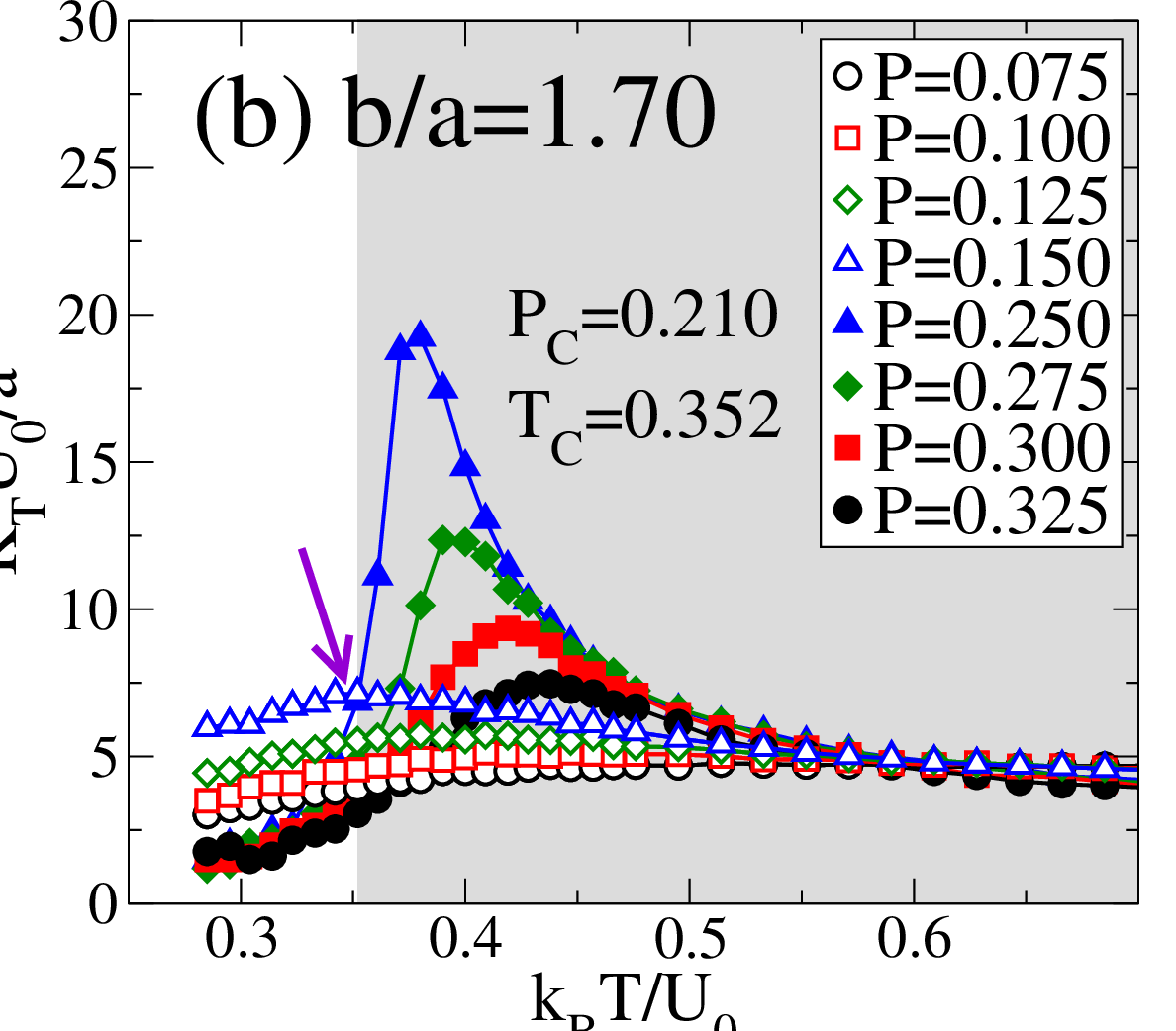}
\end{minipage}\\
\begin{minipage}[t]{0.35\linewidth}
\includegraphics[width=0.9\linewidth]{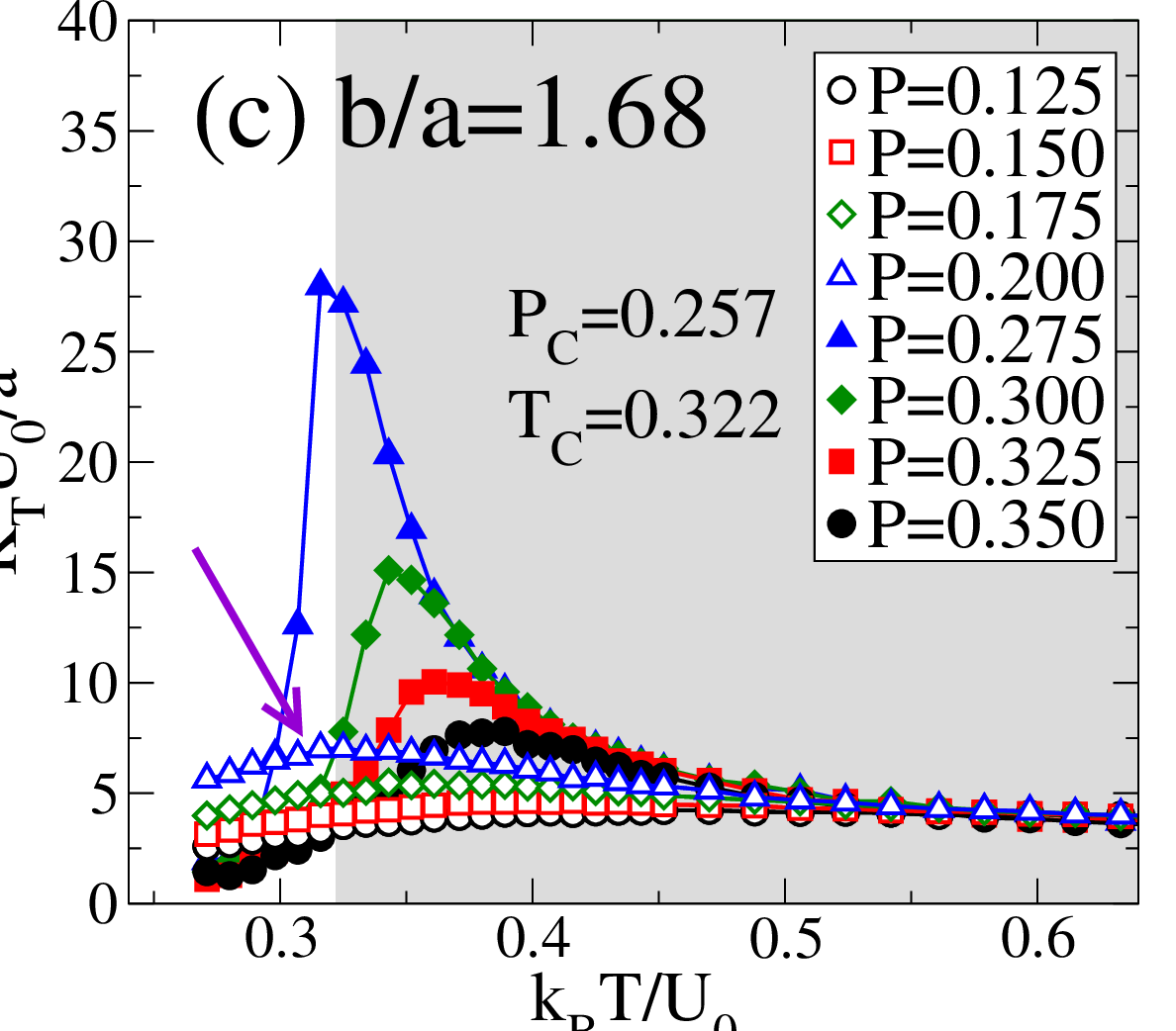}
\end{minipage}\hspace{0.1cm}
\begin{minipage}[t]{0.35\linewidth}
\includegraphics[width=0.9\linewidth]{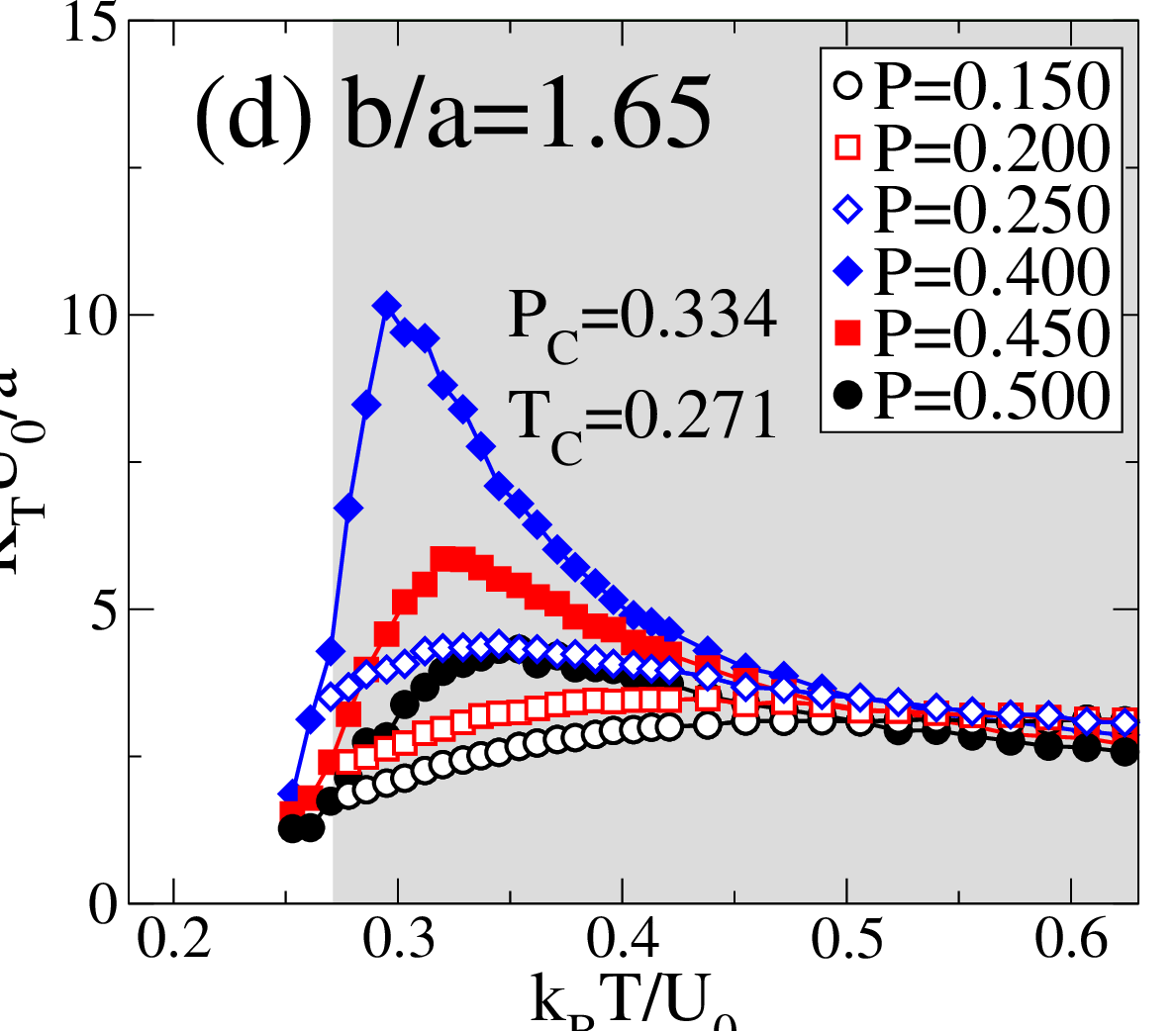}
\end{minipage}\\
\begin{minipage}[t]{0.35\linewidth}
\includegraphics[width=0.9\linewidth]{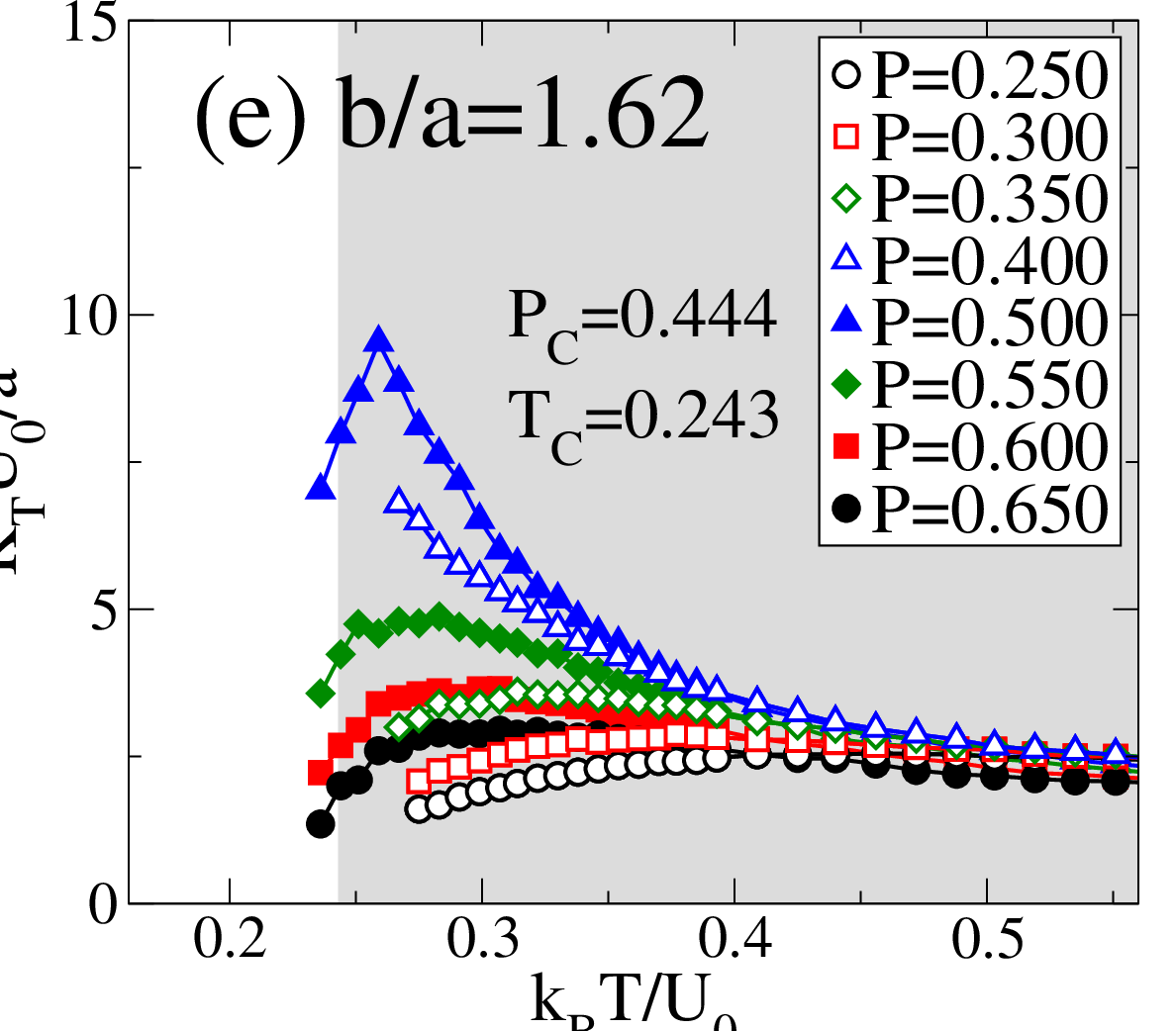}
\end{minipage}\hspace{0.1cm}
\begin{minipage}[t]{0.35\linewidth}
\includegraphics[width=0.9\linewidth]{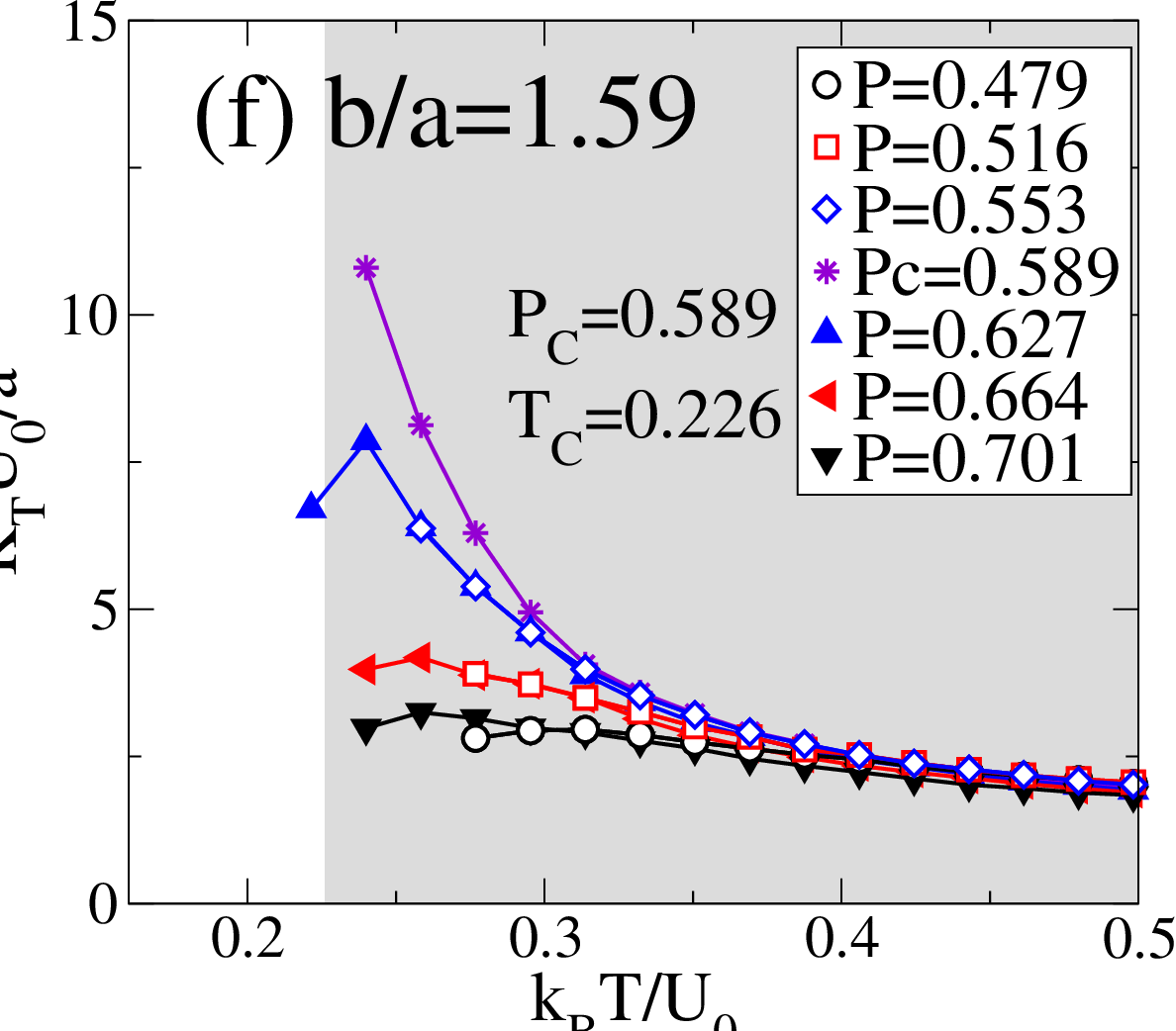}
\end{minipage}\\
\captionsetup{format=plain,justification=RaggedRight}
\caption{Compressibility $K_T$ for models with different $b/a$. The maxima in $K_T$ both for $P<P_c$ (open symbols) and for $P>P_c$ (solid symbols with arrows pointing to the peaks) are shown.
(a-e) For $b/a \leq 1.62$,the magnitudes of $K_T$ for $P>P_c$ are much larger than for $P<P_c$. 
The maxima for $P>P_c$ correspond to critical fluctuations, while the maxima for $P<P_c$ correspond to the approach to the LDL spinodal. 
(f) For $b/a=1.59$, with the horizontal coexistence line, the $K_T$ below and above $P_c$ are almost identical, with equal magnitudes of their maxima.}
\label{KT_all}
\end{figure}

Figure~\ref{KT_all} shows the behavior of $K_T$ above and below $P_c$
and contrasts it with that of $C_P$. For $b/a \geq 1.62$, when the
coexistence line slope is positive, $K_T$ shows maxima both above and
below $P_c$.  For $P>P_c$ in the one-phase region, similar to $C_P$, the
$K_T$ peaks become more prominent as the LLCP is approached
[Fig.~\ref{KT_all}(a--e)]. For a finite system, $K_T$ diverges at the
LLCP.  For $P<P_c$, we observe a second set of $K_T$ peaks with much
lower magnitudes.  For $b/a=1.59$ with a horizontal coexistence line,
$K_T$ behaves symmetrically above and below $P_c$, with equal magnitudes
of the maxima [Fig.~\ref{KT_all}(f)].

\begin{figure}[htbp]
\begin{minipage}[t]{0.52\linewidth}
\includegraphics[width=0.9\linewidth]{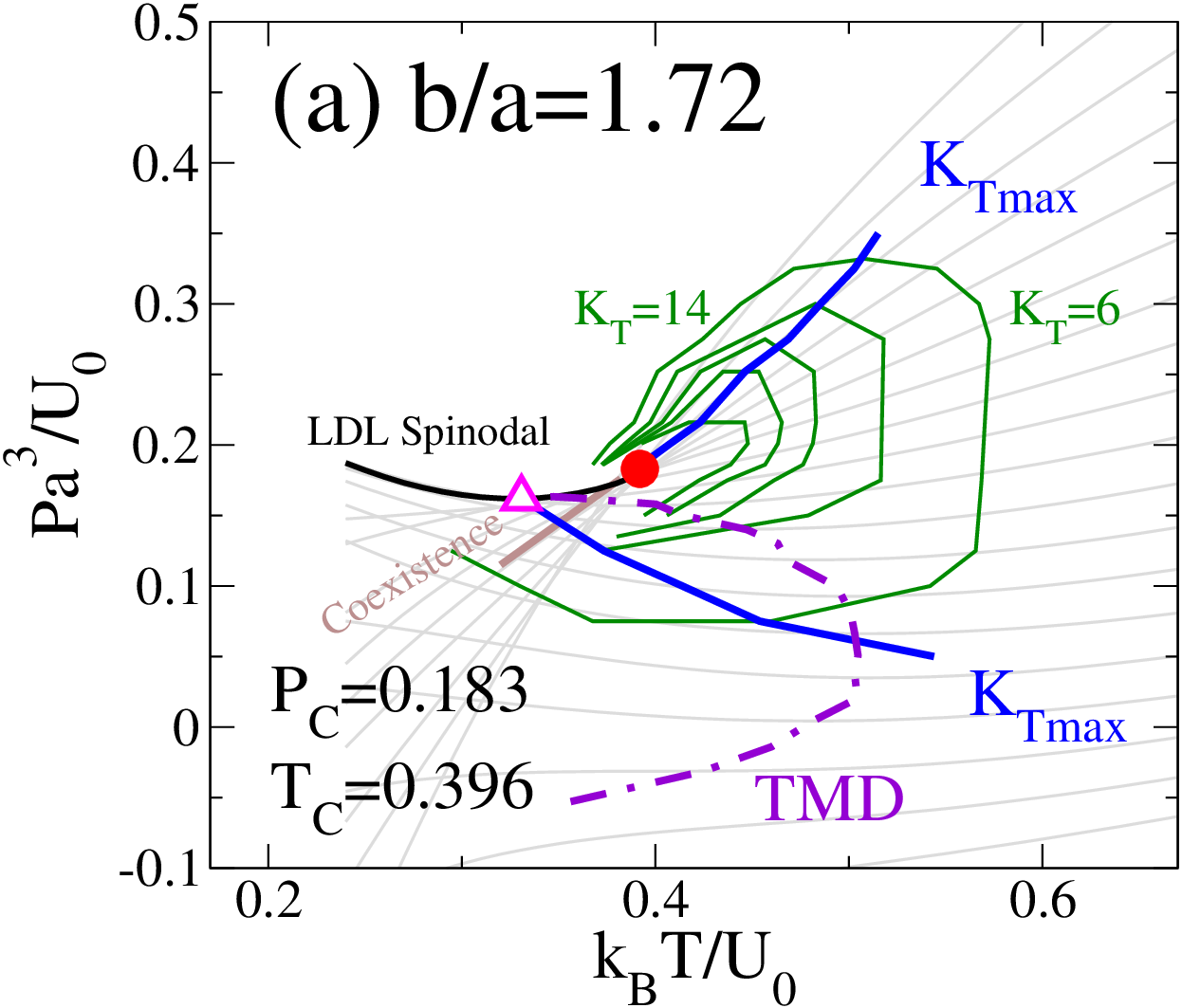}
\end{minipage}\\
\begin{minipage}[t]{0.52\linewidth}
\includegraphics[width=0.9\linewidth]{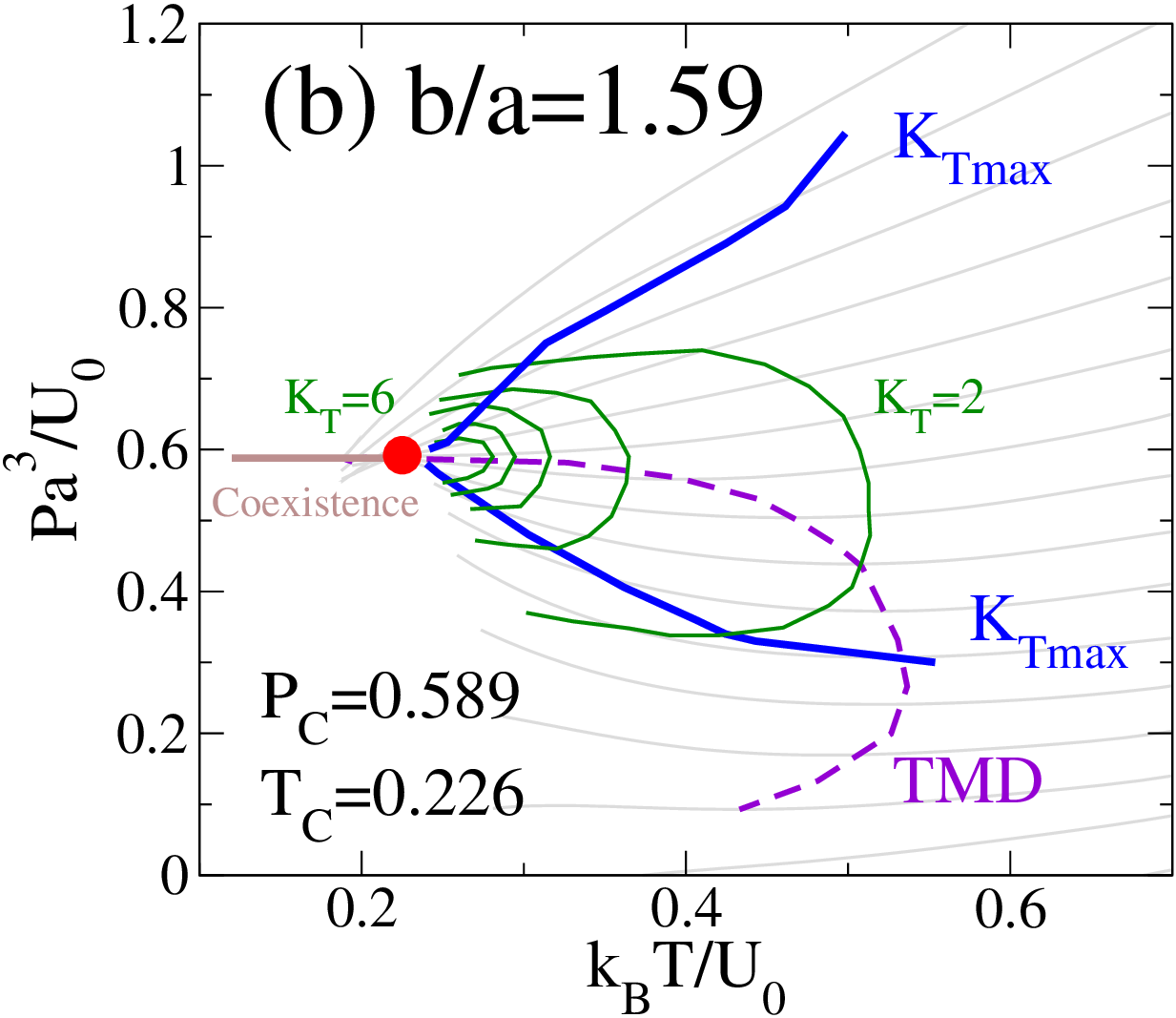}
\end{minipage}
\captionsetup{format=plain,justification=RaggedRight}
\caption{Phase diagram with compressibility $K_T$ 
  for positively sloped coexistence line case with $b/a=1.72$ 
  and horizontal coexistence line case with $b/a=1.59$.
  (a) For $b/a=1.72$, the lines of equal $K_T$ (solid green),
  changes from $K_T=6$ far away from the LLCP to $K_T=14$ close to the LLCP with interval $\Delta K_T=2$.  
  One can see that the loci of the two $K_T$ maxima cross the lines of equal $K_T$ 
  at points of their maximal and minimal pressures, and are not symmetric with respect to $P_c$. 
  The locus with higher magnitude of $K_T$ maxima, which corresponds to the critical fluctuations, 
  merges to the LLCP.
  The locus with the lower magnitude of $K_T$ maxima extends below $T_c$, 
  and terminates at the minimum pressure point of the LDL spinodal (triangle symbol), where the TMD line also terminates. 
  This brunch of $K_T$ maxima also crosses the TMD line at the point of its maximal temperature.
  (b) For $b/a=1.59$, the lines of equal $K_T$ (solid green), 
  changes from $K_T=2$ far away to $K_T=6$ close to the LLCP with interval $\Delta K_T=1$,
  The loci of equal $K_T$ form symmetric loops around the LLCP, 
  and both $K_T$ maxima lines merge to the LLCP with equal magnitudes of their maxima.
  }
\label{KT_equi}
\end{figure}

Figure \ref{KT_equi} shows the loci of $K_T$ maxima for both $b/a=1.72$
with a positive coexistence line slope and $b/a=1.59$ with a horizontal
coexistence line.  For $b/a=1.72$, the values of the $K_T$ maxima at
$P>P_c$, which corresponds to the critical fluctuations and originates
from the LLCP, have a much larger magnitude than the values of $K_T$
maxima at $P<P_c$. The second $K_T$ maxima line at $P<P_c$ corresponds
to the approach to the LDL spinodal and terminates at the lowest
pressure point of the LDL spinodal where the TMD line also terminates
\cite{XuJCP2009,sergeyReview}.  Furthermore, the $P<P_c$ $K_T$ maxima
line also crosses the TMD line at the point of its maximal temperature
\cite{SastryPRE2006}.  Similar to $C_P$, the lines of equal $K_T$ form
loops, and cross at their {\it maximal\/} pressure points with
the $P>P_c$ and $P<P_c$ $K_T$ maxima lines.  For $b/a=1.59$
[Fig.~\ref{KT_equi}], the lines of equal $K_T$ form symmetric (with
respect to $P_c$) loops around the critical point. Both $K_T$ maxima
lines merge at the LLCP.

In the case of $b/a=1.72$, from both Fig.~\ref{CP_equi} and
Fig.~\ref{KT_equi}, we can identify the Widom line as the overlapping
segment of the $C_P$ maxima and the $K_T$ maxima lines, which extends
the coexistence line into the one-phase region in the vicinity of the
LLCP. In contrast, for $b/a=1.59$ with a horizontal coexistence line,
the $C_P$ maxima line disappears, where the specific heat $C_P$ can no
longer be a good representative for critical fluctuations. Indeed, there
is no enthalpy difference between the two coexisting phases, so there
can be no contribution to enthalpy fluctuations from the critical
fluctuations. However, the density fluctuations remain well defined,
with two $K_T$ maxima lines above and below the critical pressure, both
associated with the critical fluctuations. In the vicinity of the
critical point, the two $K_T$ maxima lines merge together and can be
used to locate the critical point from measurements obtained in the
supercritical region only.

\section{Comparison with the linear scaling theory of the liquid-liquid
  critical point} 

To explain the change of behavior of the lines of response function
extrema for the horizontal coexistence line case, we adapt the linear
scaling theory of the liquid-liquid critical point developed by
M. A. Anisimov and collaborators \cite{AnisimovPRE}.

The field-dependent thermodynamic potential $\psi$ can be considered
a universal function of two scaling fields: ``ordering'' $h_1$ and
``thermal'' $h_2$. Near the critical point $\psi$ can be written
\begin{equation}
\psi \simeq h^{2 - \alpha}_{2}f(\frac{h_1}{h^{\beta+\gamma}_2}),
\end{equation}
where the critical exponents have the values for the Ising universality
class, $\alpha=0.109$, $\beta=0.326$, and $\gamma=1.239$
\cite{AnisimovPRE}.

Since our focus is on the immediate vicinity of the liquid-liquid
critical point, we neglect the curvature of the coexistence line and the
background contribution to the response functions. We assume the
scaling fields are linear analytical combinations of physical fields,
the pressure $P$, and the temperature $T$,
\begin{equation}
h_1 = a_1\Delta \hat P + a_2 \Delta \hat{T},
\end{equation}
\begin{equation}
h_2 = b_1 \Delta \hat{T} + b_2\Delta \hat P.
\end{equation}
with $\Delta \hat{P}\equiv(P - P_c)/(\rho_CRT_c)$ and $\Delta
\hat{T}\equiv(T - T_c)/T_c$, where the subscript ``$c$'' here and below
indicates the critical parameters, and $a_i$ and $b_i$ are
system-dependent coefficients.

We introduce a tuning parameter $\phi$ into the theory, and we use it to
change the slope of the coexistence line by defining
\begin{eqnarray}
a_1&\equiv &\cos \phi, \ \ a_2\equiv -\sin \phi, \nonumber \\
b_1&\equiv &\cos \phi, \ \ b_2\equiv \sin \phi.
\end{eqnarray}
Then the slope of the coexistence line is
\begin{equation}
\frac{dP}{dT}=\tan \phi.
\end{equation}
The critical (fluctuation-induced) parts of the dimensionless isobaric
specific heat and isothermal compressibility are expressed through the
scaling susceptibilities \cite{AnisimovPRE}
\begin{eqnarray}
(\hat{C}_P) &=& \hat{T}(\frac{\partial \hat{S}}{\partial \hat{T}})_{\hat{P}} \nonumber
\\ &=& \hat{T}(\sin^2 \phi \chi_1 - \sin 2\phi \chi_{12} + \cos^2 \phi \chi_2),
\\ (\hat{K}_T) &=& -\frac{1}{\hat{V}}(\frac{\partial \hat{V}}{\partial
  \hat{P}})_{\hat{T}} \nonumber 
\\ &=& \frac{1}{\hat{V}}[\cos^2 \phi \chi_1 - \sin 2\phi \chi_{12} + \sin^2 \phi \chi_2],
\\ (\hat{\alpha}_P) &=& \frac{1}{\hat{V}}(\frac{\partial
  \hat{V}}{\partial \hat{T}})_{\hat P} \nonumber 
\\ &=& \frac{1}{\hat{V}}(\frac{1}{2}\sin 2\phi \chi_1 - \cos 2\phi
\chi_{12} - \frac{1}{2}\sin 2\phi \chi_2). 
\label{ResponseF}
\end{eqnarray}
where $\hat{T}=T/T_c$, $\hat{P}=P/\rho_CRT_c$.

The scaling fields and scaling susceptibilities can be written as
functions of the ``polar'' variables $r$ and $\theta$, and two constants
$a$ and $k$, which can be obtained by fitting the experimental data,
\begin{equation}
h_1 = ar^{\beta + \gamma}\theta(1-\theta^2),
\end{equation}
\begin{equation}
h_2 = r(1 - b^2\theta^2),
\end{equation}
\begin{equation}
\chi_1 = \frac{k}{a}r^{-\gamma}c_1(\theta),
\end{equation}
\begin{equation}
\chi_{12} = kr^{\beta -1}c_{12}(\theta),
\end{equation}
\begin{equation}
\chi_2 = akr^{-\alpha}c_2(\theta),
\end{equation}
where
\begin{equation}
c_{1}(\theta) = [1-b^2\theta ^2(1-2\beta)]/c_0(\theta),
\end{equation}
\begin{equation}
c_{12}(\theta) = \beta\theta[1-\delta-\theta ^2(3-\delta)]/c_0(\theta),
\end{equation}
\begin{equation}
c_{2}(\theta) = [(1-\alpha)(1-3\theta ^2)s(\theta)-2s_2\beta\delta\theta ^2(1-\theta ^2)]/c_0(\theta),
\end{equation}
\begin{equation}
c_{0}(\theta) = (1-3\theta ^2)(1-b^2\theta ^2)+2\beta\delta b^2\theta ^2(1-\theta ^2).
\end{equation}
Here $b=\sqrt{(\gamma-2\beta)/\gamma(1-2\beta)} = 1.16679$,
$\delta=\gamma / \beta + 1 = 4.80061$, and $s(\theta)$ are known
functions,
\begin{equation}
s(\theta)=s_0+s_2\theta ^2,
\end{equation}
\begin{equation}
s_0=-(2-\alpha)f_0,
\end{equation}
\begin{equation}
s_2=-(2-\alpha)b^2(1-2\beta)f_0-\gamma f_2,
\end{equation}
with
\begin{equation}
f_0=-\frac{\beta(\delta-3)-b^2\alpha\gamma}{2b^4(2-\alpha)(1-\alpha)\alpha},
\end{equation}
\begin{equation}
f_2=\frac{\beta(\delta-3)-b^2\alpha(1-2\beta)}{2b^2(1-\alpha)\alpha}.
\end{equation}

Since all response functions are directly propositional to $k$, its
actual value is irrelevant for our study. The value of $a$ determines
the strength of the ordering field. We found that for large values of
$a$, the overlapping segments of $C_P$ and $K_T$ maxima lines are
shorter, which more adequately models the behavior of our simulations.
  
\begin{figure}[htbp]
\begin{minipage}[t]{0.4\linewidth}
\includegraphics[width=0.9\linewidth]{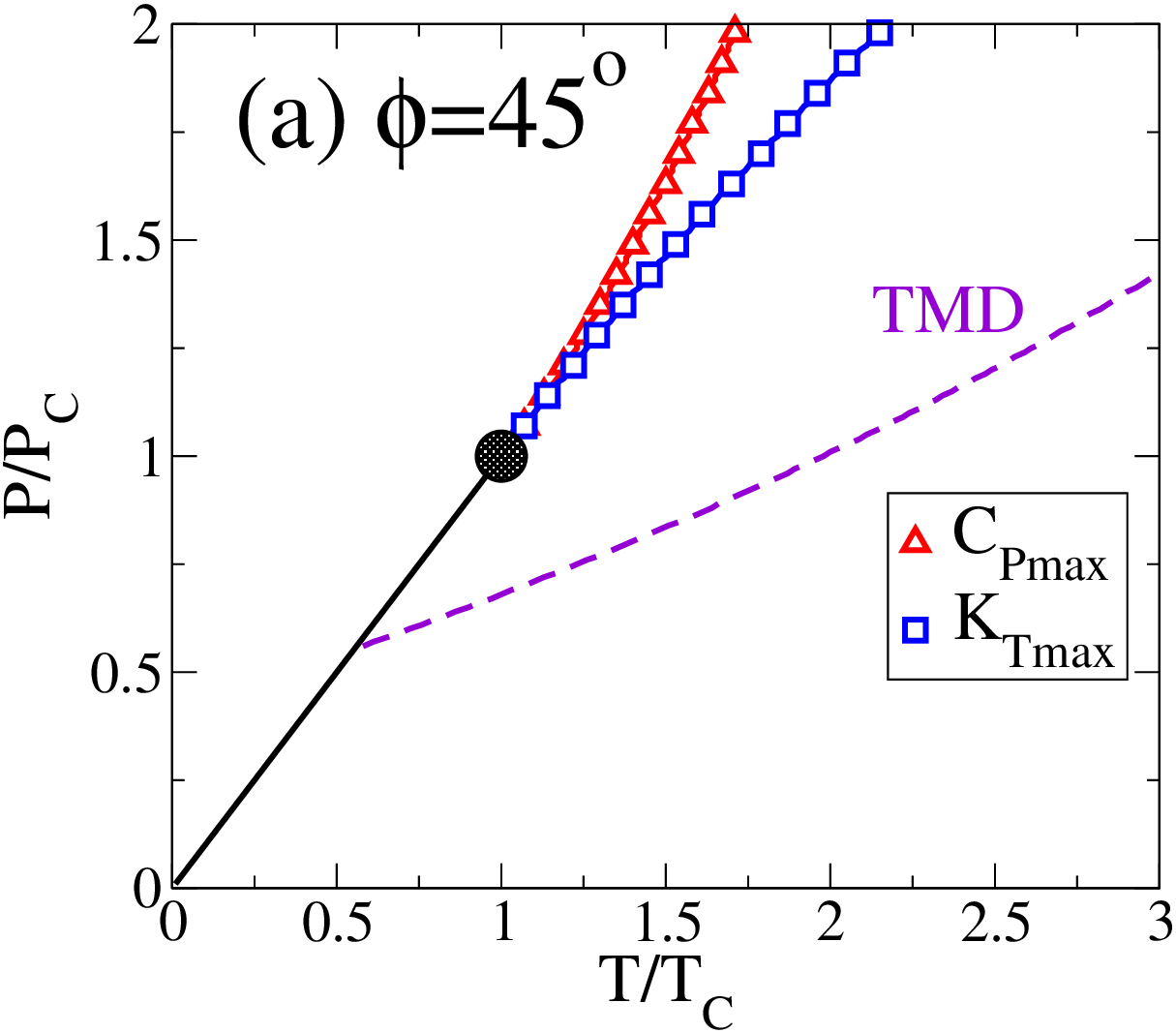}
%\subcaption{$(C_P,K_T)_{max}$ on P-T with $\phi=45^o$.}
\end{minipage}
\begin{minipage}[t]{0.4\linewidth}
\includegraphics[width=0.9\linewidth]{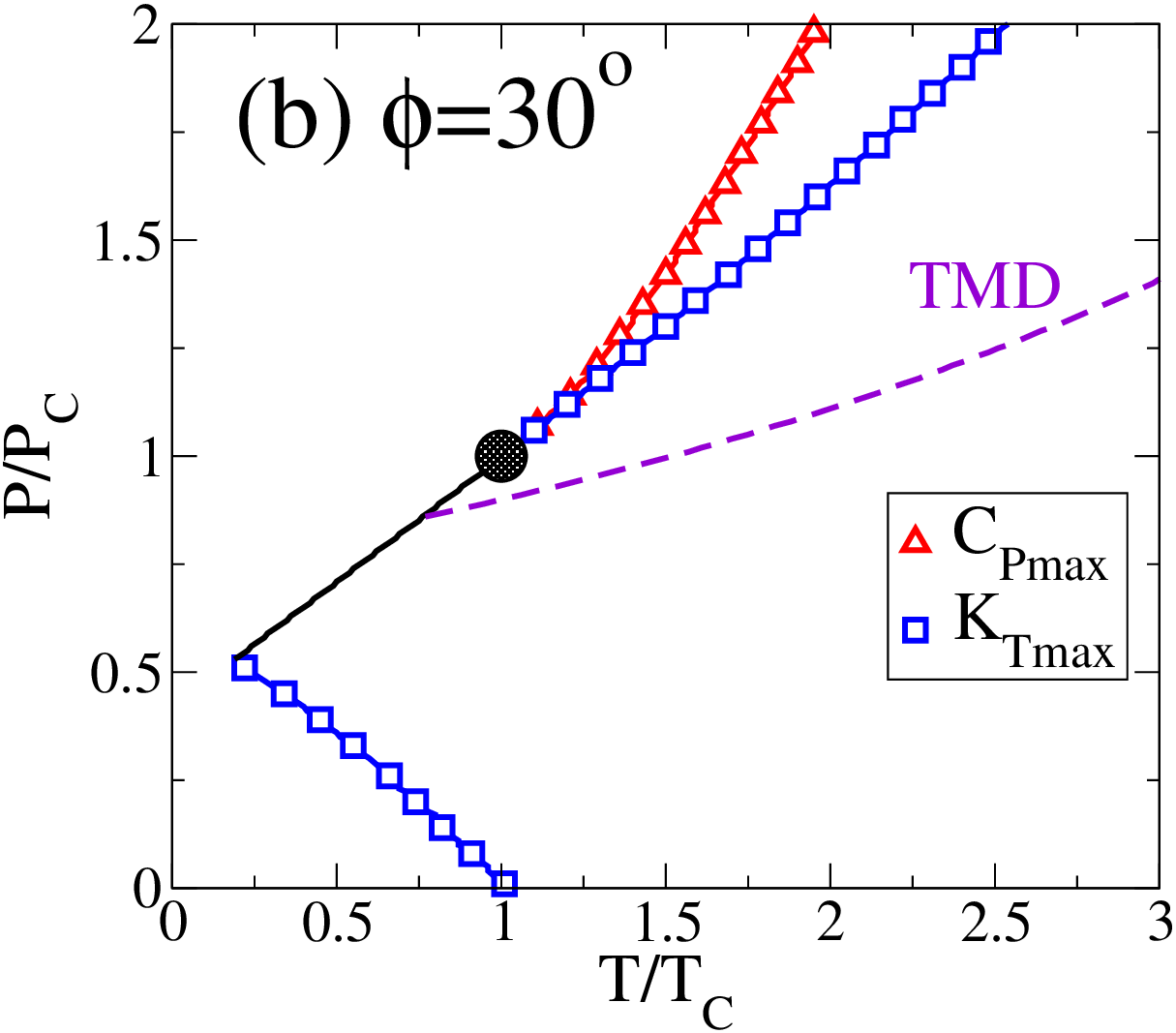}
%\subcaption{$(C_P,K_T)_{max}$ on P-T with $\phi=30^o$.}
\end{minipage}\\
\begin{minipage}[t]{0.4\linewidth}
\includegraphics[width=0.9\linewidth]{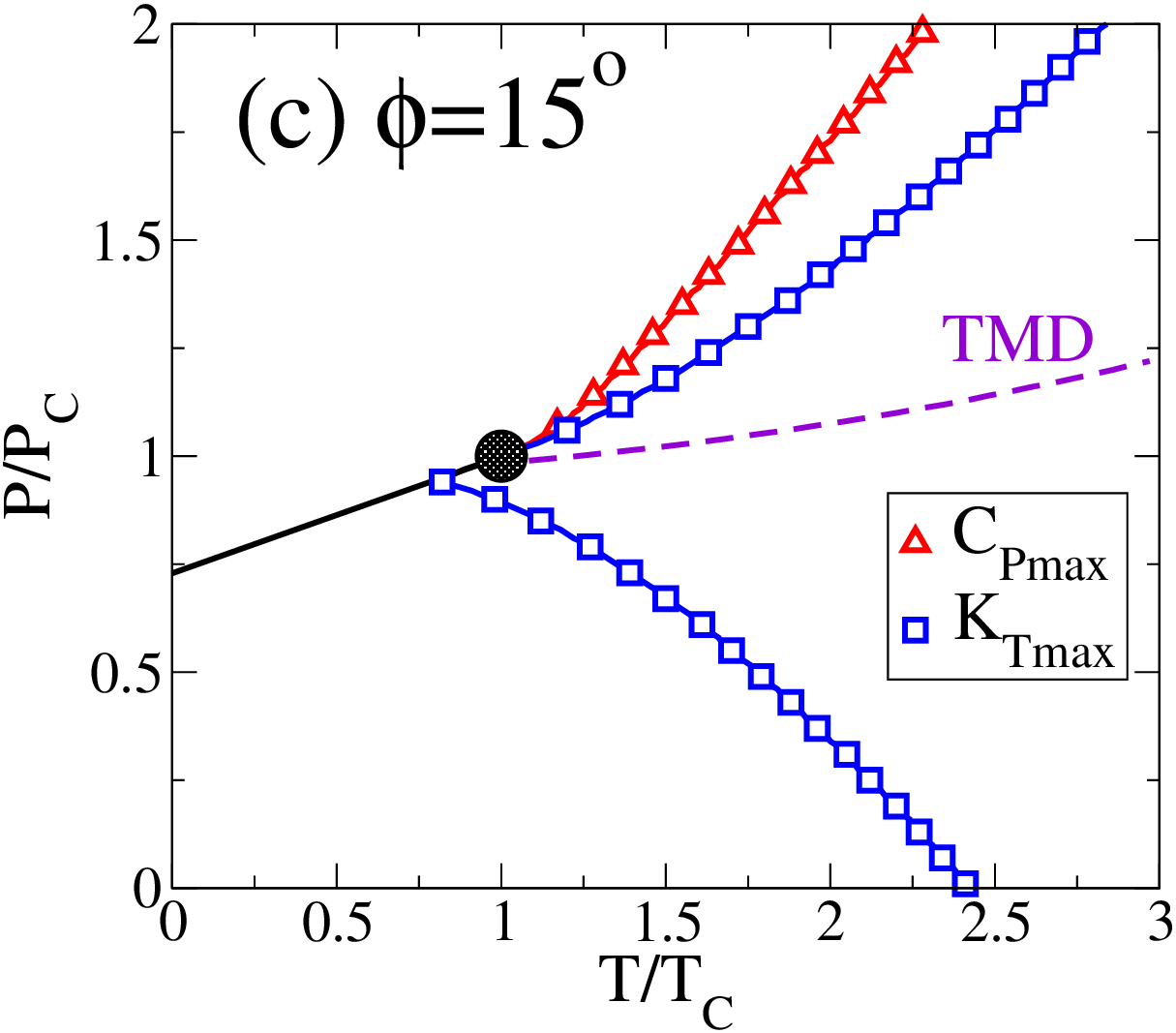}
%\subcaption{$(C_P,K_T)_{max}$ on P-T with $\phi=15^o$.}
\end{minipage}
\begin{minipage}[t]{0.4\linewidth}
\includegraphics[width=0.9\linewidth]{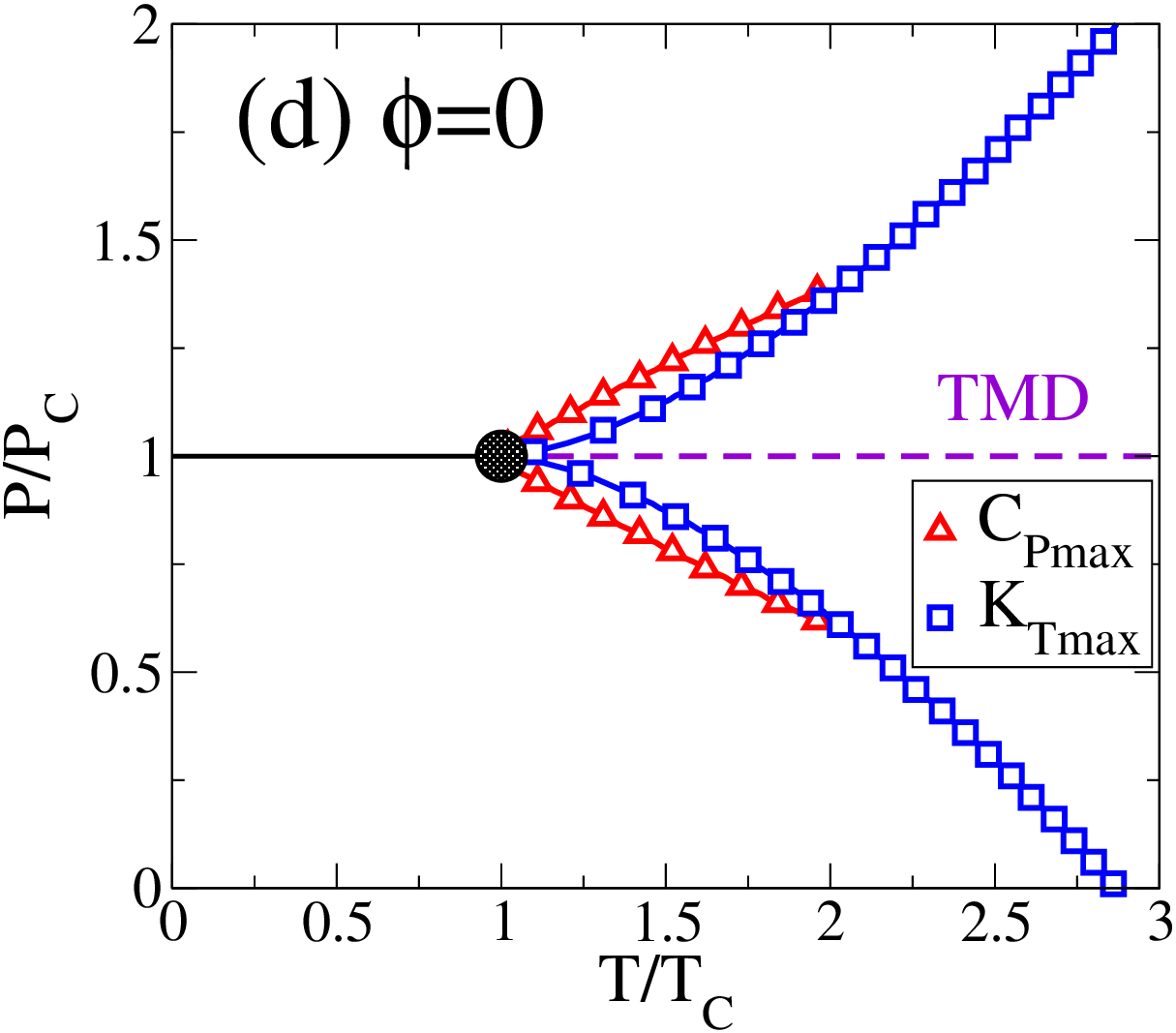}
%\subcaption{$(C_P,K_T)_{max}$ for $\phi=0$.}
\end{minipage}\\
\captionsetup{format=plain,justification=RaggedRight}
\caption{$C_P$ maxima and $K_T$ maxima obtained from the linear scaling
  theory of the critical point with $\phi=45^o, 30^o, 15^o, 0$. The
  coexistence line (solid black) has the slope $dP/dT=\tan \phi$.  One
  can see that for more positively sloped coexistence line with larger
  $\phi$, at $P>P_c$, $C_P$ maxima and $K_T$ maxima lines have longer
  overlapping segments. while for smaller $\phi$, the merged section is
  shortened, and disappears (no overlapping) when $\phi=0$, with a
  horizontal coexistence line. The second $K_T$ maxima line at $P<P_c$
  corresponding to the LDL spinodal can be seen for $\phi=30^o$ and
  $15^o$. Due to the lack of metastability of the states in the theory,
  the $K_T$ maxima at $P<P_c$ terminates at the coexistence line,
  instead in reality, it terminates at the minimum pressure point of the
  LDL spinodal.  For $\phi=0$ with horizontal coexistence line, a second
  $C_P$ maxima line at $P<P_c$ merge to the LLCP. As we know, the linear
  scaling theory can predict the critical phenomena only close to the
  LLCP, a closer study near the LLCP will follow.  Further, the TMD line
  (dash) is also shown, which is the locus of $\alpha_P=0$.  One notices
  that as the slope of coexistence line goes from positive to
  horizontal, the TMD line shifts from being below the LLCP, to right on
  top of it, which is consistent with our results in DMD
  simulation. Here $a=0.47$, and $k=0.47$. }
\label{CKmax}
\end{figure}

We study the behavior of $C_P$ and $K_T$ in systems with different
coexistence line slopes.  
Figure~\ref{CKmax} shows a plot of the loci of $C_P$ and $K_T$ maxima in
the $T$-$P$ plane.  Note that both the locus of $C_P$ maximum and the $K_T$
maxima line with the higher magnitude originate from the LLCP, and
coincide with each other close to the LLCP before they separate. The locus
of $C_P$ maximum bends towards a lower temperature than that of the
$K_T$ maximum line. As the slope of the coexistence line decreases, the
overlapping segments of the loci of $C_P$ and $K_T$ maxima shorten. When
the coexistence line is horizontal, the loci of $C_P$ and $K_T$ maxima
separate.

\begin{figure}[hbt]
\begin{minipage}[t]{0.4\linewidth}
\includegraphics[width=0.9\linewidth]{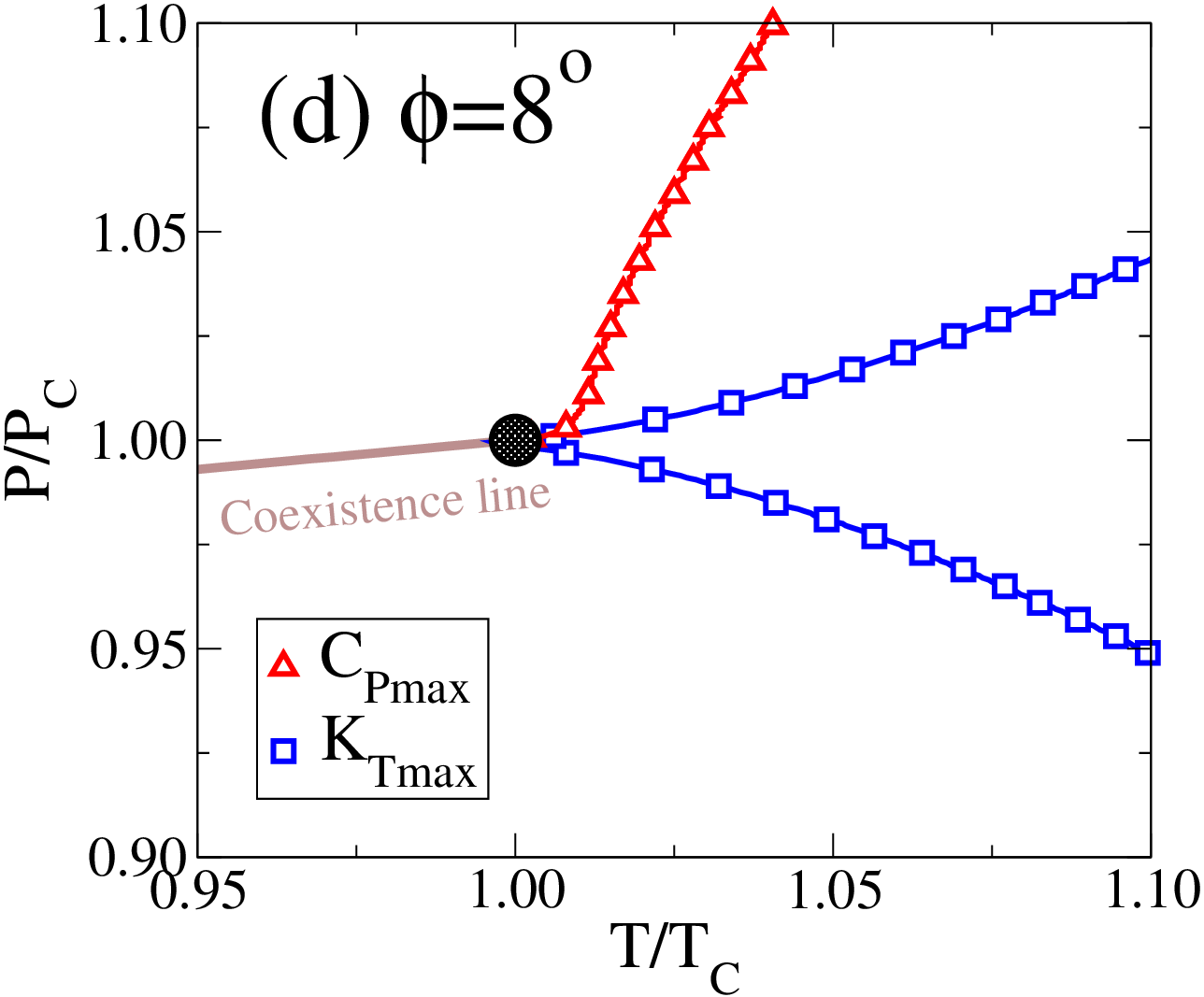}
%\subcaption{$(C_P,K_T)_{max}$ for $\phi=1.5^o$.}
\end{minipage}
\begin{minipage}[t]{0.4\linewidth}
\includegraphics[width=0.9\linewidth]{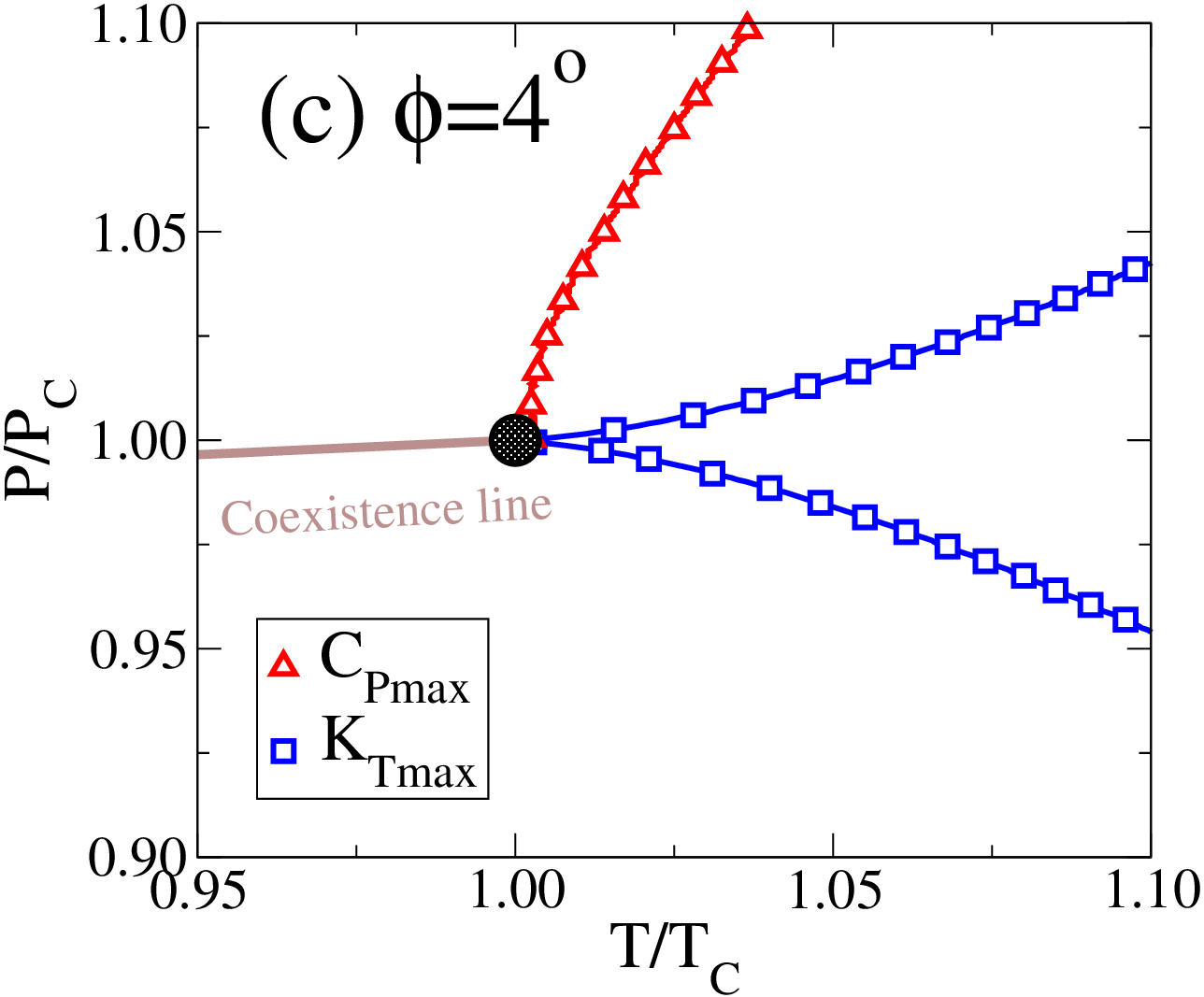}
%\subcaption{$(C_P,K_T)_{max}$ for $\phi=1^o$.}
\end{minipage}\\
\begin{minipage}[t]{0.4\linewidth}
\includegraphics[width=0.9\linewidth]{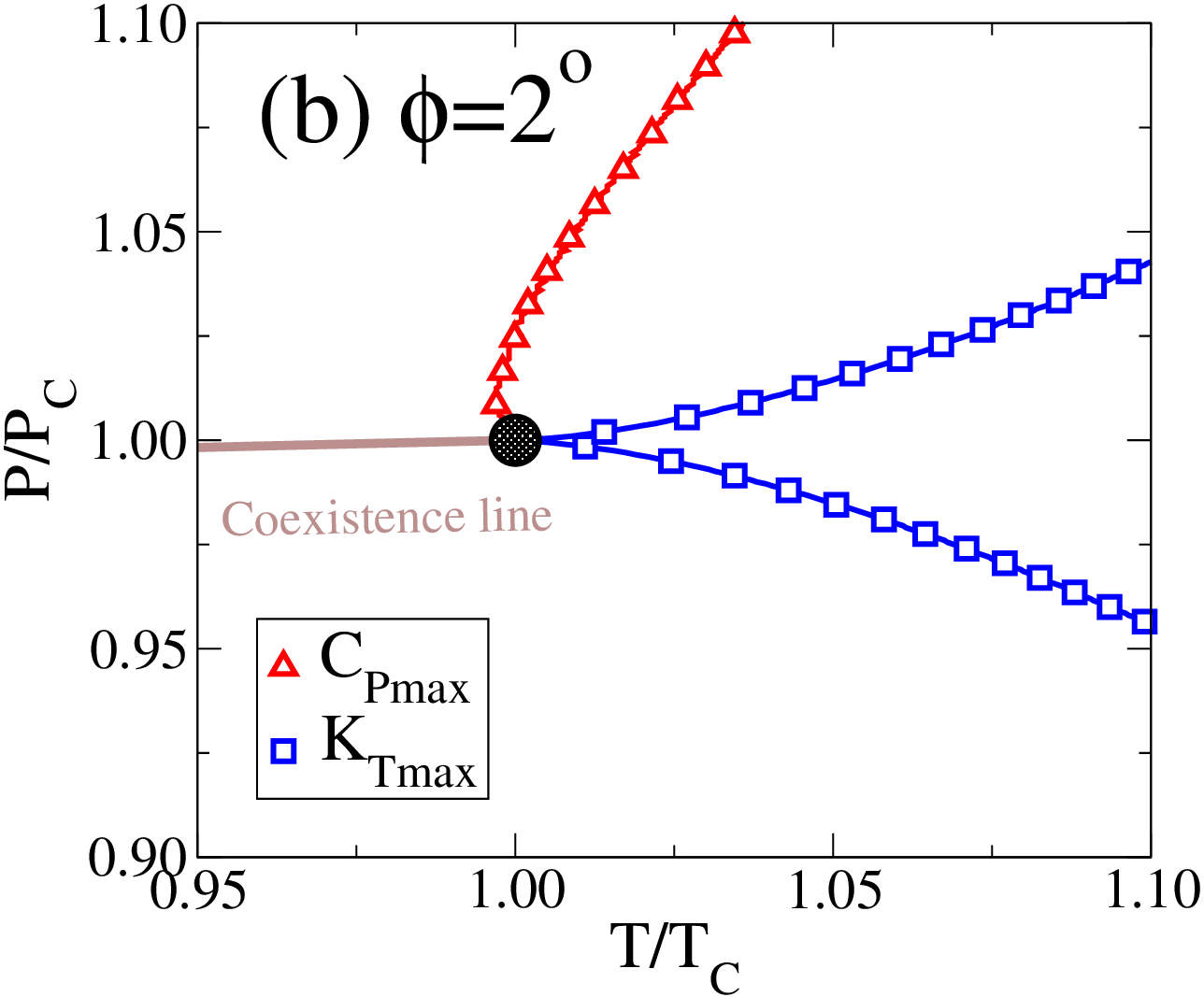}
%\subcaption{$(C_P,K_T)_{max}$ for $\phi=0.5^o$.}
\end{minipage}
\begin{minipage}[t]{0.4\linewidth}
\includegraphics[width=0.9\linewidth]{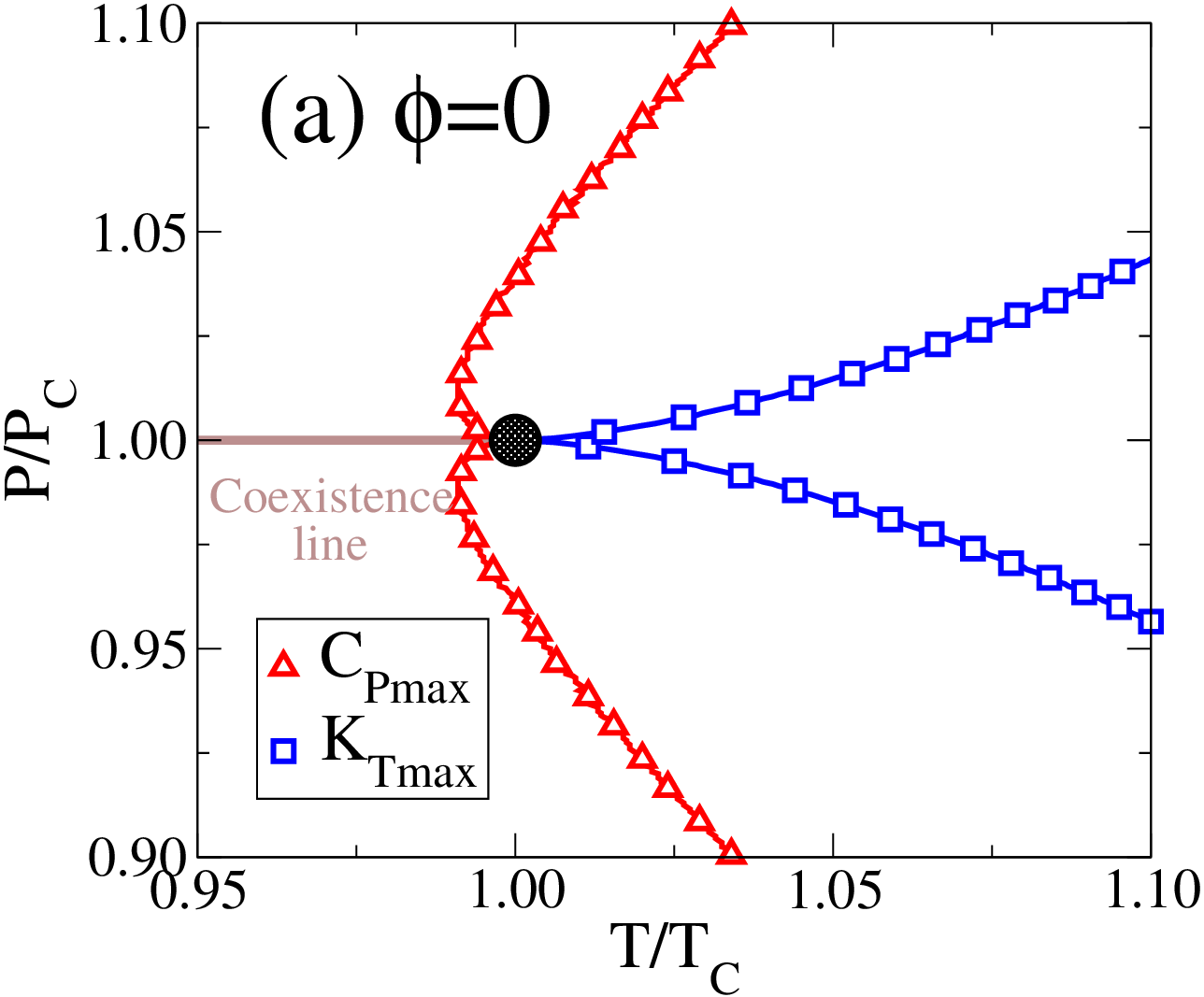}
%\subcaption{$(C_P,K_T)_{max}$ for $\phi=0$.}
\end{minipage}\\
\captionsetup{format=plain,justification=RaggedRight}
\caption{$C_P$ maxima and $K_T$ maxima loci on $T$-$P$ plane obtained
  from the linear scaling theory of the critical point, with small
  $\phi=8^{\circ}, 4^{\circ}, 6^{\circ}, 0$. One can see that as we
  decrease $\phi$, with the slope of coexistence line approaching zero,
  the $K_T$ maxima locus changes are not significant, while the $C_P$
  maxima become more and more vertical. For a horizontal slope, two
  $C_P$ maxima lines converge at $T<T_c$ and enter the LLCP horizontally
  from below. There is no overlapping between $K_T$ maxima and $C_P$
  maxima lines. Here we use $a=2$, and $k=1$, which is good to
  present the small or zero coexistence line case.}
\label{CKmax2}
\end{figure}

We then examine the systems with the slope of coexistence line very
close to zero with small $\phi$ as shown in Fig.~\ref{CKmax2}. We see
that the loci of $K_T$ maxima are similar, with two $K_T$ maxima lines
approaching the LLCP from $T>T_c$, but the $C_P$ maxima line deviates
from the $K_T$ maxima lines and its slope becomes more vertical, as
the slope of the coexistence line approaches zero.  For the horizontal
coexistence line, a second $C_P$ maxima line emerges, and converges with
the first $C_P$ maxima line at $T<T_c$. Both of the $C_P$ maxima lines enter the
critical point almost horizontally from $T<T_c$, while the $K_T$ maxima
lines enter from $T>T_c$ horizontally. 
Thus in this case, close to the LLCP at $T \geq T_c$, no $C_P$ maxima can be found. 
According to the linear scaling theory, the actual
$C_P$ maxima may exist below $T_c$ where, in our simulations, they are
buried in the crystallization region or in glassy states with no
equilibrium data for analysis.  Further, there is no convergent behavior
of the $C_P$ maxima and $K_T$ maxima near the LLCP in the horizontal
coexistence line case. The Widom line, approximated by the locus of
density fluctuation maxima, is the convergent loci of the $K_T$ maxima,
while the identification of the Widom line using the $C_P$ maxima is no
longer a fruitful method.  

\begin{figure}[htbp]
\begin{minipage}[t]{0.4\linewidth}
\includegraphics[width=0.9\linewidth]{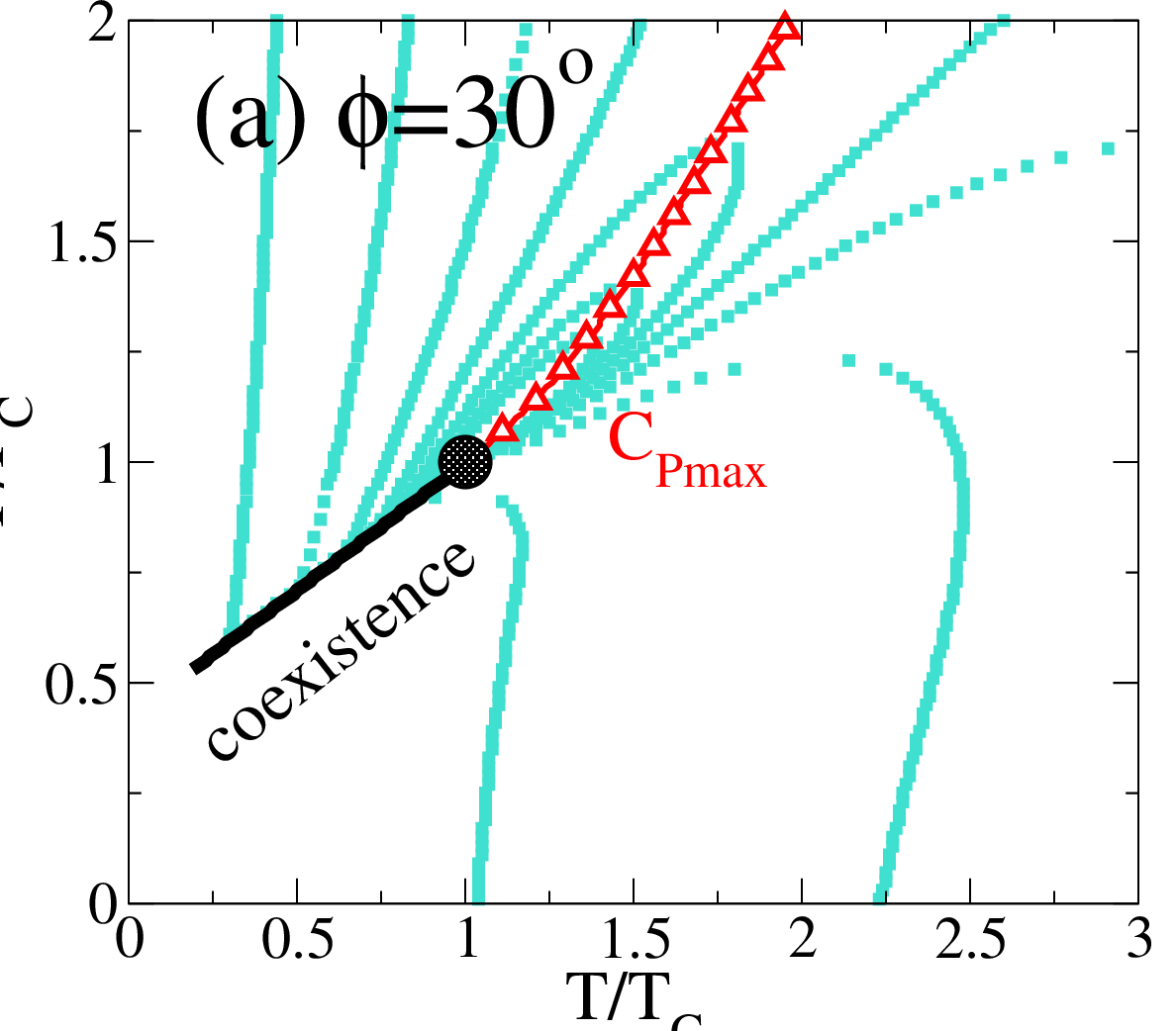}
%\subcaption{$C_P$ on P-T with $\phi=30^o$.}
\end{minipage}
\begin{minipage}[t]{0.4\linewidth}
\includegraphics[width=0.9\linewidth]{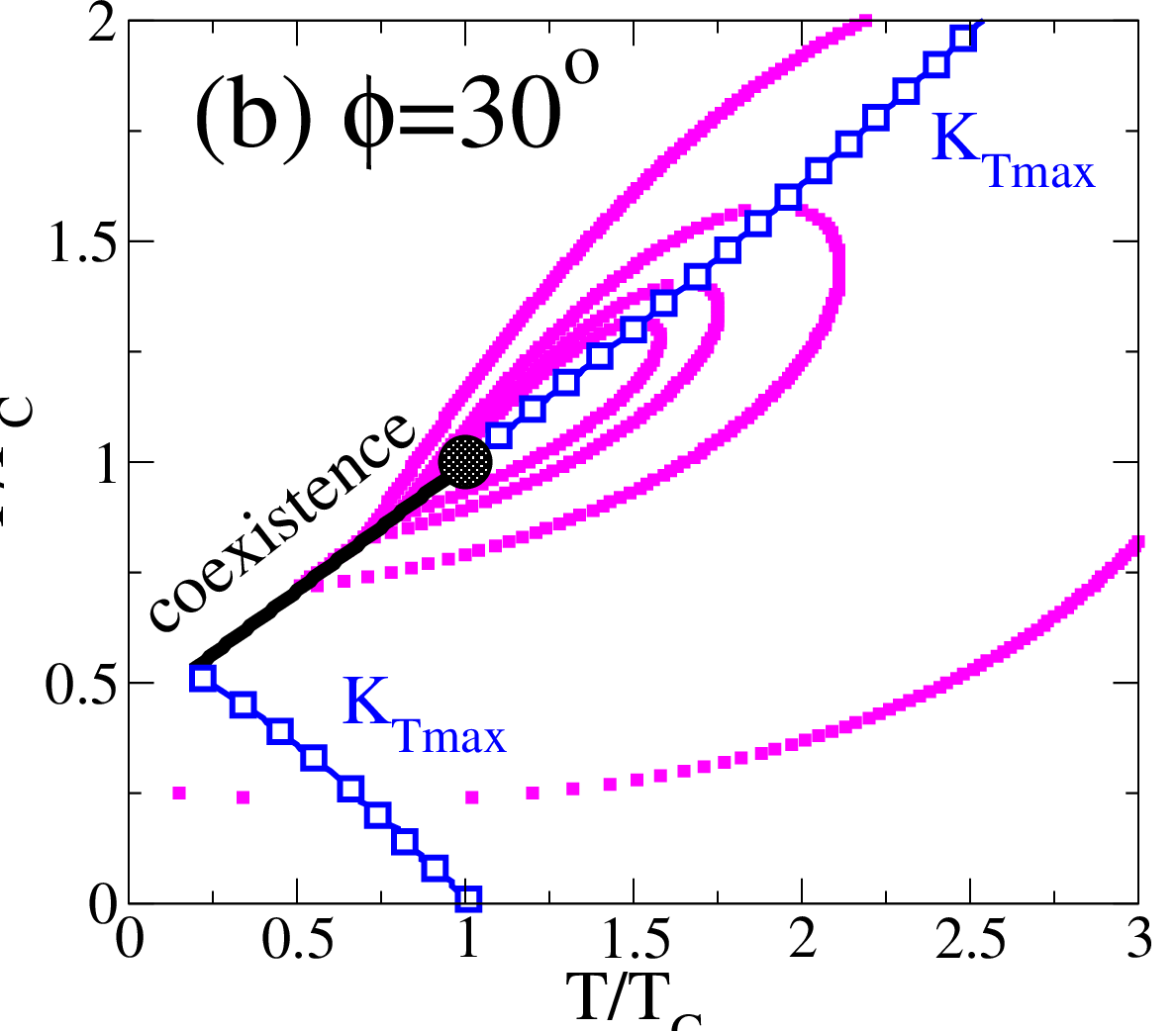}
%\subcaption{$K_T$ on P-T with $\phi=30^o$.}
\end{minipage}\\
\vspace*{0.2cm}
\begin{minipage}[t]{0.4\linewidth}
\includegraphics[width=0.9\linewidth]{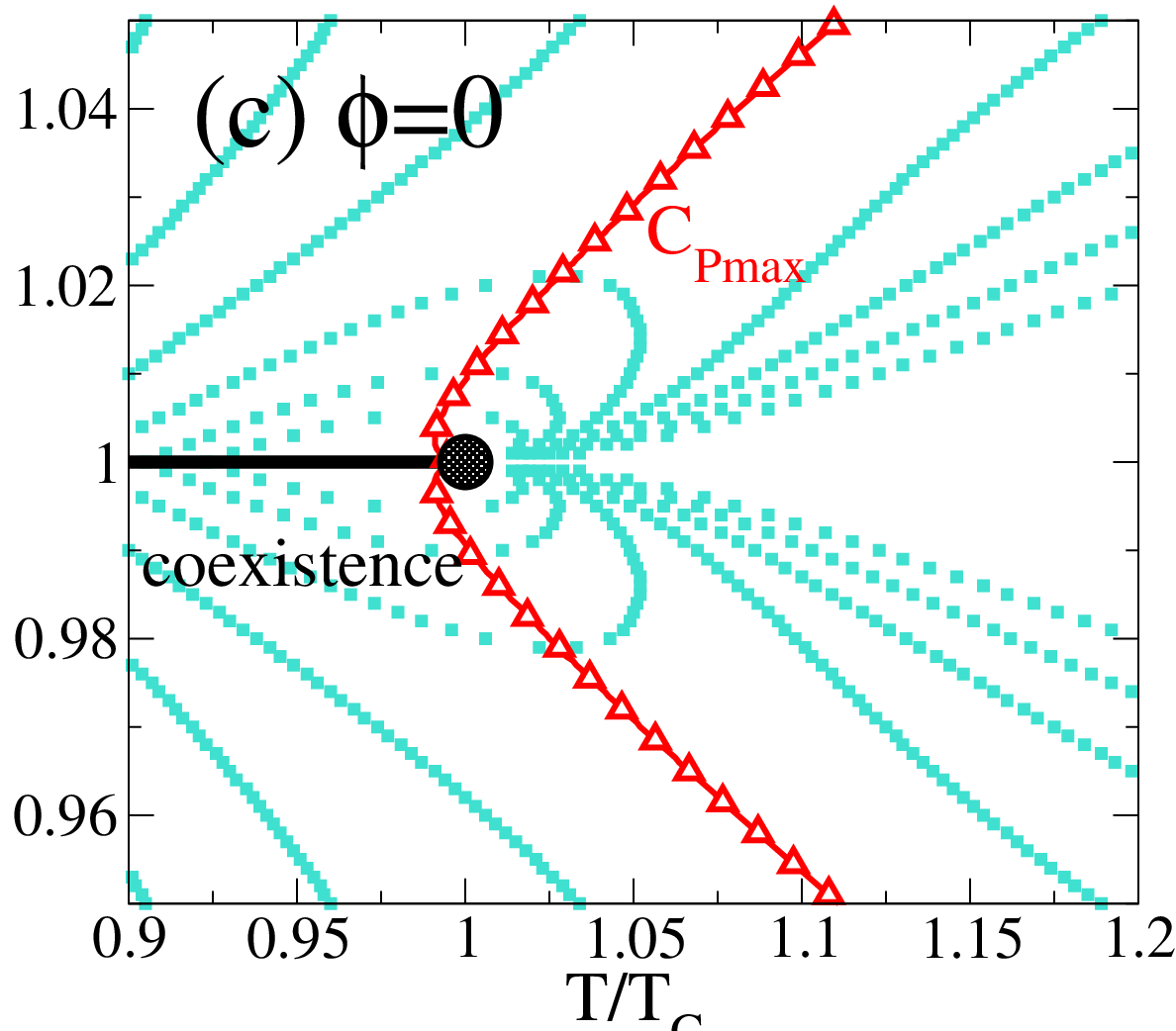}
%\subcaption{$C_P$ on P-T with $\phi=0$.}
\end{minipage}
\begin{minipage}[t]{0.4\linewidth}
\includegraphics[width=0.9\linewidth]{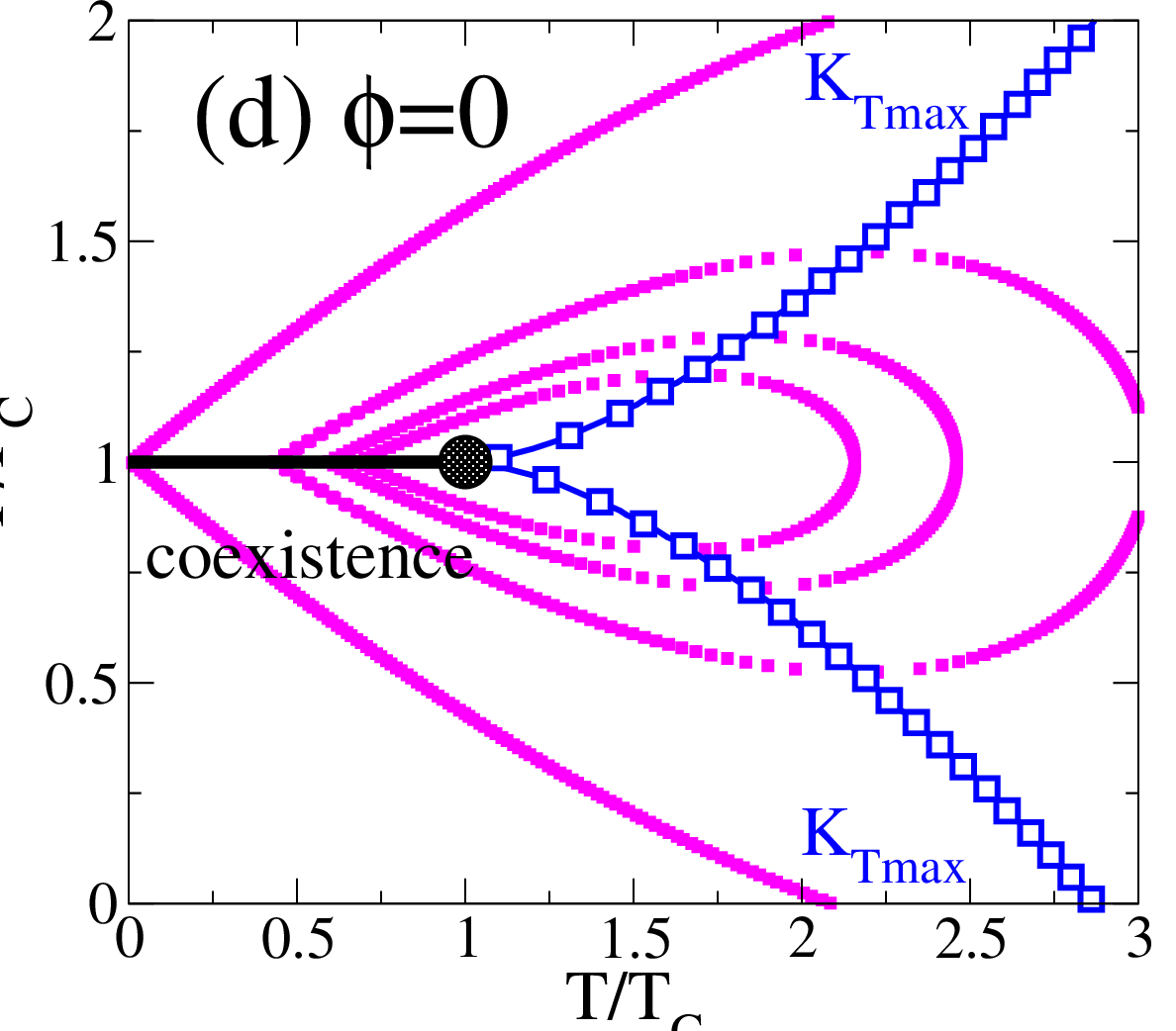}
%\subcaption{$K_T$ on P-T with $\phi=0$.}
\end{minipage}\\
\captionsetup{format=plain,justification=RaggedRight}
\caption{Comparison of $C_P$ and $K_T$ from linear scaling theory of the
  critical point.  (a,b) $C_P$ and $K_T$ for the positively sloped
  coexistence line with $\phi=30^o$, respectively. The lines of equal
  $C_P$ and $K_T$ are shown. Both $C_P$ and $K_T$ contours loop around
  LLCP, with the locus of the maxima follow the coexistence line into
  the one-phase region.  (c,d) $C_P$ and $K_T$ for the horizontal
  coexistence line with $\phi=0$, respectively.  One can see that $C_P$
  and $K_T$ are both symmetric below and above $P_c$. For clarity, (c)
  displays a narrow vicinity of the LLCP. Here we choose $a=0.47$, and
  $k=0.47$.}
\label{CK_B}
\end{figure}

We also find out that when the slope of the coexistence line
is positive, the contours of $C_P$ and $K_T$ go around the LLCP, with
the pressure maxima following the extension of the coexistence line into
the one-phase region [Fig.~\ref{CK_B}(a-f)]. When the slope of
coexistence line is horizontal, there are symmetric contours above and
below the critical pressure.  This matches well with what we find in our
simulation [Figs.~\ref{CP_equi} and \ref{KT_equi}].

\section{Glass transition}

By decreasing the $b/a$ ratio, the LLCP is pushed into a metastable
region with respect to crystallization, where the system is close to the
glass transition (GT). We hence investigate the relationship between
LLPT and GT in systems with different slopes of coexistence line.

While the inability to equilibrate in this domain was noted by Gibson
and Wilding in their seminal study \cite{gibsonWilding}, they did not discuss
the ``glass transition'' (which is not a true transition in the
thermodynamic sense). The GT is a concept useful for describing the
manner in which viscous liquid systems fall out of equilibrium on
cooling, or regain it during heating. It is better described as a
``glass transformation zone'' within which the system is neither fully
arrested nor fully equilibrated. It may be studied in simulation, as it
is in the laboratory \cite{OguniJCP1983}, by ``scanning calorimetry.''
In scanning calorimetry the enthalpy is monitored continuously as the
system attempts, and increasingly succeeds, to explore all its degrees
of freedom as temperature rises from low values where all motions except
vibrations are frozen out \cite{Angell95sr}. Typically the range of
temperature over which the transition extends is the range needed to
change the relaxation time by two orders of magnitude, so it depends on
the temperature dependence of the relaxation time \cite{Angell83jcp}. 
Being kinetic in nature, this transition is
hysteretic, as seen in our simulations. While it is usually studied by
scanning calorimetry, it can equally be studied by volumetric methods.
	
\begin{figure}[htbp]
\begin{minipage}[t]{0.32\linewidth}
\includegraphics[width=0.9\linewidth]{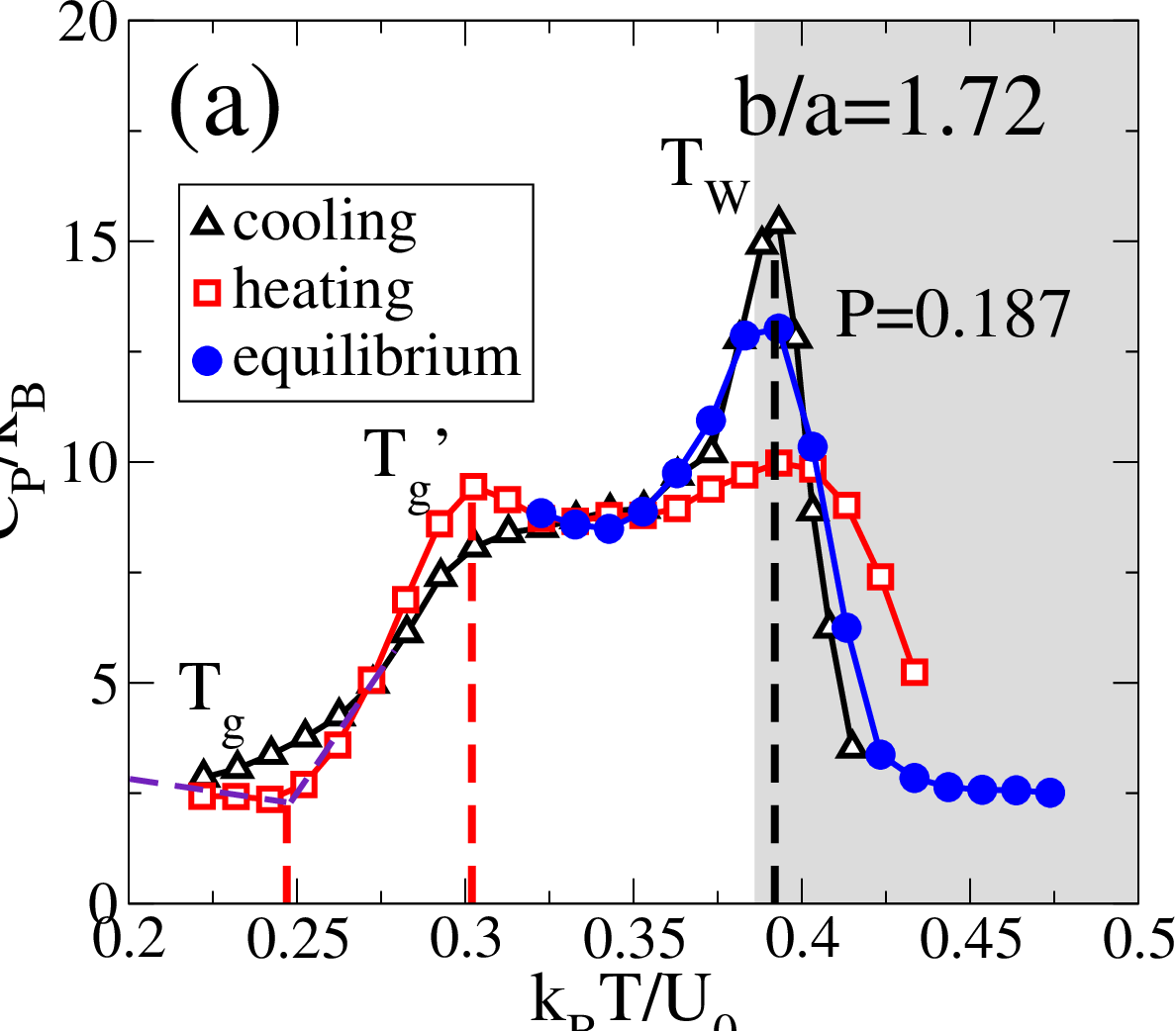}
\end{minipage}
\begin{minipage}[t]{0.32\linewidth}
\includegraphics[width=0.9\linewidth]{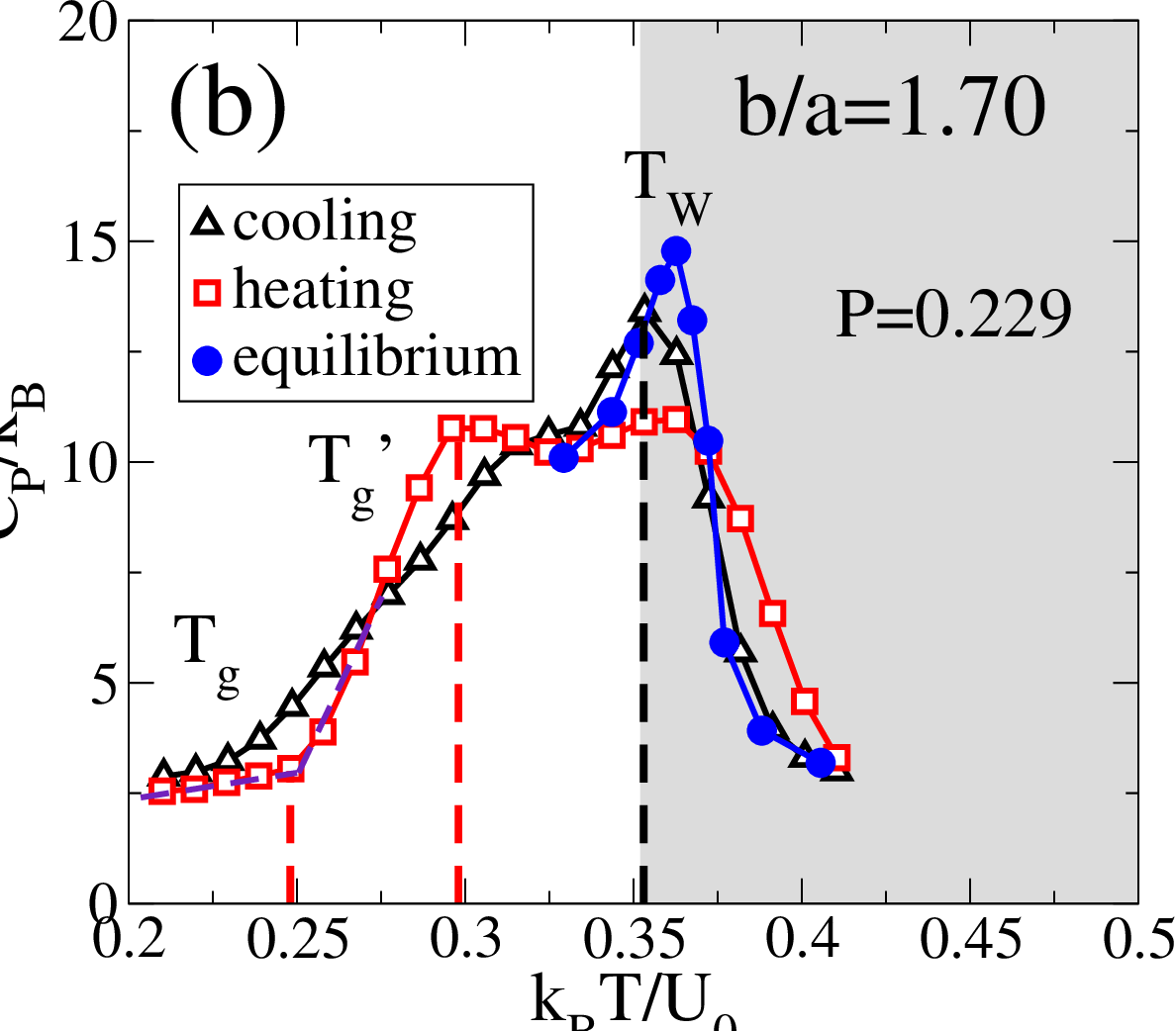}
\end{minipage}
\begin{minipage}[t]{0.32\linewidth}
\includegraphics[width=0.9\linewidth]{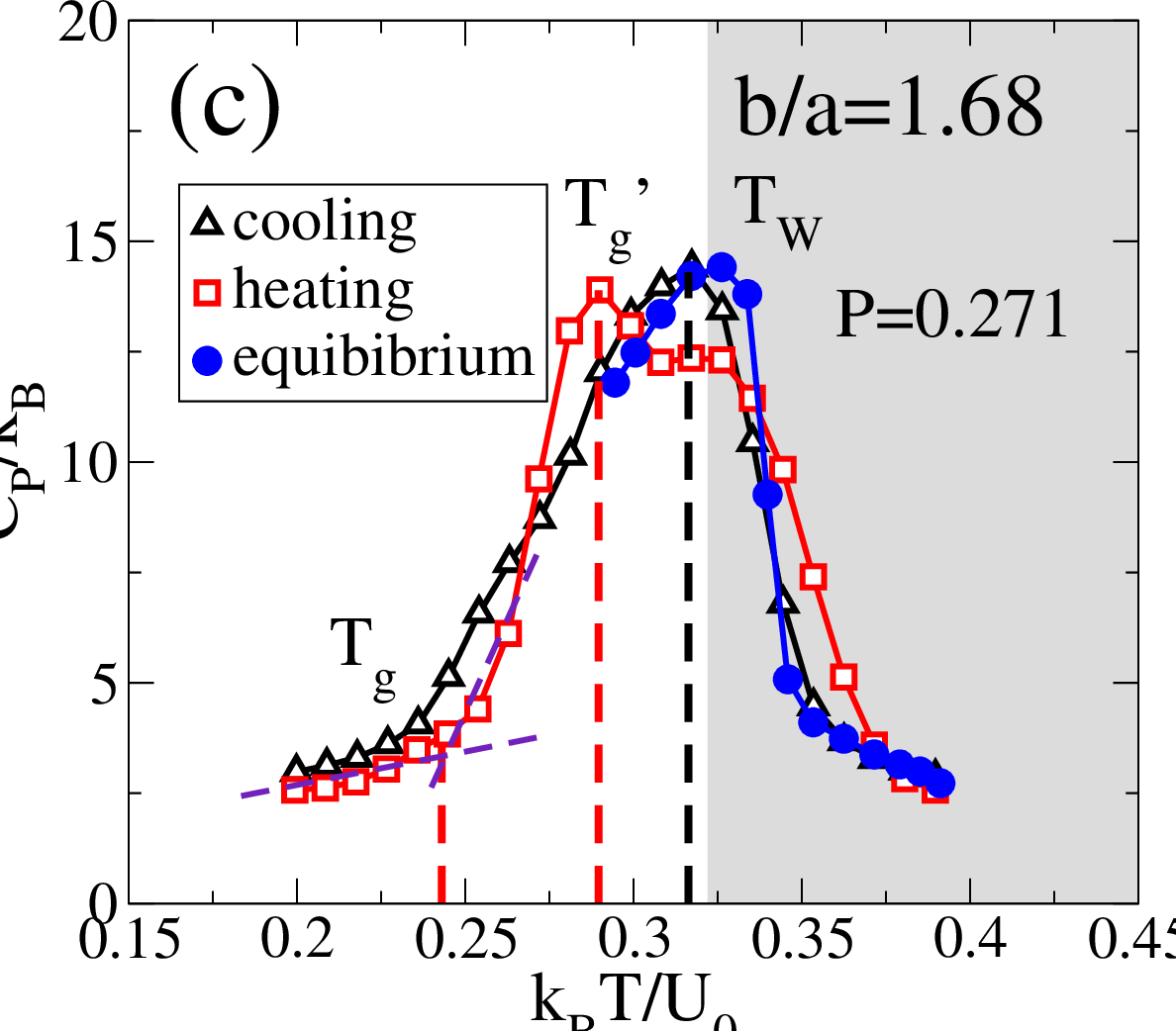}
\end{minipage}\\
\begin{minipage}[t]{0.32\linewidth}
\includegraphics[width=0.9\linewidth]{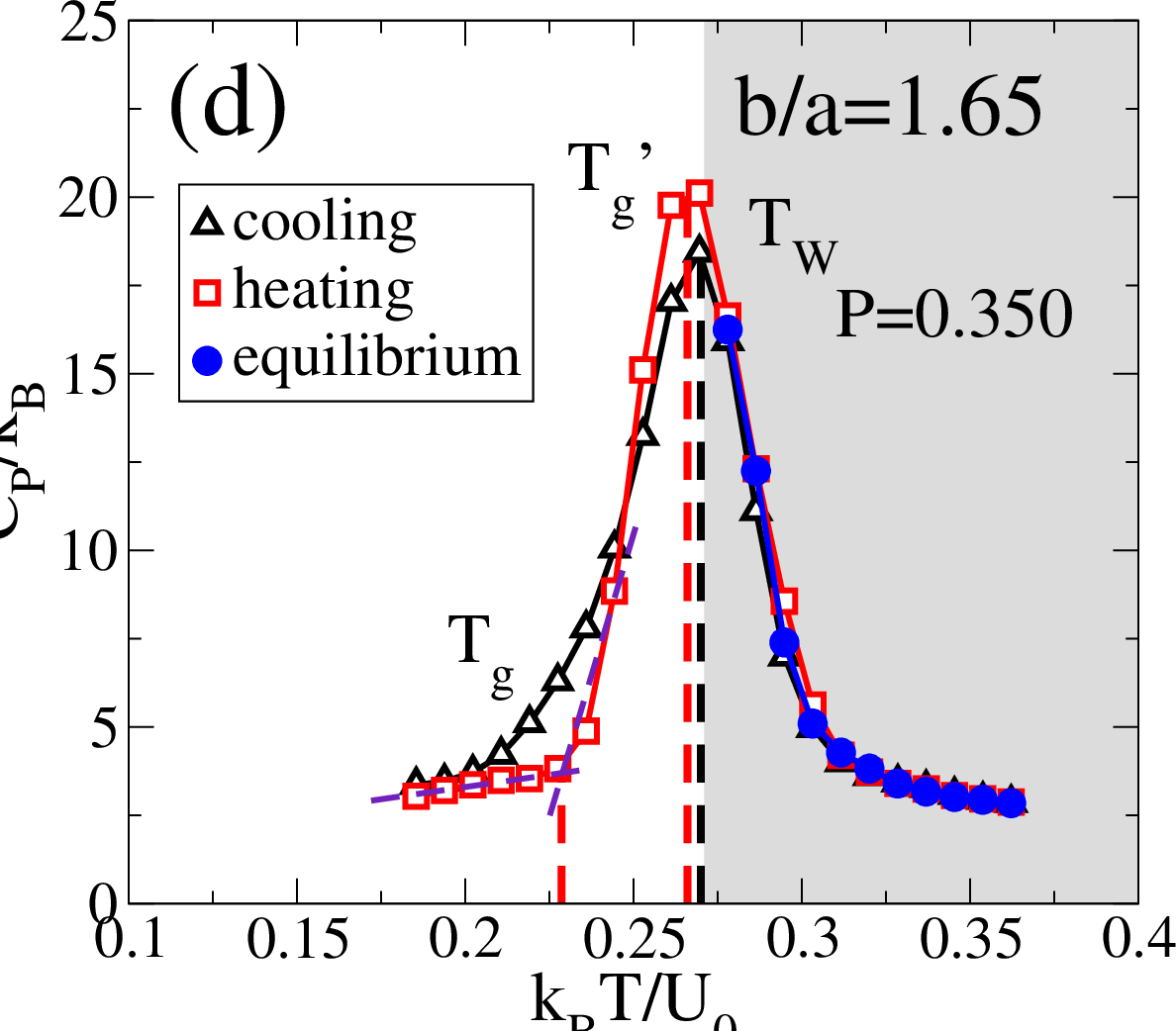}
\end{minipage}
\begin{minipage}[b]{0.32\linewidth}
\includegraphics[width=0.9\linewidth]{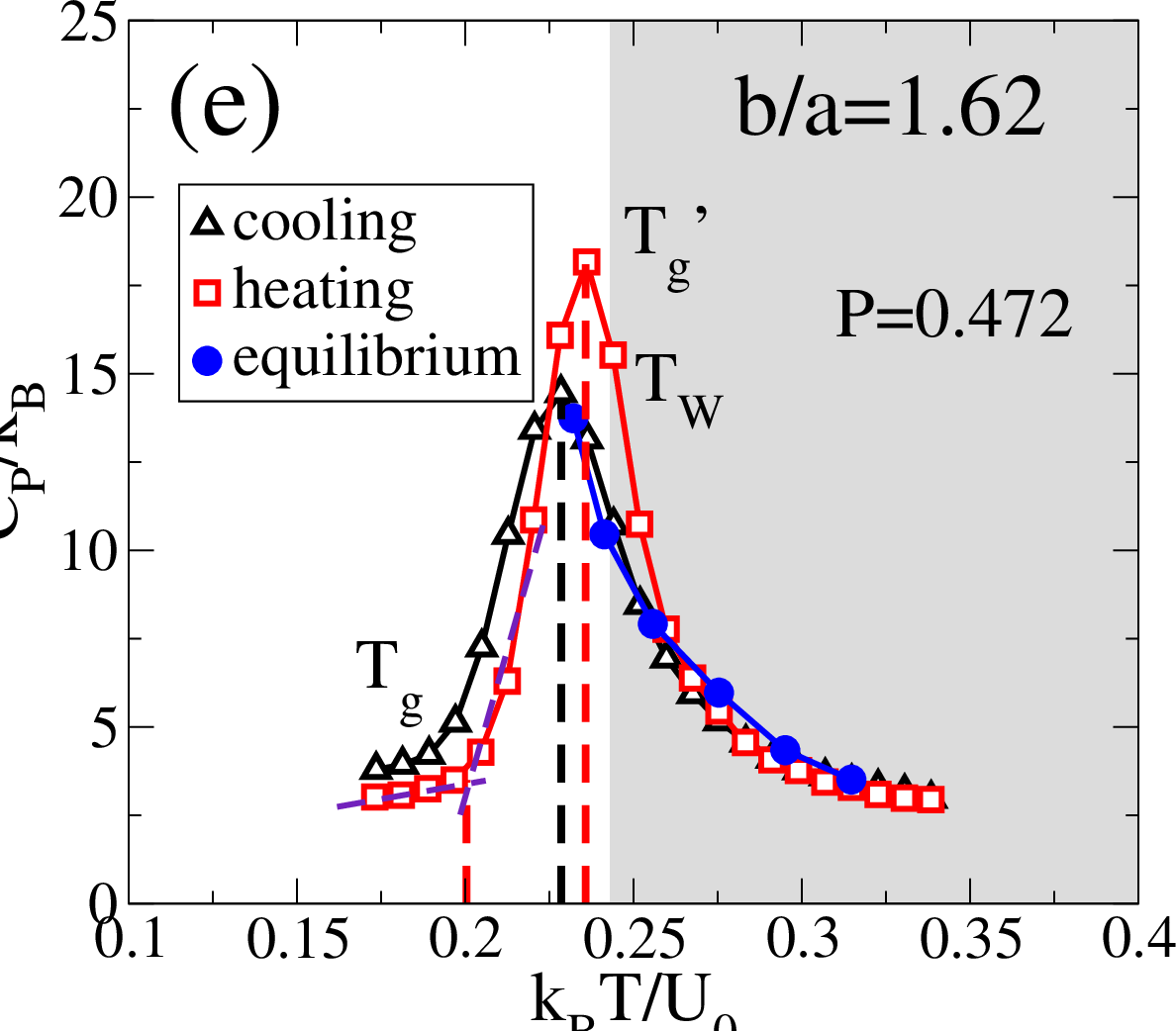}
\end{minipage}
\begin{minipage}[b]{0.32\linewidth}
\includegraphics[width=0.9\linewidth]{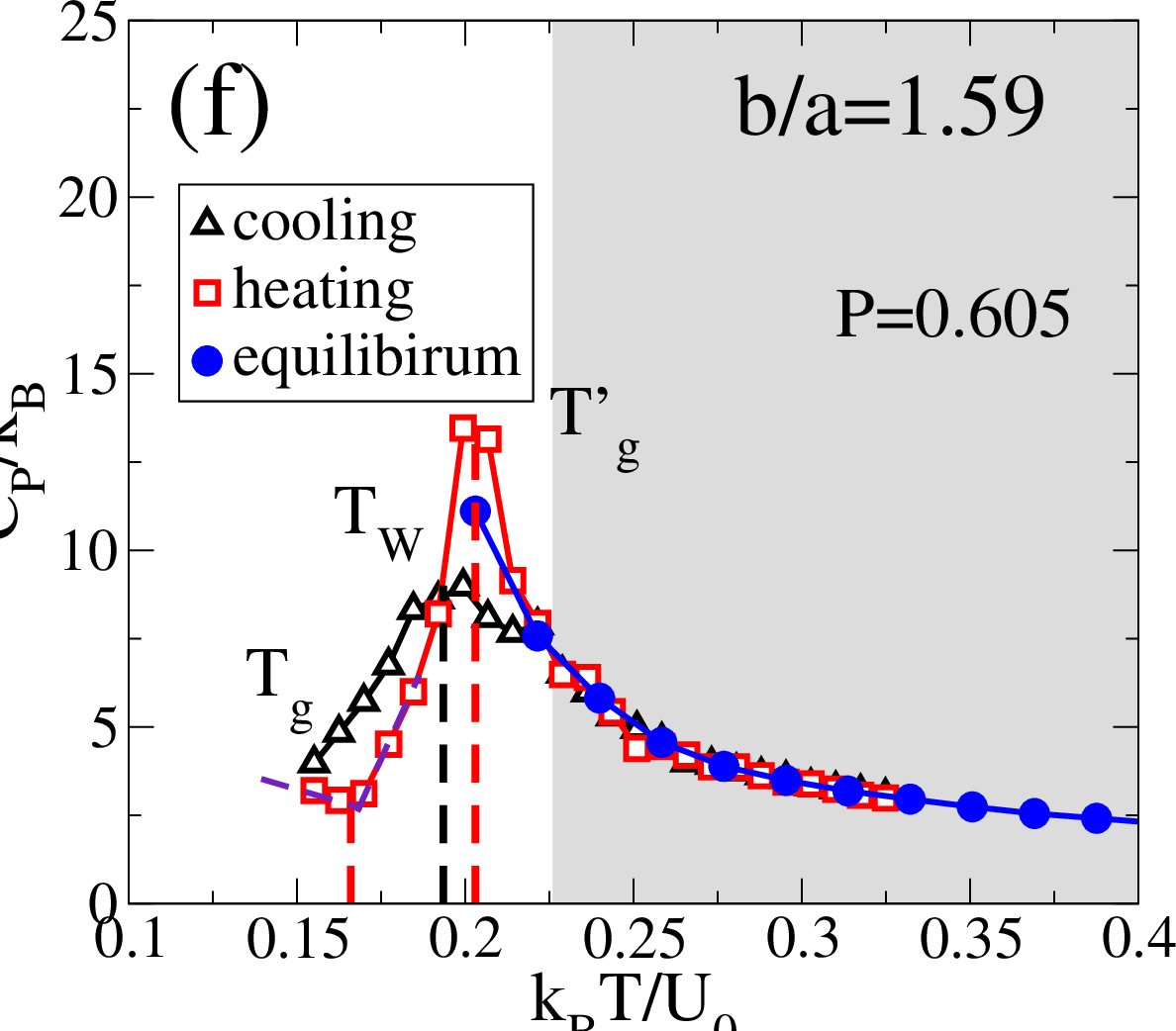}
\end{minipage}\\
\captionsetup{format=plain,justification=RaggedRight}
\caption{The comparison of $C_P$ upon slow cooling and heating for a
  selection of modified Jagla models.  The cooling/heating rate is
  $q_c=q_h=10^{-6}q_0$. An isobaric path is selected at $P$ closely
  above $P_c$ in the one-phase region. The equilibrium values of $C_P$
  are also plotted for comparison.  The gray area is the temperature
  region for $T \geq T_c$.  (a-c) For $b/a=1.72, 1.70$ and $1.68$, two
  peaks of $C_P$ upon heating can be observed.  The high temperature
  peak $T_W$, arising from the presence of the Widom line, is related to
  the LLPT.  The low temperature peak upon heating, $T'_g$, corresponds
  to the ergodicity restoration slightly above the glass transition
  $T_g$ obtained from standard construction (dash line).  One can see
  that the distance between the glass transition peak $T'_g$/$T_g$ and
  $T_W$ decreases as the $b/a$ value decreases, with the LLCP moving
  closer to the glass transition.  (d-f) For $b/a=1.65,1.62$ and $1.59$,
  upon heating, only one peak of $C_P$ can be observed, and this peak
  shifts below $T_c$, while for other cases it is well above $T_c$.  One
  also notices that system shows a prominent $K_T$ peak in the studied
  $P$ region (Fig.~5).  In these models with small coexistence line
  slope, the enthalpy fluctuations play less role in the critical
  fluctuations, while the density fluctuations is the leading term. And
  as the LLCP being pushed closer to $T_g$, the critical fluctuations
  become suppressed by the glass transition, where $C_P$ only shows one
  peak upon heating.  }
\label{CP_cHE_all}
\end{figure}

The glass temperature $T_g$ can be defined as the point at which the
uptake of configurational enthalpy begins to commence (onset $T_g$, the
value usually reported by experimentalists), or the temperature at which
equilibrium (ergodicity) is fully restored, $T'_g$. Each is defined
diagrammatically in Fig.~\ref{CP_cHE_all}, the distance between the two
amounting to about $25\%$ of the absolute value. It is a much more
diffuse phenomenon than in the laboratory where the width is
only $5\%$ of the absolute value (due to the increased temperature
dependence of the relaxation time near the laboratory $T_g$)
\cite{Angell83jcp}. 

We estimate $T_g$ and $T'_g$, as in the laboratory experiment, by
plotting the derivative of the enthalpy (apparent specific heat), during
cooling and heating the systems through GT along isobars slightly above
the critical pressure $P_c$, at constant cooling/heating rate $q_c = q_h
= 10^6q_0=10^7$K/sec [Fig.~\ref{CP_cHE_all}]. We find that for upscans
in the positively sloped coexistence line case ($b/a=1.72$), $C_P$ shows
two well-separated peaks [Fig.~\ref{CP_cHE_all}(a)]. The high
temperature peak $T_W$, is related to the fluctuations associated with
the LLPT and is used to locate the Widom line. The second peak (at the
lower temperature $T=T'_g$) is an ``overshoot'' phenomenon due to
ergodicity restoration kinetics. It is seen in most laboratory systems
(but not polymers) and is not observed during cooling, (a measure of the
hysteretic character of the glass transition). The lower $T_g$ is
obtained from the standard construction (dashed line)
\cite{AngellScience2008}.

Similar well-separated $T_W$ and $T'_g$ peaks can be observed for
$b/a=1.70$, and $1.68$ [Fig.~\ref{CP_cHE_all}(b,c)], but the temperature
difference between the two peaks shrinks as the $b/a$ value
decreases. When $b/a < 1.68$, the ``normal'' and critical fluctuations
merge because of the similarity in their time scales, but study of the
density fluctuations as reflected in the compressibility of
Fig.~\ref{KT_all} shows that indeed $T_c > T_g$, and the critical point
is not suppressed by the kinetics of ``background'' enthalpy
fluctuations (as the collected data in Fig.~\ref{CP_cHE_all} might imply
at first sight). Rather, what is happening is that the critical
fluctuations in enthalpy are losing their thermodynamic strength due to
the vanishing of any enthalpy difference between the alternative phases
that is dictated by the Clapeyron equation for horizontal
coexistence lines (see Fig.~\ref{Co_slope} for $b/a=1.59$). Thus, at
$b/a = 1.59$, the apparent specific heat plot, notwithstanding the
proximity of the critical point, is indistinguishable in character from
that previously reported for the glass transition of the low density
liquid at pressures well below $P_c$ in the earlier study of Xu et
al.~\cite{XuJCP2009}.

\begin{figure}[hbt]
\centerline{\includegraphics[width=12cm]{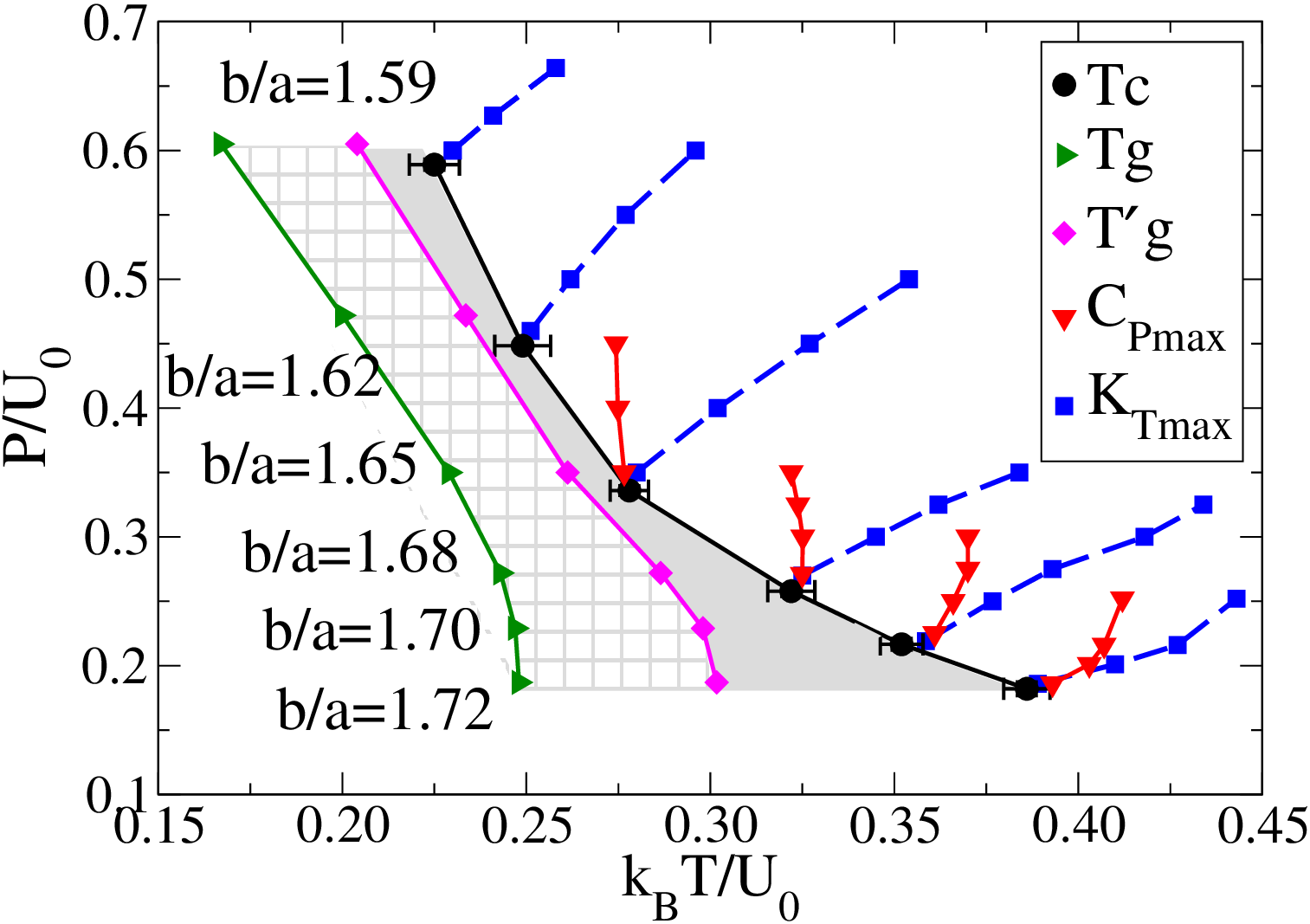}}
\captionsetup{format=plain,justification=RaggedRight}
\caption{Relative positions of the glass transition temperature $T_g$
  (onset), $T'_g$ (upper limit), critical temperature $T_c$, and the
  locus of $C_P$ and $K_T$ maxima near the LLCP, for models of different
  $b/a$. Only the high pressure branch of the $K_T$ maxima locus is
  shown.  The LLCP shifts to lower temperature and higher pressure as
  $b/a$ decreases.  $T'_g$ follows the same trend of $T_c$, but the
  temperature difference (gray area) between $T'_g$ and $T_c$ decreases
  as $b/a$ decreases. The glass transformation range (between $T_g$ and
  $T'_g$) separating glass from liquid is shown (hashed area).  For $b/a
  \leq 1.65$, $T_c$ gets pushed close to $T'_g$, where in isobaric
  cooling/heating, only one peak of $C_P$ can be found, instead of two
  well-separated maxima ($b/a=1.72-1.68$).  The locus of $C_{P}$ maxima
  increasingly separates from the locus of the $K_T$ maxima with
  decreasing $b/a$, and is no longer seen for $b/a < 1.65$. The $K_T$
  maxima line is a better representative of the Widome line for the case
  of small and zero sloped coexistence line. $T_c$ lies in the ergodic
  domain for the $b/a$ studied. For $b/a<1.59$, the LLCP cannot be
  obtained in our study by equilibrium molecular dynamics.  }
\label{Widom}
\end{figure}

Figure~\ref{Widom} shows that the critical fluctuation domain becomes
increasingly related to the slow (glassy) dynamics domain as the
repulsive potential becomes steeper (second length scale approaches the
first, more closely as $b/a$ decreases). It is unfortunate that the
increase of the equilibrium melting point, in the same $b/a$ range,
throws the system into conflict with still another, and faster kinetics,
that of crystal nucleation, so that the relation between the first two
can no longer be followed for smaller $b/a$.

Just as the mixing of Lennard-Jones (LJ) particles has made possible the
study of supercooled and glassy states of LJ, so might the mixing of
Jagla particles of different dimensions and attractive well depths, make
possible more extended studies of the critical point/glass temperature
relations. Note that in the glass-forming LJ mixtures there is no
suggestion of stable domain critical points, though specific heats in
excess of vibrational values do increase sharply as temperature
decreases.

As a final remark, it is notable that the strengths of the response
functions specific heat and compressibility in laboratory molecular
glassformers also vary in opposite directions as $T_g$ is approached,
the case of o-terphenyl being the best documented so far
\cite{AngellNaturePhys2011}. The relationship is similar to that of the
response functions maxima for the present model at small $b/a$
demonstrated in Figure ~\ref{CP_all} and ~\ref{KT_all} showing that
there is a temperature interval near the glass transition where $C_P$ is
increasing upon cooling while the $K_T$ decreasing, with the only
difference that there is no stable second critical point in the
laboratory case.

\section{Summary and Conclusions}

\begin{figure}[htbp]
\begin{minipage}[t]{0.4\linewidth}
\includegraphics[width=0.9\linewidth]{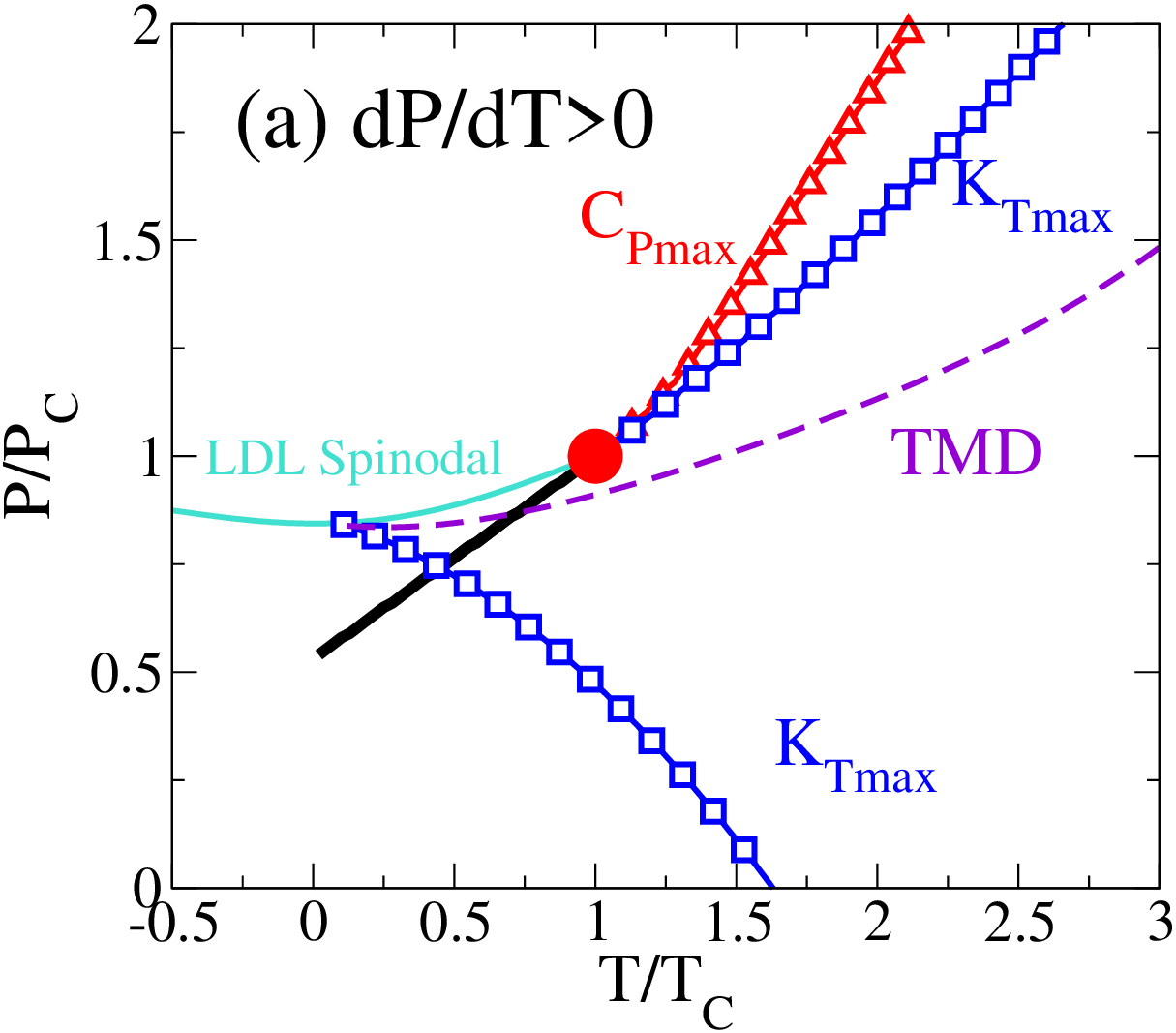}
\end{minipage}
\hspace{0.2cm} 
\begin{minipage}[t]{0.4\linewidth}
\includegraphics[width=0.9\linewidth]{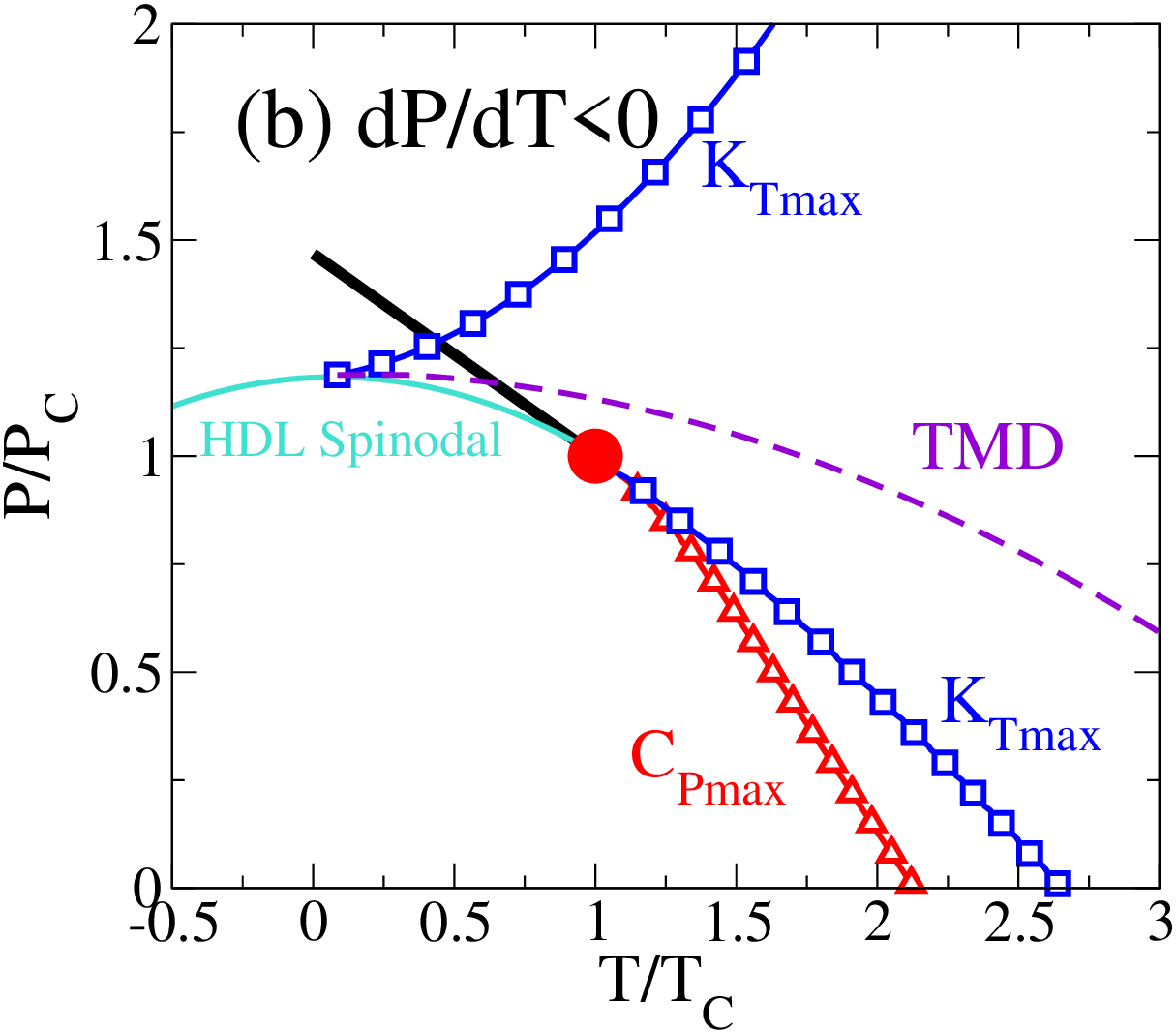}
\end{minipage}\\
\captionsetup{format=plain,justification=RaggedRight}
\caption{The relationship between the slope of the coexistence line and
  the behavior of the Widom line and the TMD line.  (a) For the case of
  positively sloped coexistence line, the $C_P$ maxima line originates
  from the LLCP and extends into the one-phase region as a continuation
  of the coexistence line, while the $K_T$ shows two maxima lines. One
  $K_T$ maxima line is related to critical fluctuations, and originates
  from the LLCP, coinciding with the $C_P$ maxima line in the vicinity
  of the critical point forming the Widom line. The other $K_T$ maxima
  line corresponds to the approach to the LDL spinodal, which terminates
  at the LDL spinodal at its lowest pressure point, where the TMD line
  also terminates (triangle symbol).  (b) For the case of negatively
  sloped coexistence line, the phase diagram is similar to the mirror
  image of the positively sloped coexistence line case, with $C_P$ and
  $K_T$ maxima lines originating from the LLCP and extending into the
  one-phase region, and overlapping in the vicinity of the critical
  point. The second $K_T$ maxima line goes above the pressure of the
  critical point, and terminates at the HDL spinodal, where the TMD line
  also terminates.  We note that in both cases, both $K_T$ extrema lines
  form a loop in the $T$-$P$ plane, which intersects the TMD line at its
  maximum temperature point of the TMD line (not shown).  In case (a)
  the critical point lies outside the density anomaly region while in
  case (b) it lies inside the density anomaly region.  Both graphs are
  constructed using the linear scaling theory of critical point with
  $\alpha=k=0.47$, and $\phi=25^0$ and $-25^0$, respectively. The lines
  of the spinodals are drawn as interpolation between the critical point
  and the extrapolated crossing point of the TMD and $K_T$ maxima lines
  beyond the coexistence line.  }
\label{PTmax}
\end{figure}

We have investigated the loci of the response function maxima in systems
with different coexistence line slopes.  We find that for the case of
positively sloped coexistence line, the $C_P$ maxima line originates
from the LLCP and extends into the one-phase region as a continuation of
the coexistence line, while compressibility $K_T$ shows two maxima lines
[Fig.~\ref{PTmax}(a)]. One of the $K_T$ maxima lines is related to
critical fluctuations, and originates from the LLCP, coinciding with the
$C_P$ maxima line in the vicinity of the critical point following the
Widom line.  This offers us a method to locate the LLCP from the high
temperature side by tracking the $C_P$ maxima line, instead of tracking
the coexistence line from the low temperature side where crystallization
and glass transition bring huge experimental obstacles.  The other $K_T$
maxima line is due to the approach to the LDL spinodal, and terminates
at the LDL spinodal at its lowest pressure point, where the TMD line
also terminates.
 
\begin{figure}[htbp]
\begin{minipage}[t]{0.4\linewidth}
\includegraphics[width=0.9\linewidth]{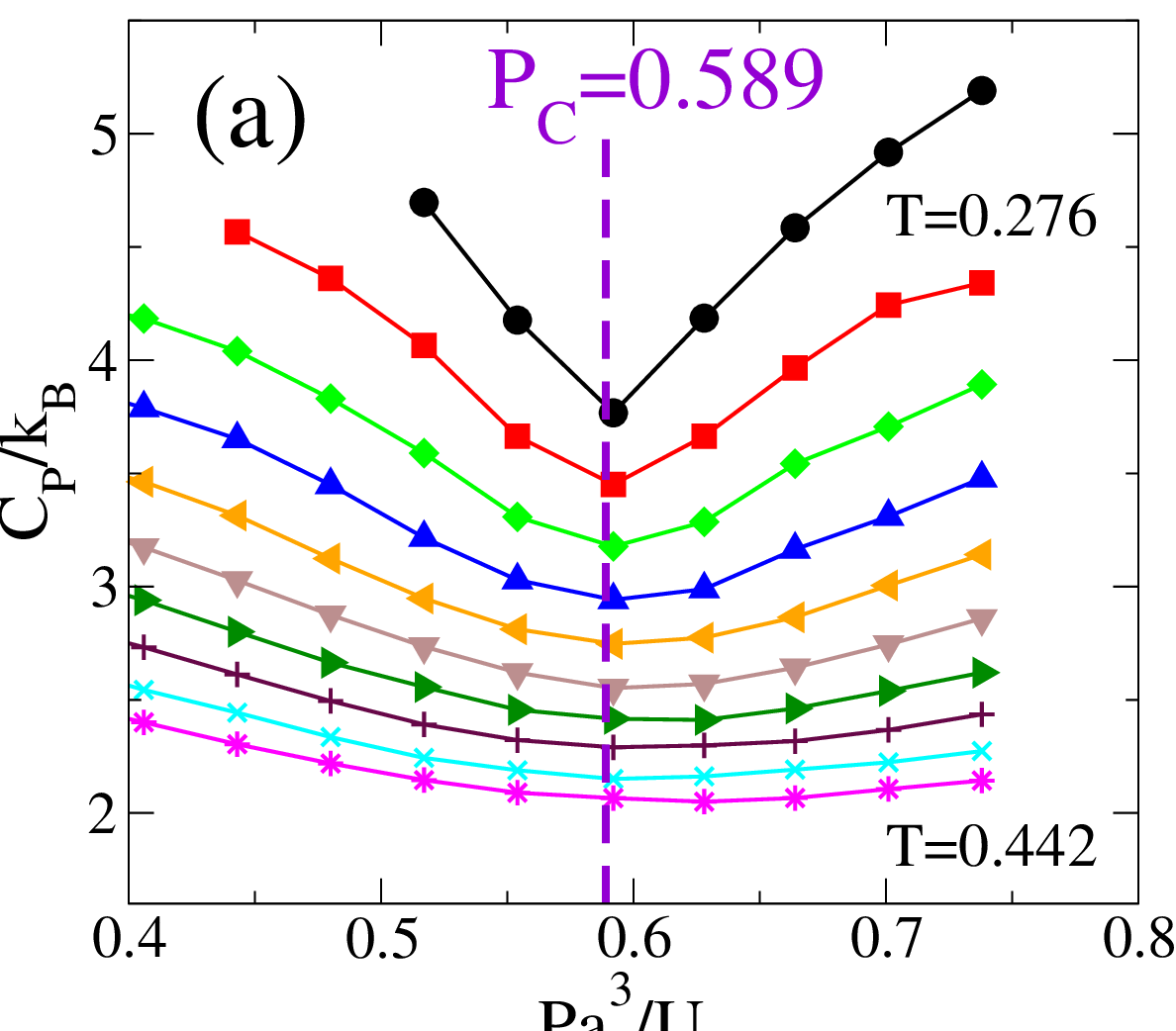}
\end{minipage}
\hspace*{0.1cm}
\begin{minipage}[t]{0.4\linewidth}
\includegraphics[width=0.9\linewidth]{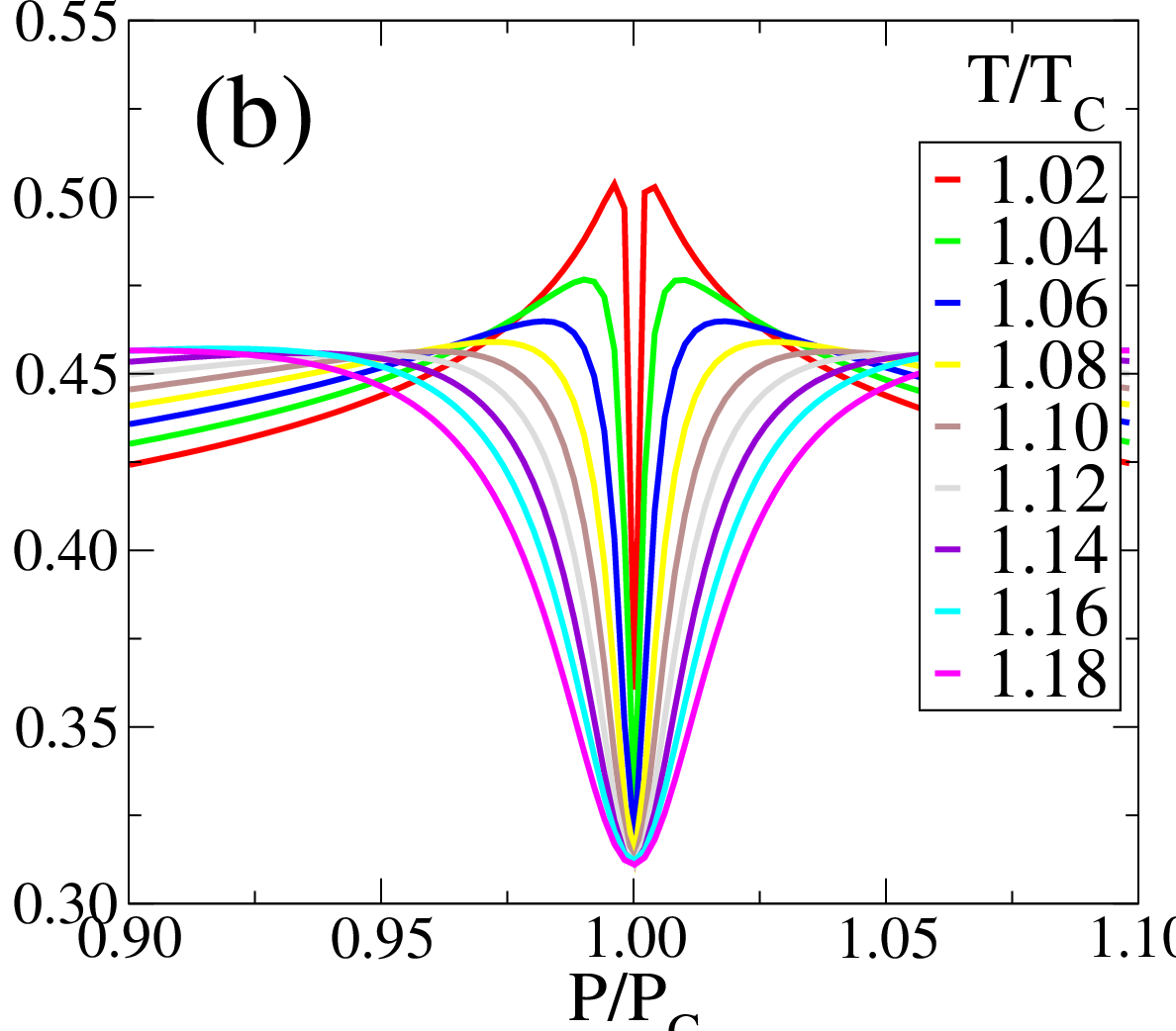}
\end{minipage}\\
\begin{minipage}[t]{0.4\linewidth}
\includegraphics[width=0.9\linewidth]{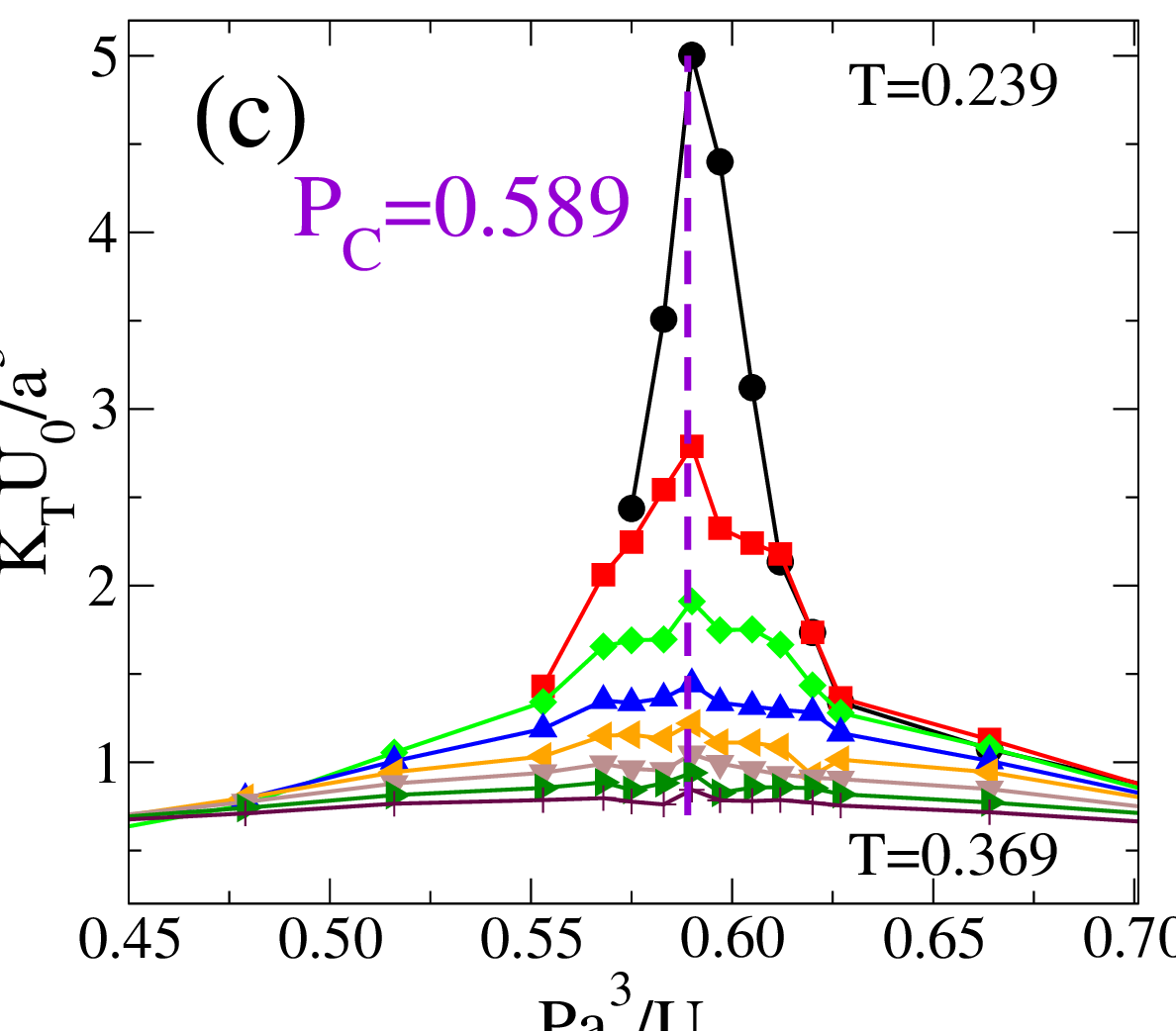}
\end{minipage}
\hspace*{0.1cm}
\begin{minipage}[t]{0.4\linewidth}
\includegraphics[width=0.9\linewidth]{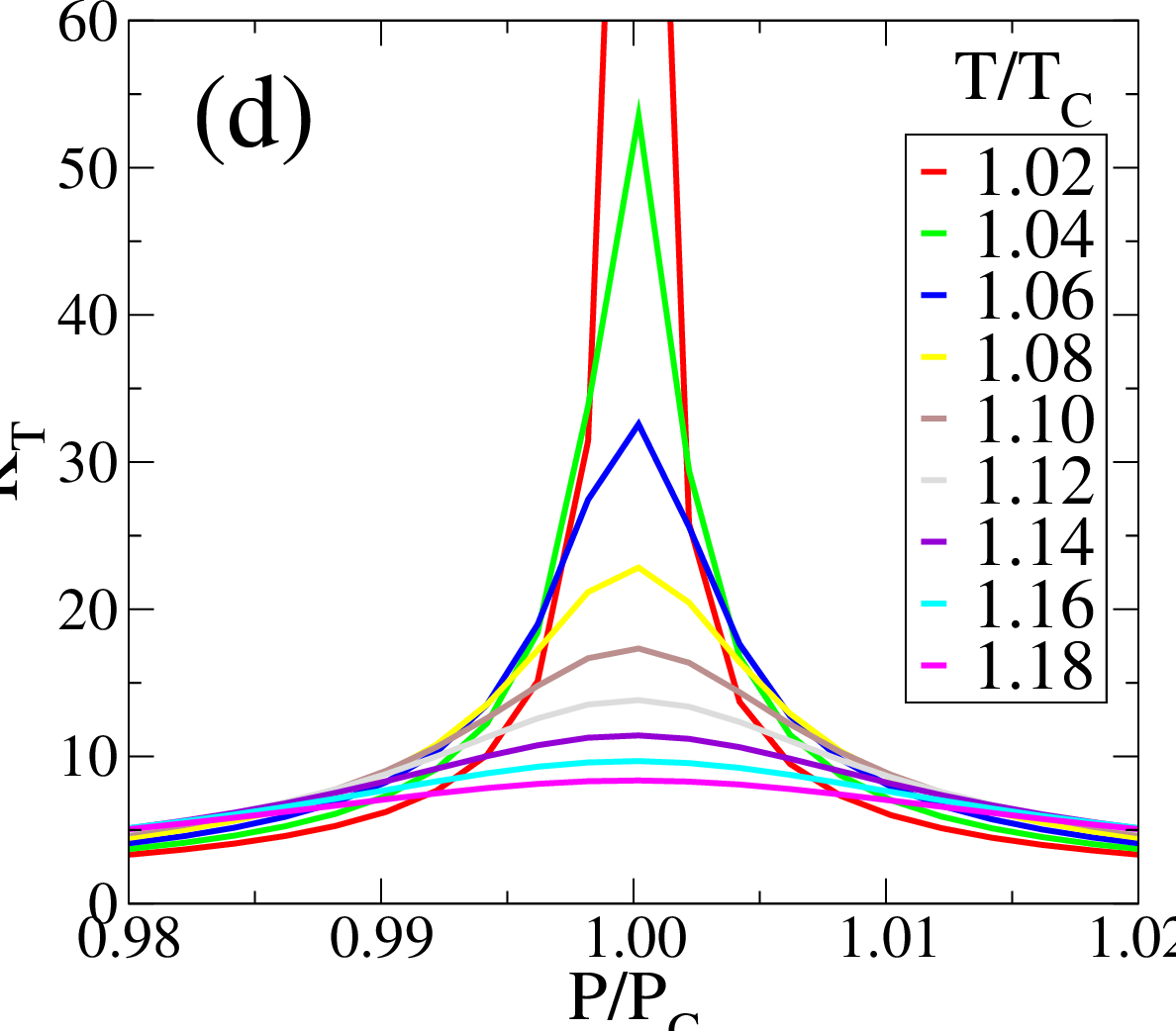}
\end{minipage}
\captionsetup{format=plain,justification=RaggedRight}
\caption{The behavior of $C_P$ and $K_T$ along different isotherms as
  functions of $P$ in simulations with a horizontal coexistence line
  $b/a=1.59$ (a,c), and the linear scaling theory with $\phi=0$ (b,d).
  (a) $C_P$ in simulations, $T$ changes from $0.276$ to $0.442$ with
  interval $\Delta T=0.180$, above $T_c=0.226$. (b) $C_P$ from linear
  scaling theory. Both results agree well with each other, that $C_P$
  shows a minimum at $P_c$ for all temperatures, and as $T \rightarrow
  T_c$, the minimum value of $C_P$ increases, and the valley of the
  minimum gets narrower. This offers a way to track the critical point
  by isotherms at equilibrium temperatures, instead of isobar cooling
  into lower temperature.  (c) $K_T$ in simulations, $T$ changes from
  $0.239$ to $0.369$ with interval $\Delta T=0.180$, above
  $T_c=0.226$. (d) $K_T$ from linear scaling theory. Again, the theory
  is consistent with the simulation results, that showing $K_T$ has
  maxima at $P_c$ for all temperatures and, as $T \rightarrow T_c$, the
  maximum value of $K_T$ increases and the peak gets sharper.  }
\label{CP_r159}
\end{figure}

As the slope of coexistence line approaches zero, $C_P$ maxima
disappears in the equilibrium region with $T \geq T_c$
[Fig.~\ref{PTmax}(b)].  However, along a constant temperature path,
$C_P$ shows a minimum at the critical pressure $P_c$
[Fig.~\ref{CP_r159}]. This is experimentally observed in water for which
$C_P$ decreases with increasing pressure \cite{OguniJCP1983}.  Hence,
for a system with horizontal coexistence line, the LLCP can still be
found by the $C_P$ minimum as function of $P$ at constant $T$.  For
$K_T$, both of the $K_T$ maxima lines as functions of $T$ are related to
critical fluctuations, and start from the LLCP and extend symmetrically
above and below $P_c$.  In addition, a third $K_T$ maxima line as a
function of $P$ at constant $T$ can be defined. All these three $K_T$
maxima lines converge at the LLCP, and together with the $C_P$ minimum
line, can be used to locate the LLCP. Since in case of coexistence line
with zero slope, the thermal field practically consides with temperature, 
this third $K_T$ maxima line gives the best approximation for the Widom line.

For negatively sloped coexistence line, the phase diagram near the
critical point is similar to the phase diagram for the positively sloped
coexistence line case reflected with respect to critical pressure
$P=P_c$.  This follows from the linear scaling theory of the critical
point since according to Eq.~(\ref{ResponseF})
$K_T(-\phi,-\theta)=K_T(\phi,\theta)$,
$C_P(-\phi,-\theta)=C_P(\phi,\theta)$, while
$\alpha_P(-\phi,-\theta)=-\alpha_P(\phi,\theta)$, which comes from the
symmetry of the functions with $\chi_i(\theta)$, $\chi_1(\theta)$, and
$\chi_2(\theta)$ being even, but $\chi_{12}(\theta)$ being odd.  The
$C_P$ maxima line and one of the $K_T$ maxima lines both originate from
the LLCP and extend into the one-phase region, while overlapping with
each other in the vicinity of the critical point
[Fig.~\ref{PTmax}(b)]. The second $K_T$ maxima line goes above the
pressure of the critical point, and terminates at the point of maximum
pressure of the HDL spinodal, where the TMD line also terminates.

Note that the location of the critical point with respect to the TMD
line is related to the slope of the coexistence line.  When the slope of
the coexistence line is positive, the critical point stays outside the
density anomaly region; when it is negative, the critical point is
inside the density anomaly region; when it is horizontal, the TMD line
terminates right at the LLCP [Fig.~\ref{PTmax}].  Indeed, when the slope
is positive, the volume of the low temperature phase is smaller than the
volume of the high temperature phase. Thus if we connect these two
phases by the isobar with $P>P_c$, the volume along this isobar
decreases with $T$, the region above the critical point corresponds to
the $\alpha_P>0$ and, because $\alpha_P$ is a continuous function
everywhere except at the LLCP, $\alpha_P$ also remains positive in the
one-phase region for pressure below the LLCP. Accordingly, the LLCP lies
outside the region of density anomaly.  Analogous considerations show
that, when the slope of the coexistence line is negative, the LLCP
remains inside the density anomaly region, as is surely the case for
water.

We note that, in all three cases, the two $K_T$ maxima lines, merging
with the $K_T$ minima lines, form a loop and cross the TMD line at its
highest temperature point \cite{SastryPRE2006}. 

By changing the parameters of the Jagla interacting potential, we can
obtain systems with different slopes of the LLPT coexistence line.  We
find that, when the slope of the coexistence line is small, the
identification of the Widom line is no longer possible by tracing the
$C_P$ maxima.  As the slope of the coexistence line approaches zero, the
$C_P$ maxima lines become increasingly vertical and, when the slope of
the coexistence line is horizontal, it cannot be observed in
simulations. The study of $C_P$ maxima is best reserved for systems in
which the slope of the coexistence line is strongly positive or
negative. However, the response function maxima in terms of density
fluctuations are still well defined, and it is possible to identify the
Widom line by following the loci of $K_T$ maxima.  These results are in
good agreement with the linear scaling theory.

\section{Acknowledgments}

We thank D. Corradini, and E. Lascaris for helpful discussions.  JL
and HES thank the NSF Chemistry Division for support (grants CHE
0911389 and CHE 0908218). XLM thanks the NSFC grant 11174006 and
2012CB921404 for support. SVB thanks the Office of the Academic
Affairs of Yeshiva University for funding the Yeshiva University
high-performance computer cluster and acknowledges the partial support
of this research through the Dr. Bernard W. Gamson Computational
Science Center at Yeshiva College. CAA acknowledges support from
NSF-CHE Grant No. 0909120.

\newpage

\begin{table}
\begin{center}
\begin{tabular}{c|c|c}
\hline
$\ \ \ \ \ b/a\ \ \ \ \ $ & $\ \ \ \ \ c/a\ \ \ \ \ $ & $\ \ \ U_R/U_0\ \ \ $ 
\\
\hline
\hline
1.72 & 3.000 & 3.478 \\
\hline
1.70 & 2.93 & 3.293 \\
\hline
1.68 & 2.86 & 3.126 \\
\hline
1.65 & 2.76 & 2.906 \\
\hline
1.62 & 2.67 & 2.715 \\
\hline
1.60 & 2.62 & 2.601 \\
%\hline
%1.595 & 2.60 & 2.574 \\
\hline
1.59 & 2.59 & 2.547 \\
\hline
\end{tabular}
\caption{Renormalized parameters for modified Jagla potential \cite{gibsonWilding}.}
\label{TableModel}
\end{center}
\end{table}

\newpage

\newpage

\newpage

\newpage

\end{document}